%% file: main.tex
\pdfoutput=1

\documentclass[12pt,a4paper]{article}

\usepackage{ifthen} 
\usepackage{rotating}
\usepackage{dashrule}
\usepackage{array} 
\usepackage{dcolumn}
\usepackage{comment}
\usepackage{xcolor}

\usepackage{graphpap} 
\usepackage{rotating} 
\usepackage{tikz}
\usepackage{xcolor}
\usepackage{longtable} 

\usepackage{lscape}
\usepackage{multirow} 
\usepackage{mathrsfs}
\usepackage{mathtools} 
\usetikzlibrary{patterns}
\usepackage{nicefrac} 
\usepackage{transparent}
\usepackage{afterpage} 
\usepackage{comment} 
\usepackage[normalem]{ulem}
\usepackage{cancel} 

\newboolean{pdflatex}
\setboolean{pdflatex}{true} 

\newboolean{articletitles}
\setboolean{articletitles}{true} 

\newboolean{uprightparticles}
\setboolean{uprightparticles}{True} 

\def\paperauthors{LHCb collaboration} 
\def\paperasciititle{Study of BsjpsipipiKK decays} 
\def\papertitle{Study of \mbox{$\decay{\Bs}{\jpsi\pip\pim\Kp\Km}$}~decays} 
\def\paperkeywords{{High Energy Physics}, {LHCb}} 
\def\papercopyright{\the\year\ CERN for the benefit of the LHCb collaboration} 
\def\paperlicence{CC BY 4.0 licence}
\def\paperlicenceurl{https://creativecommons.org/licenses/by/4.0/}

\makeatletter
\g@addto@macro\bfseries{\boldmath}
\makeatother

\input{preamble}
\usepackage{longtable} 

\newcolumntype{d}[1]{D{,}{\,\pm\,}{#1} }
\newcolumntype{f}[1]{D{,}{.}{#1} }

\usepackage{tikz}

\begin{document}

\renewcommand{\thefootnote}{\fnsymbol{footnote}}
\setcounter{footnote}{1}

\input{title-LHCb-PAPER}


\renewcommand{\thefootnote}{\arabic{footnote}}
\setcounter{footnote}{0}



\pagestyle{plain} 
\setcounter{page}{1}
\pagenumbering{arabic}


%


\input{introduction}

\input{detector}
\input{selection}

\input{xcc}

\input{XKK}

\input{kstar}

\input{mass}

\input{Xpeak}

\input{systematics}

\input{results}

\input{acknowledgements}





\usetikzlibrary{patterns}

\clearpage
\addcontentsline{toc}{section}{References}
\bibliographystyle{LHCb}
\bibliography{main,standard,LHCb-PAPER,LHCb-CONF,LHCb-DP,LHCb-TDR}
 

\newpage
\input{LHCb_Authorship_07-Sep-2020}

\end{document}

%% file: preamble.tex
\usepackage[top=1in, bottom=1.25in, left=1in, right=1in]{geometry}

\usepackage{tikz}

\usepackage{microtype}
\usepackage{lineno}  
\usepackage{xspace} 
\usepackage{caption} 

\usepackage{graphicx}  
\usepackage{color}
\usepackage{colortbl}
\graphicspath{{./figs/}} 
\DeclareGraphicsExtensions{.pdf,.PDF,png,.PNG}

\usepackage{amsmath} 
\usepackage{amssymb}
\usepackage{amsfonts}
\usepackage{upgreek} 

\newcommand*\patchAmsMathEnvironmentForLineno[1]{%
\expandafter\let\csname old#1\expandafter\endcsname\csname #1\endcsname
\expandafter\let\csname oldend#1\expandafter\endcsname\csname
end#1\endcsname
 \renewenvironment{#1}%
   {\linenomath\csname old#1\endcsname}%
   {\csname oldend#1\endcsname\endlinenomath}%
}
\newcommand*\patchBothAmsMathEnvironmentsForLineno[1]{%
  \patchAmsMathEnvironmentForLineno{#1}%
  \patchAmsMathEnvironmentForLineno{#1*}%
}
\AtBeginDocument{%
\patchBothAmsMathEnvironmentsForLineno{equation}%
\patchBothAmsMathEnvironmentsForLineno{align}%
\patchBothAmsMathEnvironmentsForLineno{flalign}%
\patchBothAmsMathEnvironmentsForLineno{alignat}%
\patchBothAmsMathEnvironmentsForLineno{gather}%
\patchBothAmsMathEnvironmentsForLineno{multline}%
\patchBothAmsMathEnvironmentsForLineno{eqnarray}%
}

\usepackage{hyperxmp}

\usepackage[pdftex,
            pdfauthor={\paperauthors},
            pdftitle={\paperasciititle},
            pdfkeywords={\paperkeywords},
            pdfcopyright={Copyright (C) \papercopyright},
            pdflicenseurl={\paperlicenceurl}]{hyperref}

\usepackage[all]{hypcap} 

\input{lhcb-symbols-def} 

\input{local-symbols-list} 

\usepackage{cite} 
\usepackage{mciteplus}

%% file: lhcb-symbols-def.tex
\usepackage{xspace} 
\usepackage{upgreek}

\def\lhcb   {\mbox{LHCb}\xspace}

\def\MagUp {\mbox{\em Mag\kern -0.05em Up}\xspace}

\ifthenelse{\boolean{uprightparticles}}%
{

 \def\Peta        {\ensuremath{\upeta}\xspace}

 \def\Pmu         {\ensuremath{\upmu}\xspace}

 \def\Ppi         {\ensuremath{\uppi}\xspace}                 
                  
 \def\Prho        {\ensuremath{\uprho}\xspace}

 \def\Pphi        {\ensuremath{\upphi}\xspace}                 
                  
 \def\Pchi        {\ensuremath{\upchi}\xspace}                 
 \def\Ppsi        {\ensuremath{\uppsi}\xspace}

 \def\PDelta      {\ensuremath{\Delta}\xspace}                 
 \def\PXi         {\ensuremath{\Xi}\xspace}                 
 \def\PLambda     {\ensuremath{\Lambda}\xspace}                 
 \def\PSigma      {\ensuremath{\Sigma}\xspace}                 
 \def\POmega      {\ensuremath{\Omega}\xspace}                 
 \def\PUpsilon    {\ensuremath{\Upsilon}\xspace}

 \def\PB      {\ensuremath{\mathrm{B}}\xspace}                 
                  
 \def\PD      {\ensuremath{\mathrm{D}}\xspace}

 \def\PJ      {\ensuremath{\mathrm{J}}\xspace}                 
 \def\PK      {\ensuremath{\mathrm{K}}\xspace}

 \def\PS      {\ensuremath{\mathrm{S}}\xspace}

 \def\PX      {\ensuremath{\mathrm{X}}\xspace}

 \def\Pb      {\ensuremath{\mathrm{b}}\xspace}                 
 \def\Pc      {\ensuremath{\mathrm{c}}\xspace}

 \def\Pi      {\ensuremath{\mathrm{i}}\xspace}

 \def\Pp      {\ensuremath{\mathrm{p}}\xspace}

 \def\Ps      {\ensuremath{\mathrm{s}}\xspace}

 \def\thebaroffset{0.0em}
}
{

 \def\Peta        {\ensuremath{\eta}\xspace}

 \def\Pmu         {\ensuremath{\mu}\xspace}

 \def\Ppi         {\ensuremath{\pi}\xspace}                 
                  
 \def\Prho        {\ensuremath{\rho}\xspace}

 \def\Pphi        {\ensuremath{\phi}\xspace}                 
                  
 \def\Pchi        {\ensuremath{\chi}\xspace}                 
 \def\Ppsi        {\ensuremath{\psi}\xspace}                 
                  
 \mathchardef\PDelta="7101
 \mathchardef\PXi="7104
 \mathchardef\PLambda="7103
 \mathchardef\PSigma="7106
 \mathchardef\POmega="710A
 \mathchardef\PUpsilon="7107
                  
 \def\PB      {\ensuremath{B}\xspace}                 
                  
 \def\PD      {\ensuremath{D}\xspace}

 \def\PJ      {\ensuremath{J}\xspace}                 
 \def\PK      {\ensuremath{K}\xspace}

 \def\PS      {\ensuremath{S}\xspace}

 \def\PX      {\ensuremath{X}\xspace}

 \def\Pb      {\ensuremath{b}\xspace}                 
 \def\Pc      {\ensuremath{c}\xspace}

 \def\Pi      {\ensuremath{i}\xspace}

 \def\Pp      {\ensuremath{p}\xspace}

 \def\Ps      {\ensuremath{s}\xspace}

 \def\thebaroffset{0.18em}
}
\newcommand{\offsetoverline}[2][\thebaroffset]{\kern #1\overline{\kern -#1 #2}}%

\makeatletter
\ifcase \@ptsize \relax
  \newcommand{\miniscule}{\@setfontsize\miniscule{4}{5}}
\or
  \newcommand{\miniscule}{\@setfontsize\miniscule{5}{6}}
\or
  \newcommand{\miniscule}{\@setfontsize\miniscule{5}{6}}
\fi
\makeatother

\DeclareRobustCommand{\optbar}[1]{\shortstack{{\miniscule (\rule[.5ex]{1.25em}{.18mm})}
  \\ [-.7ex] $#1$}}

\def\mup        {{\ensuremath{\Pmu^+}}\xspace}
\def\mun        {{\ensuremath{\Pmu^-}}\xspace} 

\def\mumu       {{\ensuremath{\Pmu^+\Pmu^-}}\xspace}

\def\squark    {{\ensuremath{\Ps}}\xspace}
\def\squarkbar {{\ensuremath{\overline \squark}}\xspace}

\def\cquark    {{\ensuremath{\Pc}}\xspace}
\def\cquarkbar {{\ensuremath{\overline \cquark}}\xspace}
\def\ccbar     {{\ensuremath{\cquark\cquarkbar}}\xspace}
\def\bquark    {{\ensuremath{\Pb}}\xspace}

\def\pion   {{\ensuremath{\Ppi}}\xspace}

\def\pip    {{\ensuremath{\pion^+}}\xspace}
\def\pim    {{\ensuremath{\pion^-}}\xspace}

\def\kaon    {{\ensuremath{\PK}}\xspace}
\def\Kbar    {{\ensuremath{\offsetoverline{\PK}}}\xspace}

\def\KorKbar {\kern \thebaroffset\optbar{\kern -\thebaroffset \PK}{}\xspace}
\def\Kz      {{\ensuremath{\kaon^0}}\xspace}

\def\Kp      {{\ensuremath{\kaon^+}}\xspace}
\def\Km      {{\ensuremath{\kaon^-}}\xspace}

\def\KS      {{\ensuremath{\kaon^0_{\mathrm{S}}}}\xspace}

\def\Kstarz  {{\ensuremath{\kaon^{*0}}}\xspace}
\def\Kstarzb {{\ensuremath{\Kbar{}^{*0}}}\xspace}
\def\Kstar   {{\ensuremath{\kaon^*}}\xspace}

\def\Dbar    {{\ensuremath{\offsetoverline{\PD}}}\xspace}
\def\D       {{\ensuremath{\PD}}\xspace}

\def\DorDbar {\kern \thebaroffset\optbar{\kern -\thebaroffset \PD}\xspace}
\def\Dz      {{\ensuremath{\D^0}}\xspace}

\def\Dp      {{\ensuremath{\D^+}}\xspace}
\def\Dm      {{\ensuremath{\D^-}}\xspace}

\def\DpDm    {\ensuremath{\Dp {\kern -0.16em \Dm}}\xspace}
\def\Dstar   {{\ensuremath{\D^*}}\xspace}

\def\Dstarzb {{\ensuremath{\Dbar{}^{*0}}}\xspace}

\def\Dstarp  {{\ensuremath{\D^{*+}}}\xspace}

\def\Ds      {{\ensuremath{\D^+_\squark}}\xspace}

\def\B       {{\ensuremath{\PB}}\xspace}

\def\BorBbar {\kern \thebaroffset\optbar{\kern -\thebaroffset \PB}\xspace}

\def\Bd      {{\ensuremath{\B^0}}\xspace}

\def\BdorBdbar {\kern \thebaroffset\optbar{\kern -\thebaroffset \Bd}\xspace}
\def\Bu      {{\ensuremath{\B^+}}\xspace}

\def\Bp      {{\ensuremath{\Bu}}\xspace}

\def\Bs      {{\ensuremath{\B^0_\squark}}\xspace}

\def\BsorBsbar {\kern \thebaroffset\optbar{\kern -\thebaroffset \Bs}\xspace}

\def\jpsi     {{\ensuremath{{\PJ\mskip -3mu/\mskip -2mu\Ppsi}}}\xspace}
\def\psitwos  {{\ensuremath{\Ppsi{(2\PS)}}}\xspace}

\def\chiczero {{\ensuremath{\Pchi_{\cquark 0}}}\xspace}
\def\chicone  {{\ensuremath{\Pchi_{\cquark 1}}}\xspace}
\def\chictwo  {{\ensuremath{\Pchi_{\cquark 2}}}\xspace}

\def\Y#1S{\ensuremath{\PUpsilon{(#1S)}}\xspace}

\def\proton      {{\ensuremath{\Pp}}\xspace}
\def\antiproton  {{\ensuremath{\overline \proton}}\xspace}

\def\Lz          {{\ensuremath{\PLambda}}\xspace}

\def\LorLbar     {\kern \thebaroffset\optbar{\kern -\thebaroffset \PLambda}\xspace}

\def\Lb           {{\ensuremath{\Lz^0_\bquark}}\xspace}

\def\BF         {{\ensuremath{\mathcal{B}}}\xspace}
\def\BR         {\BF}

\newcommand{\decay}[2]{\ensuremath{#1\!\to #2}\xspace} 

\def\to                 {\ensuremath{\rightarrow}\xspace}

\def\CP                {{\ensuremath{C\!P}}\xspace}

\def\AT#1     {\ensuremath{A_{\mathrm{T}}^{#1}}\xspace}           

\def\C#1      {\ensuremath{\mathcal{C}_{#1}}\xspace}                       
\def\Cp#1     {\ensuremath{\mathcal{C}_{#1}^{'}}\xspace}                    
\def\Ceff#1   {\ensuremath{\mathcal{C}_{#1}^{\mathrm{(eff)}}}\xspace}        
\def\Cpeff#1  {\ensuremath{\mathcal{C}_{#1}^{'\mathrm{(eff)}}}\xspace}       
\def\Ope#1    {\ensuremath{\mathcal{O}_{#1}}\xspace}                       
\def\Opep#1   {\ensuremath{\mathcal{O}_{#1}^{'}}\xspace}                    

\newcommand{\nospaceunit}[1]{\ensuremath{\text{#1}}}       
\newcommand{\aunit}[1]{\ensuremath{\text{\,#1}}}       

\newcommand{\tev}{\aunit{Te\kern -0.1em V}\xspace}
\newcommand{\gev}{\aunit{Ge\kern -0.1em V}\xspace}
\newcommand{\mev}{\aunit{Me\kern -0.1em V}\xspace}
\newcommand{\kev}{\aunit{ke\kern -0.1em V}\xspace}
\newcommand{\ev}{\aunit{e\kern -0.1em V}\xspace}
 
\newcommand{\mevc}{\ensuremath{\aunit{Me\kern -0.1em V\!/}c}\xspace}
\newcommand{\gevc}{\ensuremath{\aunit{Ge\kern -0.1em V\!/}c}\xspace}
\newcommand{\mevcc}{\ensuremath{\aunit{Me\kern -0.1em V\!/}c^2}\xspace}
\newcommand{\gevcc}{\ensuremath{\aunit{Ge\kern -0.1em V\!/}c^2}\xspace}

\def\mm   {\aunit{mm}\xspace}

\def\mum  {\ensuremath{\,\upmu\nospaceunit{m}}\xspace}

\def\fb   {\ensuremath{\aunit{fb}}\xspace}
\def\invfb   {\ensuremath{\fb^{-1}}\xspace}

\def\gsim{{~\raise.15em\hbox{$>$}\kern-.85em
          \lower.35em\hbox{$\sim$}~}\xspace}
\def\lsim{{~\raise.15em\hbox{$<$}\kern-.85em
          \lower.35em\hbox{$\sim$}~}\xspace}

\def\sPlot{\mbox{\em sPlot}\xspace}

\def\pt         {\ensuremath{p_{\mathrm{T}}}\xspace}

\def\evtgen     {\mbox{\textsc{EvtGen}}\xspace}

\def\geant      {\mbox{\textsc{Geant4}}\xspace}

\def\photos     {\mbox{\textsc{Photos}}\xspace}

\def\pythia     {\mbox{\textsc{Pythia}}\xspace}

\def\tell1  {TELL1\xspace}
\def\ukl1   {UKL1\xspace}

\newcommand{\ie}{\mbox{\itshape i.e.}\xspace}



%% file: local-symbols-list.tex
\def\BsTopsitwophi     {\decay{\Bs}{\psitwos \Pphi}}

\def\BsTochiophi     {\decay{\Bs}{\chicone(3872) \Pphi}}
\def\BsTokstkst     {\decay{\Bs}{\jpsi \Kstarz \Kstarzb}}

\def\KstKpi {\decay{\Kstarz}{\Kp \pim}}
\def\psipipi {\decay{\psitwos}{\jpsi \pip \pim}}

\def\chipipi {\decay{\chio}{\jpsi \pip \pim}}

\def\BsTopsiphtwopi     {\decay{\Bs}{\jpsi \pip \pim  \Pphi}}

\def\chio     {\chicone(3872)}

\def\fsX {\jpsi \Pphi}

\newcommand{\kevcc}{\ensuremath{\aunit{ke\kern -0.1em V\!/}c^2}\xspace}

\def\chiconex  {\ensuremath{\chicone(3872)}\xspace}

\newcommand{\Rchi} {\ensuremath{\mathcal{R}^{\chicone(3872)\Pphi}_{\psitwos\Pphi}}\xspace}

\newcommand{\Rkst}{\ensuremath{\mathcal{R}^{\jpsi\Kstarz\Kstarzb}_{\psitwos\Pphi}}\xspace}

\newcommand{\Rnonphi}{\ensuremath{\mathcal{R}_{\Kp\Km }}\xspace}

\usepackage{boldline}
\usepackage{bm}

%% file: title-LHCb-PAPER.tex

\begin{titlepage}
\pagenumbering{roman}

\vspace*{-1.5cm}
\centerline{\large EUROPEAN ORGANIZATION FOR NUCLEAR RESEARCH (CERN)}
\vspace*{1.5cm}
\noindent
\begin{tabular*}{\linewidth}{lc@{\extracolsep{\fill}}r@{\extracolsep{0pt}}}
\ifthenelse{\boolean{pdflatex}}
{\vspace*{-1.5cm}\mbox{\!\!\!\includegraphics[width=.14\textwidth]{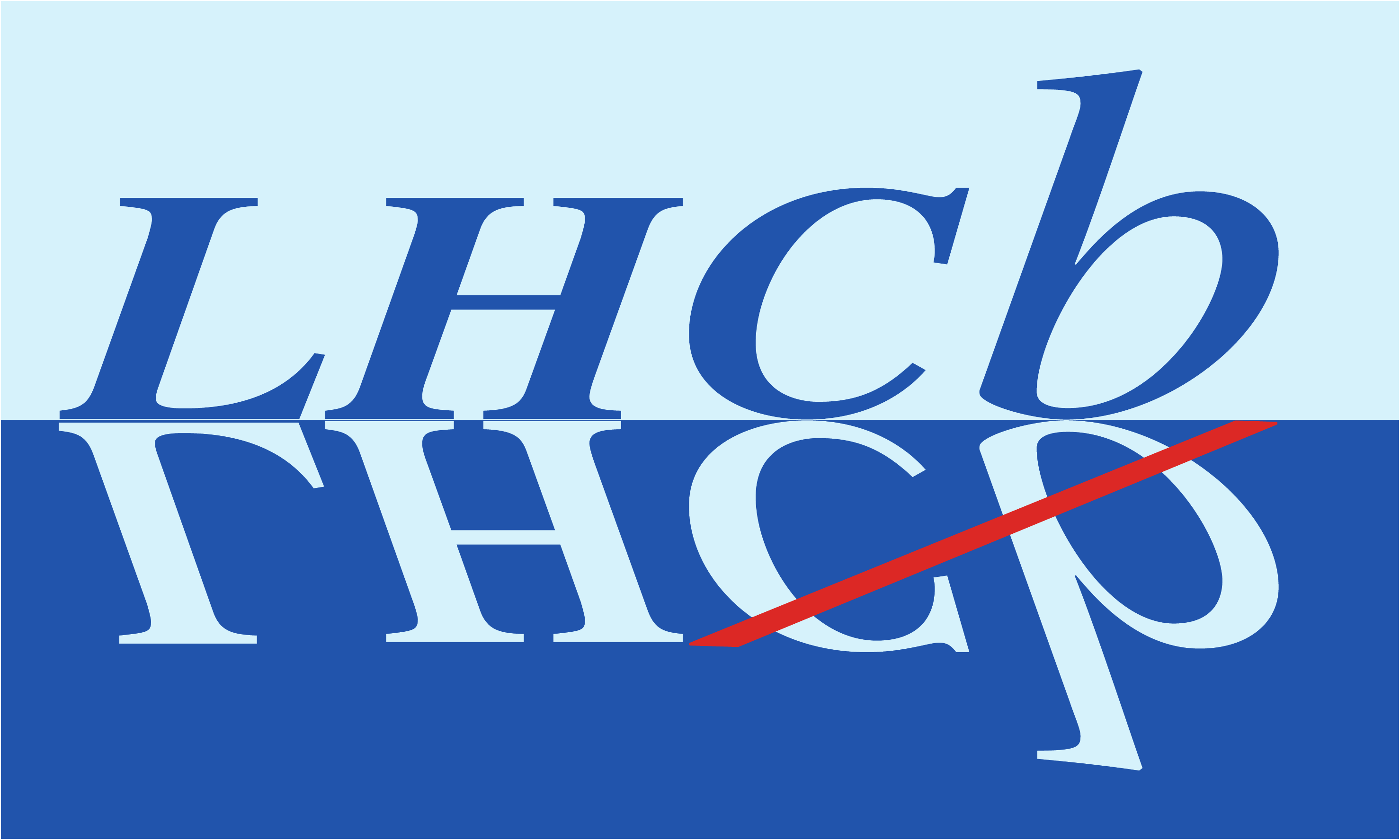}} & &}%
{\vspace*{-1.2cm}\mbox{\!\!\!\includegraphics[width=.12\textwidth]{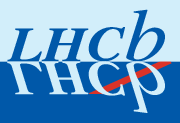}} & &}%
\\
 & & CERN-EP-2020-192\\  
 & & LHCb-PAPER-2020-035 \\  
  & & November 3, 2020 \\ 
\end{tabular*}

\vspace*{0.2cm}

{\normalfont\bfseries\boldmath\huge
\begin{center}
  \papertitle 
\end{center}
}

\vspace*{0.2cm}

\begin{center}
\paperauthors\footnote{Authors are listed at the end of this paper.}
\end{center}

\vspace{\fill}

\begin{abstract}
\noindent 
The decays $\decay{\Bs}{\jpsi\pip\pim\Kp\Km}$ are studied
using  a~data set corresponding 
to an~integrated luminosity of 9\invfb, 
collected with the~LHCb detector 
in proton\nobreakdash-proton collisions
at centre\nobreakdash-of\nobreakdash-mass 
energies of 7, 8 and 13\tev.
The~decays
\mbox{$\decay{\Bs}{\jpsi\Kstarz\Kstarzb}$}
and 
\mbox{$\decay{\Bs}
{\chicone(3872)}\Kp\Km$},
where the~$\Kp\Km$~pair does not originate from 
a~\Pphi~meson, 
are 
observed for the~first time. 
Precise measurements of the~ratios of branching fractions 
between intermediate 
$\chicone(3872)\Pphi$,
$\jpsi\Kstarz\Kstarzb$,
$\psitwos\Pphi$ and 
$\chicone(3872)\Kp\Km$~states
are  reported.
A~structure, denoted
as $\PX(4740)$,  
is observed  in the~$\jpsi\Pphi$~mass spectrum
and, assuming a~Breit\nobreakdash--Wigner parameterisation, 
its~mass and width are determined to be  
\begin{eqnarray*}
m_{\PX(4740)}      
&  = &  4741 \pm 6  \phantom{0}\pm  6  \phantom{0}\mevcc \,, \\
\Gamma_{\PX(4740)} 
& = &  \phantom{00}53 \pm  15  \pm  11   \mev \,,
\end{eqnarray*}
where the~first uncertainty is statistical 
and the~second is  systematic.
In~addition, 
the~most precise single measurement of  
the~mass of the~\Bs~meson is performed 
and gives a~value of
\begin{equation*}
    m_{\Bs} = 5366.98 \pm 0.07 \pm 0.13 \mevcc\,.
\end{equation*}

\end{abstract}

\vspace*{0.5cm}

\begin{center}
  Published in JHEP 02 (2021) 024 
\end{center}

\vspace{\fill}

{\footnotesize 
\centerline{\copyright~\papercopyright. \href{\paperlicenceurl}{\paperlicence}.}}
\vspace*{2mm}

\end{titlepage}


\newpage
\setcounter{page}{2}
\mbox{~}
%

\cleardoublepage

%% file: introduction.tex
\section{Introduction}
\label{sec:Introduction}

Decays of beauty hadrons to final states with charmonia
provide a~unique laboratory to study 
the~properties of charmonia and 
charmonium\nobreakdash-like states. 
A~plethora of new states has been observed in such 
decays, 
including the~$\chicone(3872)$~particle~\cite{Choi:2003ue},
pentaquark~\cite{LHCb-PAPER-2015-029,
LHCb-PAPER-2016-009,
LHCb-PAPER-2019-014,
LHCb-PAPER-2016-015}
and numerous
tetraquark~\cite{
Choi:2007wga,
Mizuk:2009da,
Chilikin:2013tch,
LHCb-PAPER-2014-014,
LHCb-PAPER-2015-038,
LHCb-PAPER-2016-015,
LHCb-PAPER-2016-018,
LHCb-PAPER-2016-019,
LHCb-PAPER-2018-034,
LHCb-PAPER-2018-043}
candidates 
as well as conventional charmonium states, such as 
the~tensor D\nobreakdash-wave 
$\Ppsi_2(3823)$~meson~\cite{Bhardwaj:2013rmw,
LHCb-PAPER-2020-009}. 
The~nature of many exotic charmonium\nobreakdash-like  candidates 
remains unclear. 
A~comparison of production rates 
with respect to those  
of conventional  charmonium states 
in decays 
of beauty hadrons can shed light on 
their~production mechanisms~\cite{Maiani:2017kyi}.
For~example, 
the~$\Dstar\Dbar$~rescattering 
mechanism~\cite{Artoisenet:2010va,Braaten:2019yua}
would give a~large contribution 
to the~$\chicone(3872)$~production
and affect the~pattern of decay 
rates of beauty hadrons. 
A~modified pattern 
is also expected 
for 
a~compact\nobreakdash-tetraquark 
interpretation
of the~\mbox{$\chicone(3872)$}~state~\cite{Maiani:2020zhr}.

The decay  
chain $\decay{\Bs}{\left( \decay{\chicone(3872)}{\jpsi\pip\pim}\right) 
\left(\decay{\Pphi}{\Kp\Km}\right)}$
is experimentally 
easiest to study in (quasi)\,two\nobreakdash-body decays
of a~\Bs~meson with a~\chiconex~particle  in the~final state.
This decay has recently been studied by 
the~CMS collaboration, 
which found the~ratio of branching fractions 
for the~ $\decay{\Bs}{\chicone(3872)\Pphi}$ 
and  $\decay{\Bd}{\chicone(3872)\Kz}$ decays  
to be compatible with  unity, and two times smaller     than   the~ratio of 
branching   fractions for $\decay{\Bs}{\chicone(3872)\Pphi}$ 
and  
$\decay{\Bu}{\chicone(3872)\Kp}$~decays~\cite{Sirunyan:2020qir}. 

The~decay of the~\Bs~meson into 
the~${\jpsi\pip\pim\Kp\Km}$~final state
allows the~mass spectrum of the~$\jpsi\Pphi$~system
to  be studied. Four  tetraquark candidates have been observed by the~LHCb collaboration
using an~amplitude analysis of~\mbox{$\decay{\Bu}{\jpsi\Pphi}\Kp$}~decays~\cite{LHCb-PAPER-2016-019}. 
These states are denoted  by the~PDG as~$\chicone(4140)$, 
$\chicone(4274)$,
$\chiczero(4500)$ and 
$\chiczero(4700)$~\cite{PDG2020}.
In~\mbox{$\decay{\Bs}{\jpsi\pip\pim\Pphi}$}~decays, the~$\jpsi\Pphi$ mass can be probed 
up to approximately 300\mevcc above 
the~allowed kinematic limit in~\mbox{$\decay{\Bu}{\jpsi\Pphi}\Kp$}~decays. 

The $\decay{\Bs}{\jpsi\Pphi\Pphi}$ decay has been observed  
by the~LHCb collaboration~\cite{LHCb-PAPER-2015-033} using 
a~data set collected in 2011\nobreakdash--2012 LHC 
data taking. 
While   the~energy  release  
in this decay
is small, 
the~measured branching fraction is large,
possibly indicating 
a~non\nobreakdash-trivial decay 
dynamics that enhances the~decay rate.
A~comparison 
of the~rate of 
the~$\decay{\Bs}{\jpsi\Pphi\Pphi}$~decay with that of  $\decay{\Bs}{\jpsi\Peta^{\prime}\Pphi}$,
$\decay{\Bs}{\jpsi\Peta^{\prime}\Peta^{\prime}}$ and 
$\decay{\Bs}{\jpsi\Kstarz\Kstarzb}$ decay modes 
could clarify the~dynamics: 
the~first two modes have similar quark content
while the~third is formed by three vector mesons
and results in the~$\jpsi\Kp\Km\pip\pim$~final state.

In this paper, a~sample of $\decay{\Bs}{\jpsi\pip\pim\Kp\Km}$~decays is analysed, 
with the~$\jpsi$~meson   reconstructed in the~$\mup\mun$~final state.
The~study is based on proton-proton\,($\proton\proton$) collision   data, 
corresponding to 
integrated luminosities of 1, 2 and 6\invfb, 
collected with the~LHCb detector at centre\nobreakdash-of\nobreakdash-mass  
energies of 7, 8 and 13\tev, respectively. 
This~data sample  is used to measure 
the~rates of the~$\decay{\Bs}{ \chicone(3872) \Pphi}$,
$\decay{\Bs}{\jpsi\Kstarz\Kstarzb}$ decays,
where $\Kstarz$~denotes the~\mbox{$\kaon^{\ast}(892)^0$}~resonance,
and  
\mbox{$\decay{\Bs}{ \chicone(3872)  \Kp\Km}$}~decays,
where 
the~$\Kp\Km$~pair
does not originate from a $\Pphi$~meson.
The~presence of~\mbox{$\decay{\Bs}{ \left( \decay{\psitwos}{\jpsi\pip\pim} \right) \left( 
\decay{\Pphi}{\Kp\Km}\right) }$}~decays in the~same sample 
provides a~convenient mode for normalising the~observed rates
of the~different final states since the~branching fraction 
of this~decay is known~\cite{PDG2020}. 
This~paper presents measurements 
of the~following ratios of branching fractions\,($\BR$), 
\begin{subequations}\label{eq:r}
\begin{eqnarray}
\Rchi 
& \equiv  &  
\dfrac{ \BR\left(\BsTochiophi\right) \times \BR\left(\chipipi\right)}
{\BR\left(\BsTopsitwophi\right) \times \BR\left(\psipipi\right)}\,, \label{eq:r_chi} \\ 
\Rkst 
& \equiv & 
 \dfrac{\BR\left(\BsTokstkst\right) \times \left(\BR\left(\KstKpi\right)\right)^2}
 {\BR\left(\BsTopsitwophi\right) \times 
 \BR\left(\psipipi\right)
 \times \BR\left(\decay{\Pphi}{\Kp\Km}\right) 
 }\,, \label{eq:r_kstar} \\ 
\Rnonphi 
 & \equiv & 
\dfrac
{\BR(\decay{\Bs}{ \chicone(3872) \left( \Kp\Km\right)_{\mathrm{\text{non-}}{\Pphi}} }) }
{\BR\left(\decay{\Bs}{ \chicone(3872)  \Pphi }\right)
\times \BR\left( \decay{\Pphi}{\Kp\Km}\right)}\,. \label{eq:rnonphi}
\end{eqnarray}
\end{subequations}
The~$\jpsi\Pphi$~mass spectrum 
from~\mbox{$\decay{\Bs}{\jpsi\pip\pim\Pphi}$}~decays
is investigated to search for resonant contributions. 
The~large size of the analysed sample and the low level of background also  
allows for  a~precise determination of the~mass of~the~\Bs~meson. 
The~mass is measured using a~subsample
enriched in~\mbox{$\decay{\Bs}{\psitwos\Pphi}$}~decays, 
which have a~small energy release.

%% file: detector.tex
\section{Detector and simulation}
\label{sec:Detector}

The \lhcb detector~\cite{LHCb-DP-2008-001,LHCb-DP-2014-002} is a single-arm forward
spectrometer covering the~pseudorapidity range $2<\eta <5$,
designed for the study of particles containing $\bquark$~or~$\cquark$~quarks. 
The~detector includes a high-precision tracking system consisting of a 
silicon-strip vertex detector surrounding the \proton\proton interaction
region~\cite{LHCb-DP-2014-001}, a large-area silicon-strip detector located
upstream of a dipole magnet with a bending power of about $4{\mathrm{\,Tm}}$,
and three stations of silicon-strip detectors and straw drift tubes~\cite{LHCb-DP-2013-003,LHCb-DP-2017-001} placed downstream of the magnet. 
The tracking system provides a measurement of the momentum of charged particles
with a relative uncertainty that varies from $0.5\%$ at low momentum to $1.0\%$~at~$200 \gevc$. 
The~momentum scale is calibrated using samples of $\decay{\jpsi}{\mumu}$ 
and $\decay{\Bu}{\jpsi\Kp}$~decays collected concurrently
with the~data sample used for this analysis~\cite{LHCb-PAPER-2012-048,LHCb-PAPER-2013-011}. 
The~relative accuracy of this
procedure is estimated to be $3 \times 10^{-4}$ using samples of other
fully reconstructed $\bquark$~hadrons, $\PUpsilon$~and
$\KS$~mesons.
The~minimum distance of a track to a primary $\proton\proton$\nobreakdash-collision vertex\,(PV), 
the~impact parameter\,(IP), 
is~measured with a~resolution of $(15+29/\pt)\mum$, where \pt is the component 
of the~momentum transverse to the beam, in\,\gevc. Different types of charged hadrons
are distinguished using information from two ring-imaging Cherenkov detectors\,(RICH)~\cite{LHCb-DP-2012-003}. Photons,~electrons and hadrons are identified 
by a~calorimeter system consisting of scintillating\nobreakdash-pad 
and preshower detectors, 
an electromagnetic and 
a~hadronic calorimeter~\cite{LHCb-DP-2020-001}. Muons are~identified by a~system 
composed of alternating layers of iron and multiwire proportional chambers~\cite{LHCb-DP-2012-002}.

The online event selection is performed by a trigger~\cite{LHCb-DP-2012-004}, 
which consists of a hardware stage, based on information from the calorimeter and muon systems,
followed by a~software stage, which applies a~full event reconstruction. 
The hardware trigger selects muon candidates with high transverse momentum or dimuon candidates with a~high value of 
the~product
of the~$\pt$ of the~muons. 
In~the~software trigger two 
oppositely charged muons are required to form 
a~good\nobreakdash-quality
vertex that is significantly displaced from every~PV,
with a~dimuon mass exceeding~$2.7\gevcc$.

Simulated events are used to describe signal  
shapes
and to~compute the efficiencies needed to determine 
the~branching fraction ratios.
In~the~simulation, \proton\proton collisions are generated 
using \pythia~\cite{Sjostrand:2007gs}  with a~specific \lhcb configuration~\cite{LHCb-PROC-2010-056}. 
Decays of unstable particles are described by 
the~\evtgen 
package~\cite{Lange:2001uf}, 
in which final-state radiation is generated using \photos~\cite{Golonka:2005pn}. 
The~\mbox{$\decay{\chicone(3872)}{\jpsi\pip\pim}$}~decays are 
simulated  proceeding
via an~S\nobreakdash-wave $\jpsi\Prho^0$ 
intermediate state~\cite{LHCb-PAPER-2015-015}. 
The~model described in 
Refs.~\cite{Gottfried:1977gp,Voloshin:1978hc,Peskin:1979va,Bhanot:1979vb} is used to describe the~\psitwos decays.
The simulation is corrected to reproduce the~transverse momentum and 
rapidity distributions of the~$\Bs$ observed in data.
The~interaction of the~generated particles with the~detector, 
and its response, are implemented using
the~\geant toolkit~\cite{Allison:2006ve, *Agostinelli:2002hh} 
as described in Ref.~\cite{LHCb-PROC-2011-006}.
To~account for imperfections in the~simulation of
charged\nobreakdash-particle reconstruction, 
the~track reconstruction efficiency
determined from simulation 
is corrected using data-driven techniques~\cite{LHCb-DP-2013-002}.

%% file: selection.tex
\section{Event selection}
\label{sec:evt_sel}
Candidate $\decay{\Bs}{\jpsi\pip\pim\Kp\Km}$~decays 
are reconstructed 
using 
similar selection criteria 
to those used in Refs.~\cite{LHCb-PAPER-2013-047,
LHcb-PAPER-2015-060,
LHCb-PAPER-2019-023}.
Muon and hadron candidates are identified 
using combined information from 
the~RICH, calorimeter and muon detectors~\cite{LHCb-PROC-2011-008}. 
They are required to have a~transverse momentum larger 
than $550$, $200$ and  $400\mevc$ 
for muon, pion and kaon candidates, respectively. 
To~ensure that
the~particles can be efficiently 
separated by the~RICH detectors,
kaons and pions are required to have a~momentum 
between $3.2$~and $150\gevc$. 
To~reduce the combinatorial background due to particles 
produced promptly
in the~$\proton\proton$~interaction, 
only tracks that are inconsistent with originating from 
a~primary vertex are used. 
Pairs of oppositely charged muons consistent with originating 
from a~common vertex are combined to form $\jpsi$ candidates. 
The~mass of the~dimuon candidate is required to be between $3.05$ and $3.15\gevcc$.

Selected $\jpsi$ candidates are combined with 
two oppositely charged 
kaons as well as two oppositely charged pions 
to form~$\decay{\Bs}{\jpsi\pip\pim\Kp\Km}$~candidates.
A~requirement on the~quality 
of the~common six\nobreakdash-prong vertex is imposed.   
To~improve the~mass resolution for 
the~\Bs~candidates, 
the~mass of the~\mumu~pair 
is constrained 
to the~known mass of the~\jpsi~meson~\cite{PDG2020}
and the~\Bs~candidate is constrained to originate from 
its associated~PV.\footnote{The~associated
PV is the~one that is most consistent with the~flight direction of the~\Bs~candidate.}
Finally, the~decay time of the~$\Bs$ candidates is required to be 
 between 0.2 and $2.0\mm/c$. 
 The~lower limit is used to reduce 
 background from particles coming 
 from the~PV
 while the~upper limit suppresses poorly 
 reconstructed candidates.

\begin{figure}[t]
  \setlength{\unitlength}{1mm}
  \centering
  \begin{picture}(150,120)
    %
    \put(  0, 0){ 
      \includegraphics*[width=150mm,height=120mm,%
       ]{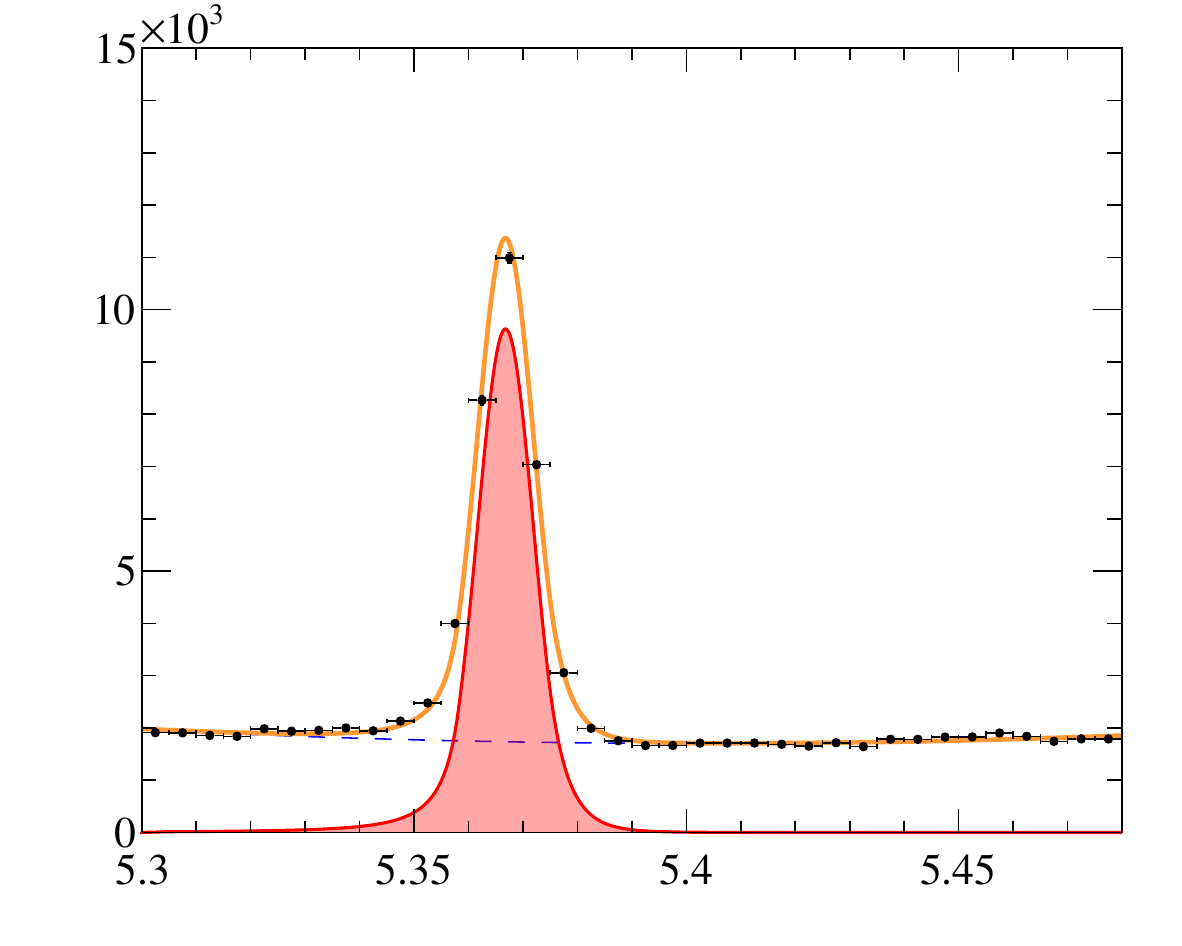}
    }
    \put(5,85){\rotatebox[]{90}{\Large{Candidates$/(5\mevcc)$}}}
	\put(120,103){\Large\lhcb}
		\put(75,92) {\begin{tikzpicture}[x=1mm,y=1mm]\filldraw[fill=red!35!white,draw=red,thick]  (0,0) rectangle (10,3);\end{tikzpicture} }
	\put(75,85){\color[RGB]{85,83,246}     {\hdashrule[0.0ex][x]{10mm}{2.0pt}{3.0mm 0.3mm} } }
	\put(75,78){\color[RGB]{255,153,51} {\rule{10mm}{4.0pt}}}
	\put( 87,92.5){{$\decay{\Bs}{\jpsi\pip\pim\Kp\Km}$ }}
	\put( 87,85.0){{background}}
	\put( 87,78){{total}}
  \put( 65,  2) { \large{$m_{\jpsi\pip\pim\Kp\Km}$}} 
  \put( 125,  2) { \large{$\left[\!\gevcc\right]$}}
  \end{picture}
	\caption {\small 
	   Distribution of the~$\jpsi\pip\pim\Kp\Km$ mass 
	   of selected $\Bs$ candidates
	   shown as points with error bars. 
	   A~fit, described in the~text, is overlaid. 
	   }
	\label{fig:Bs_can}
\end{figure}
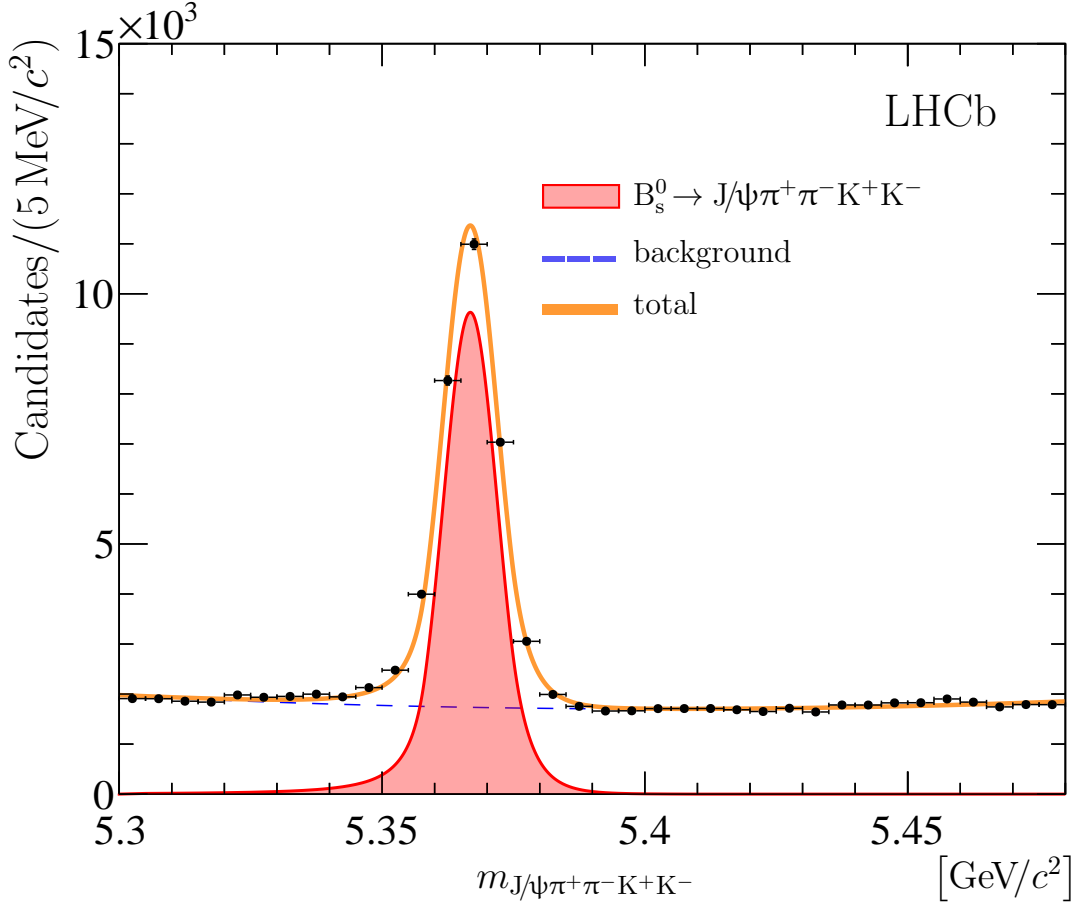

A~possible background    
from~\mbox{$\decay{\Lb}{\jpsi\pip\pim\proton\Km}$}
and  
\mbox{$\decay{\Bd}{\jpsi\pip\pim\Kp\pim}$}~decays, 
with the~proton or a~pion misidentified as a~kaon, 
is suppressed using a~veto. 
After assigning the~proton or pion mass
to one of the~kaons, 
only candidates outside 
the~mass intervals \mbox{$5.606<m_{\jpsi\pip\pim\proton\Km}<5.632\gevcc$} 
and 
\mbox{$5.266<m_{\jpsi\pip\pim\Kp\pim}<5.288\gevcc$}
are retained in the~analysis. 
The~mass distribution of the~selected 
$\decay{\Bs}{\jpsi\pip\pim\Kp\Km}$~candidates is shown 
in Fig.~\ref{fig:Bs_can}.
The~data are fit with the~sum
 of a~modified Gaussian function
with power-law tails on both sides of 
the~distribution~\cite{LHCb-PAPER-2011-013,Skwarnicki:1986xj} 
and a~linear 
combinatorial background component.
The \Bs~signal yield 
is \mbox{$(26.5\pm0.2)\times10^3$}~candidates.

%% file: xcc.tex

\section{
$\decay{\Bs}{\chicone(3872)\Pphi}$
and $\decay{\Bs}{\psitwos\Pphi}$~decays}
\label{sec:xcc_phi}

The yields of~\mbox{$\decay{\Bs}{\PX_{\ccbar}\Pphi}$}~decays, 
where $\PX_{\ccbar}$ denotes either the $\psitwos$ or 
the~$\chicone(3872)$~state,
are determined using a~three\nobreakdash-dimensional 
unbinned  extended maximum\nobreakdash-likelihood fit 
to the~$\jpsi\pip\pim\Kp\Km$~mass ($m_{\jpsi\pip\pim\Kp\Km}$)
the~$\jpsi\pip\pim$~mass~($m_{\jpsi\pip\pim}$) and  
the~$\Kp\Km$~mass ($m_{\Kp\Km}$) distributions.
The~fit is performed simultaneously in 
two separate regions of the~$m_{\jpsi\pip\pim}$, 
\mbox{$3.67<m_{\jpsi\pip\pim}<3.70\gevcc$} and  
\mbox{$3.85<m_{\jpsi\pip\pim}<3.90\gevcc$}, 
corresponding to~\mbox{$\decay{\Bs}{\psitwos\Pphi}$}
and \mbox{$\decay{\Bs}{\chicone(3872)\Pphi}$}
signals, respectively.
Only candidates 
with 
\mbox{$0.995<m_{\Kp\Km}<1.060\gevcc$} and  
\mbox{$5.30<m_{\jpsi\pip\pim\Kp\Km}<5.48\gevcc$}
are considered.  
To~improve the~resolution on the~$\jpsi\pip\pim$~mass
and to eliminate a~small correlation between the
$m_{\jpsi\Kp\Km\pip\pim}$ and $m_{\jpsi\pip\pim}$~variables, 
the~$m_{\jpsi\pip\pim}$ variable is computed
using a~kinematic fit~\cite{Hulsbergen:2005pu} 
that constrains the~mass of the~\Bs~candidate 
to its known value~\cite{PDG2020}.
In~each region, the~three\nobreakdash-dimensional fit model is defined 
as a~sum  of eight components.
Four~of these components correspond 
to decays of \Bs~mesons:
\begin{enumerate}
\item a~signal $\decay{\Bs}{\PX_{\ccbar}}\Pphi$~component, 
described by the product of~\Bs, $\PX_{\ccbar}$ and $\Pphi$~signal templates, 
discussed in detail 
in the~next paragraph; 
\item 
a~component  corresponding 
to~\mbox{$\decay{\Bs}{\PX_{\ccbar}\Kp\Km}$}~decays,
where the~$\Kp\Km$~pair does not originate from
a~\Pphi~meson, 
parameterised by 
the~product of~\Bs~and
$\PX_{\ccbar}$~signal~templates and 
a~slowly varying template  describing the~non-resonant 
$\Kp\Km$~distribution, 
referred to below as the  non\nobreakdash-resonant 
$\Kp\Km$~function;
\item 
a~component corresponding 
to~\mbox{$\decay{\Bs}{ \jpsi\pip\pim\Pphi}$}~decays,
parameterised as a~product of the~\Bs~and $\Pphi$~signal templates and 
a~slowly varying template describing the~non-resonant 
$\jpsi\pip\pim$~mass distribution, 
referred to as the~non\nobreakdash-resonant 
$\jpsi\pip\pim$~function hereafter;
\item a~component corresponding to 
the~decay \mbox{$\decay{\Bs}{  \jpsi\pip\pim 
\Kp\Km } $} 
with no narrow resonance in either 
the~$\jpsi\pip\pim$ or the $\Kp\Km$~systems, described by the~product 
of the~\Bs~signal  template and 
a~slowly varying
function 
$f_{\mathrm{bkg}}\left(m_{\jpsi\pip\pim},
m_{\Kp\Km}\right)$, described below.   
\end{enumerate} 
Four additional components correspond to 
random 
\mbox{$\PX_{\ccbar}\Pphi$},
\mbox{$\PX_{\ccbar}\Kp\Km$},
\mbox{$\jpsi\pip\pim\Pphi$}
and  
\mbox{$\jpsi\pip\pim\Kp\Km$}
combinations. 
Their parameterisation uses 
a~second\nobreakdash-order 
polynomial function
in $m_{\jpsi\Kp\Km\pip\pim}$, 
denoted as $\mathcal{F}_{\Bs}$.
 These~four background components are: 
 \begin{enumerate}
 \item a~component corresponding 
 to random combinations of $\PX_{\ccbar}$ and $\Pphi$~signals,
 parameterised as a~product of 
 the~$\mathcal{F}_{\Bs}$~function and
 the~$\PX_{\ccbar}$ 
 and $\Pphi$~signal templates;
 \item a~component corresponding to random combinations of 
 an~$\PX_{\ccbar}$~signal
 with a~non\nobreakdash-resonant $\Kp\Km$~pair,
 parameterised as a~product of 
 the~$\mathcal{F}_{\Bs}$~function, 
 the~signal $\PX_{\ccbar}$~template and   
 the~non\nobreakdash-resonant
 $\Kp\Km$~function;
\item a~component corresponding to random combinations of 
 a~$\Pphi$~signal 
 with a non\nobreakdash-resonant $\jpsi\pip\pim$~combination, 
 parameterised as a~product of 
 the~$\mathcal{F}_{\Bs}$~function, 
 the~signal $\Pphi$~template and   
 the~non\nobreakdash-resonant
 $\jpsi\pip\pim$~function;
 \item a component corresponding to random 
 $\jpsi\pip\pim\Kp\Km$~combinations 
 parameterised as 
 a~product of  
 the~$\mathcal{F}_{\Bs}$~function and 
 the~function $f_{\mathrm{bkg}}\left(m_{\jpsi\pip\pim},
m_{\Kp\Km}\right)$.
\end{enumerate} 

In~the~$m_{\jpsi\pip\pim\Kp\Km}$ distribution, 
the~$\Bs$ signal shape is modelled with a~modified Gaussian function with 
power\nobreakdash-law tails on both sides of 
the~distribution~\cite{LHCb-PAPER-2011-013,Skwarnicki:1986xj}.
The~tail parameters
are fixed from~simulation, 
while the~mass of the~\Bs~meson is allowed to vary.
The~detector resolution taken from simulation 
is corrected 
by a~scale factor, $s_{\Bs}$, 
that accounts for a~small discrepancy 
 between data and simulation~\cite{LHCb-PAPER-2020-009} and 
is allowed to vary.  
The~$\Pphi$ and $\PX_{\ccbar}$~signal templates are 
modelled with relativistic 
P\nobreakdash-wave and S\nobreakdash-wave 
Breit\nobreakdash--Wigner functions, respectively, 
convolved with the~detector resolution functions described below. 
Due~to the~proximity of 
the~$\chicone(3872)$~state to 
the~$\Dz\Dstarzb$~threshold,
modelling this component with a~Breit\nobreakdash--Wigner 
function may not be adequate~\cite{Hanhart:2007yq,
Stapleton:2009ey,
Kalashnikova:2009gt,
Artoisenet:2010va,
Hanhart:2011jz}. 
However, the~analyses in Refs.~\cite{LHCb-PAPER-2020-008,LHCb-PAPER-2020-009} 
demonstrate that a~good description of data 
is obtained with a~Breit\nobreakdash--Wigner 
line shape when the~detector resolution is included. 
The~mass of the~\psitwos~state 
is allowed to vary, 
while the~width is fixed to its known value~\cite{PDG2020}.
The~width of the~$\chicone(3872)$~state
and the~mass difference 
\mbox{$m_{\psitwos}-m_{\chicone(3872)}$} 
are constrained to 
their~known values~\cite{LHCb-PAPER-2020-008,LHCb-PAPER-2020-009}  
using Gaussian constraints.
The~detector resolution is described by 
a~symmetric modified Gaussian function with power\nobreakdash-law tails 
on both sides of the~distribution~\cite{LHCb-PAPER-2011-013,Skwarnicki:1986xj}, 
with all parameters determined from simulation. 
The resolution functions for the $\PX_{\ccbar}$~templates
are corrected 
by a~common  scale factor,
$s_{\PX_{\cquark\cquarkbar}}$, 
to~account for a~small discrepancy  
in the~detector resolution  
between data and 
simulation~\cite{LHCb-PAPER-2020-008,LHCb-PAPER-2020-009}. 
This factor is determined from data.
The~non\nobreakdash-resonant
$\Kp\Km$ and 
$\jpsi\pip\pim$~distributions 
are modelled by the product of
a~linear function and  
two\nobreakdash-body,  
$\Phi_{2,5}\left(m_{\Kp\Km}\right) $, 
and 
three\nobreakdash-body,  
$\Phi_{3,5}\left(m_{\jpsi\pip\pim}\right) $
phase\nobreakdash-space distributions 
for five\nobreakdash-body \Bs~decays~\cite{Byckling}.\footnote{
The~phase\nobreakdash-space mass distribution of 
an~$l$\nobreakdash-body combination of
particles from a~$n$\nobreakdash-body decay is approximated by 
$\Phi_{l,n}(x) \propto x_{\ast}^{ (3l-5)/2}\left(1-x_{\ast}\right)^{3(n-l)/2-1}$,
where 
$x_{\ast}\equiv ( x-x_{\mathrm{min}} ) /(x_{\mathrm{max}}-x_{\mathrm{min}})$, 
and 
$x_{\mathrm{min}}$, $x_{\mathrm{max}}$ denote 
the~minimal   and maximal values of $x$, respectively,
}
The~function $f_{\mathrm{bkg}}$ 
is parameterised by  
\begin{equation}
f_{\mathrm{bkg}}\left(m_{\jpsi\pip\pim},
m_{\Kp\Km}\right) \equiv 
\Phi_{3,5}\left(m_{\jpsi\pip\pim}\right)  
\Phi_{2,5}\left(m_{\Kp\Km}\right)
P_{\mathrm{bkg}}\left(m_{\jpsi\pip\pim},
m_{\Kp\Km}\right)\,, 
\end{equation}
where $P_{\mathrm{bkg}}$ is 
a~polynomial function that 
is linear in one variable for 
each fixed value of the~other variable. 

\begin{figure}[t]
  \setlength{\unitlength}{1mm}
  \centering
  \begin{picture}(160,130)
    %
    \put(  0, 65){ 
      \includegraphics*[width=80mm,height=65mm,%
      ]{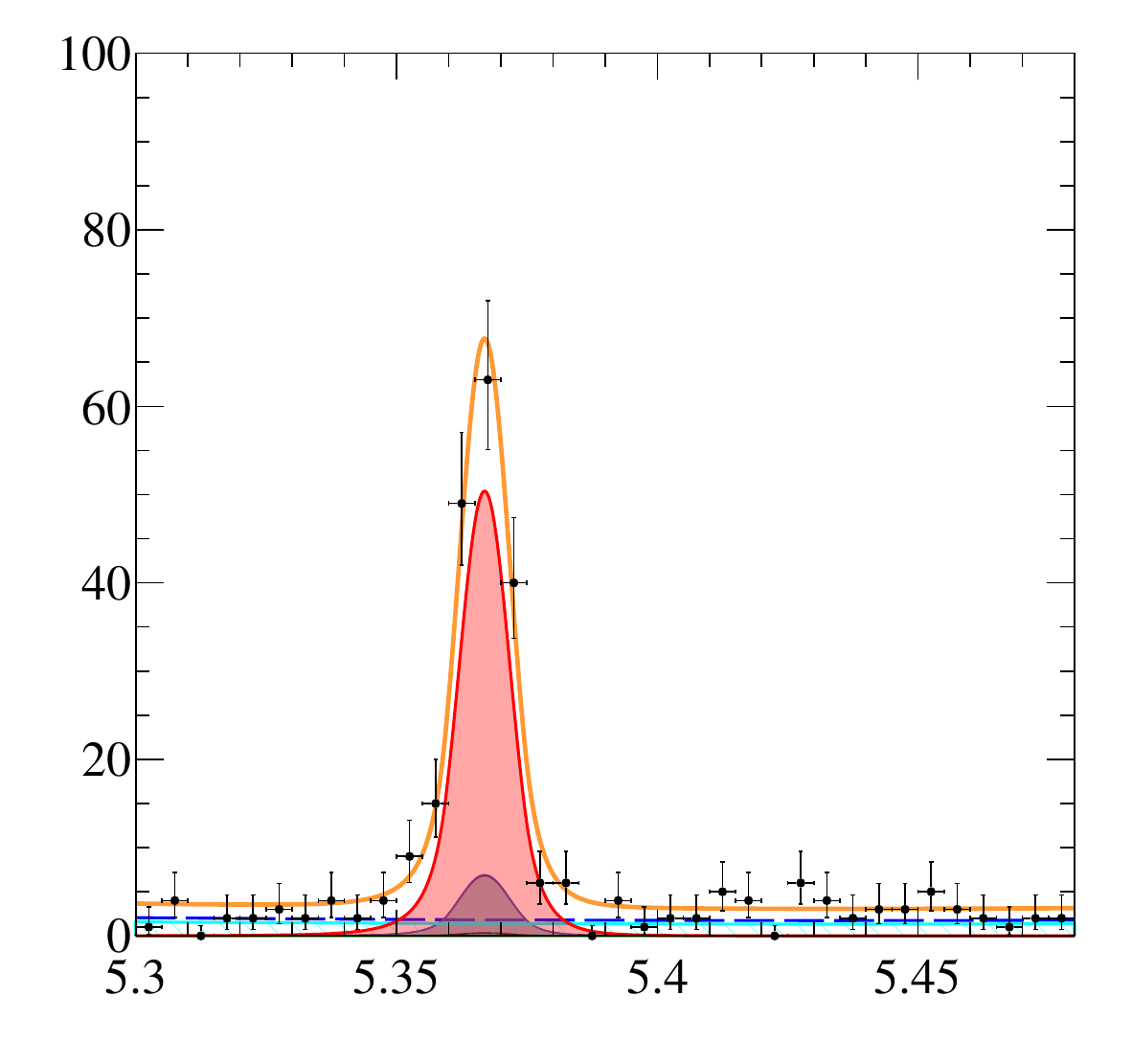}
    }
     \put( 80, 65){ 
      \includegraphics*[width=80mm,height=65mm,%
      ]{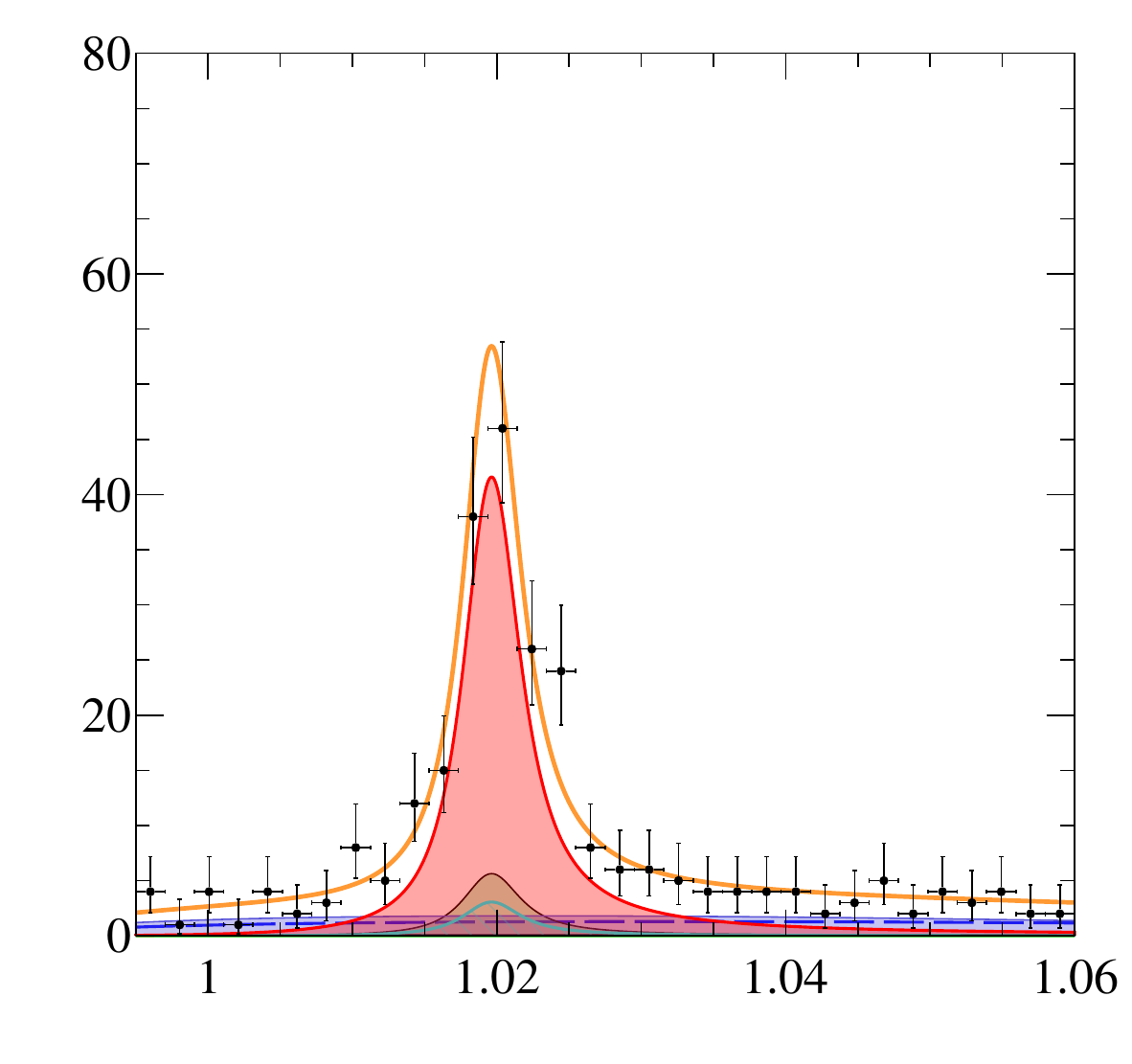}
    }
    \put(  0,  0){ 
      \includegraphics*[width=80mm,height=65mm,%
      ]{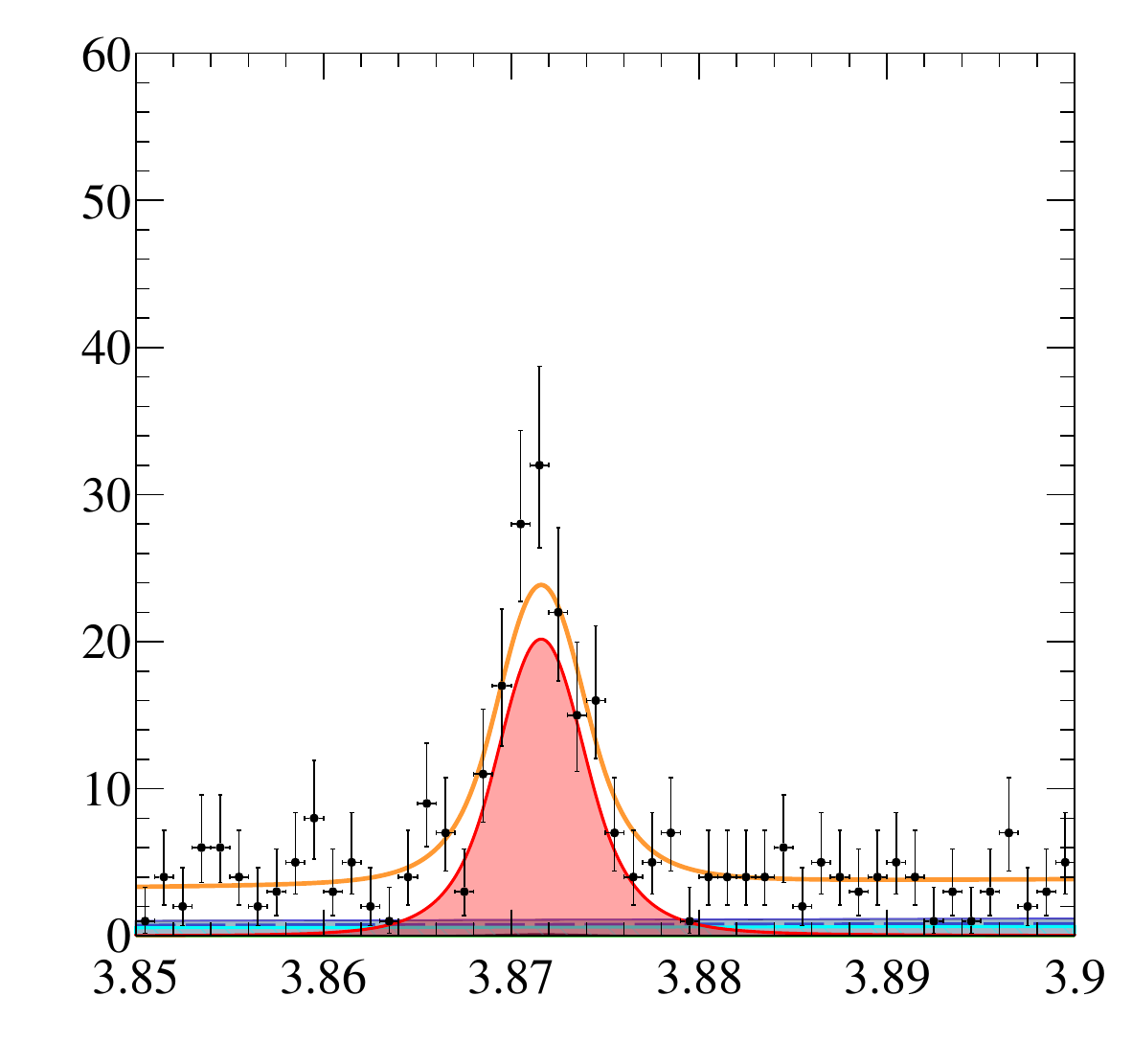}
    }
    \put(  0 , 105){\rotatebox[]{90} {Candidates$/(5\mevcc)$} }
    \put( 80, 105) {\rotatebox[]{90}{Candidates$/(2\mevcc)$} }
    \put(  0,  40) {\rotatebox[]{90}{Candidates$/(1\mevcc)$} }
    \put( 30, 65) { $m_{\jpsi\pip\pim\Kp\Km}$} 
    \put( 62, 65) { $\left[\!\gevcc\right]$} 
    \put(115, 65) { $m_{\Kp\Km}$} 
    \put(142,65) { $\left[\!\gevcc\right]$} 
    \put( 35, 0) { $m_{\jpsi\pip\pim}$} 
    \put( 62, 0) { $\left[\!\gevcc\right]$} 
	\put( 14,52.5){\scriptsize$5.350<m_{\jpsi\pip\pim\Kp\Km}<5.384\gevcc$}
	\put( 14,56.5){\scriptsize$1.01<m_{\Kp\Km}<1.03\gevcc$}
	
	\put( 14,117.5){\scriptsize$3.864<m_{\jpsi\pip\pim}<3.880\gevcc$}
	\put( 14,121.5){\scriptsize$1.01<m_{\Kp\Km}<1.03\gevcc$}

	\put( 94,117.5){\scriptsize$5.350<m_{\jpsi\pip\pim\Kp\Km}<5.384\gevcc$}
	\put( 94,121.5){\scriptsize$3.864<m_{\jpsi\pip\pim}<3.880\gevcc$}
	
	\put( 64, 57){\small\lhcb}
	\put( 64,122){\small\lhcb}
	\put(144,122){\small\lhcb}

	\definecolor{gr}{rgb}{0.35, 0.83, 0.33}
	\definecolor{br}{rgb}{0.43, 0.98, 0.98}
	\definecolor{vi}{rgb}{0.39, 0.37, 0.96}

	\definecolor{db}{rgb}{0.1, 0.08, 0.41}
	 \put(90,50) {\begin{tikzpicture}[x=1mm,y=1mm]\filldraw[fill=red!35!white,draw=red,thick]  (0,0) rectangle (8,3);\end{tikzpicture} }
    \put(90,45){\begin{tikzpicture}[x=1mm,y=1mm]\filldraw[fill=gr!35!white,draw=gr,thick]  (0,0) rectangle (8,3);\end{tikzpicture} }
    \put(90,40){\begin{tikzpicture}[x=1mm,y=1mm]\filldraw[fill=br!35!white,draw=br,thick]    (0,0) rectangle (8,3);\end{tikzpicture} }
    \put(90,35) {\begin{tikzpicture}[x=1mm,y=1mm]    \filldraw[fill=vi!35!white,draw=vi,thick]  (0,0) rectangle (8,3);\end{tikzpicture} }
	\put(90,30) {\begin{tikzpicture}[x=1mm,y=1mm] \draw[thin,vi,pattern=north east lines, pattern color=vi] (0,0) rectangle (8,3);\end{tikzpicture} }
	\put(90,25)
	{\begin{tikzpicture}[x=1mm,y=1mm]	\draw[thin,br,pattern=north west lines, pattern color=br]    (0,0) rectangle (8,3);\end{tikzpicture} }
	\put(90,20) {\begin{tikzpicture}[x=1mm,y=1mm]	\draw[thin,gr,pattern=north west lines, pattern color=gr]    (0,0) rectangle (8,3);\end{tikzpicture} }
	\put(90,15) {\color[RGB]{36,70,246}     {\hdashrule[0.0ex][x]{8mm}{1.0pt}{2.0mm 0.3mm} } }
	\put(90,10){\color[RGB]{255,153,51} {\rule{8mm}{2.0pt}}}
	
	\put( 100,50){\small{\decay{\Bs}{\chicone(3872)\Pphi}}}

	\put( 100,45){\small{\decay{\Bs}{\jpsi\pip\pim\Pphi}}}
	\put( 100,40){\small{\decay{\Bs}{\chicone(3872)\Kp\Km}}}
	\put( 100,35){\small{\decay{\Bs}{\jpsi\pip\pim\Kp\Km }}}

	\put( 100,30){\small{comb. $\chicone(3872)\Pphi$}}

	\put( 100,25){\small{comb. $\jpsi\pip\pim\Pphi$}}
	\put( 100,20){\small{comb. $\chicone(3872)\Kp\Km$}}

	\put( 100,15){\small{comb. $\jpsi\pip\pim\Kp\Km$}} 
	\put( 100,10){\small{total}}
  \end{picture}
  \caption { \small
   Distributions 
   of the~(top left)~\mbox{$\jpsi\pip\pim\Kp\Km$},
   (top right)~\mbox{$\Kp\Km$} and 
   (bottom left)~\mbox{$\jpsi\pip\pim$}~mass
   of selected \mbox{$\decay{\Bs}{\chicone(3872)\Pphi}$}~candidates
   shown as points with error bars. 
   A~fit, described in the~text, is overlaid.
  }
  \label{fig:fit_sim_chi}
\end{figure}
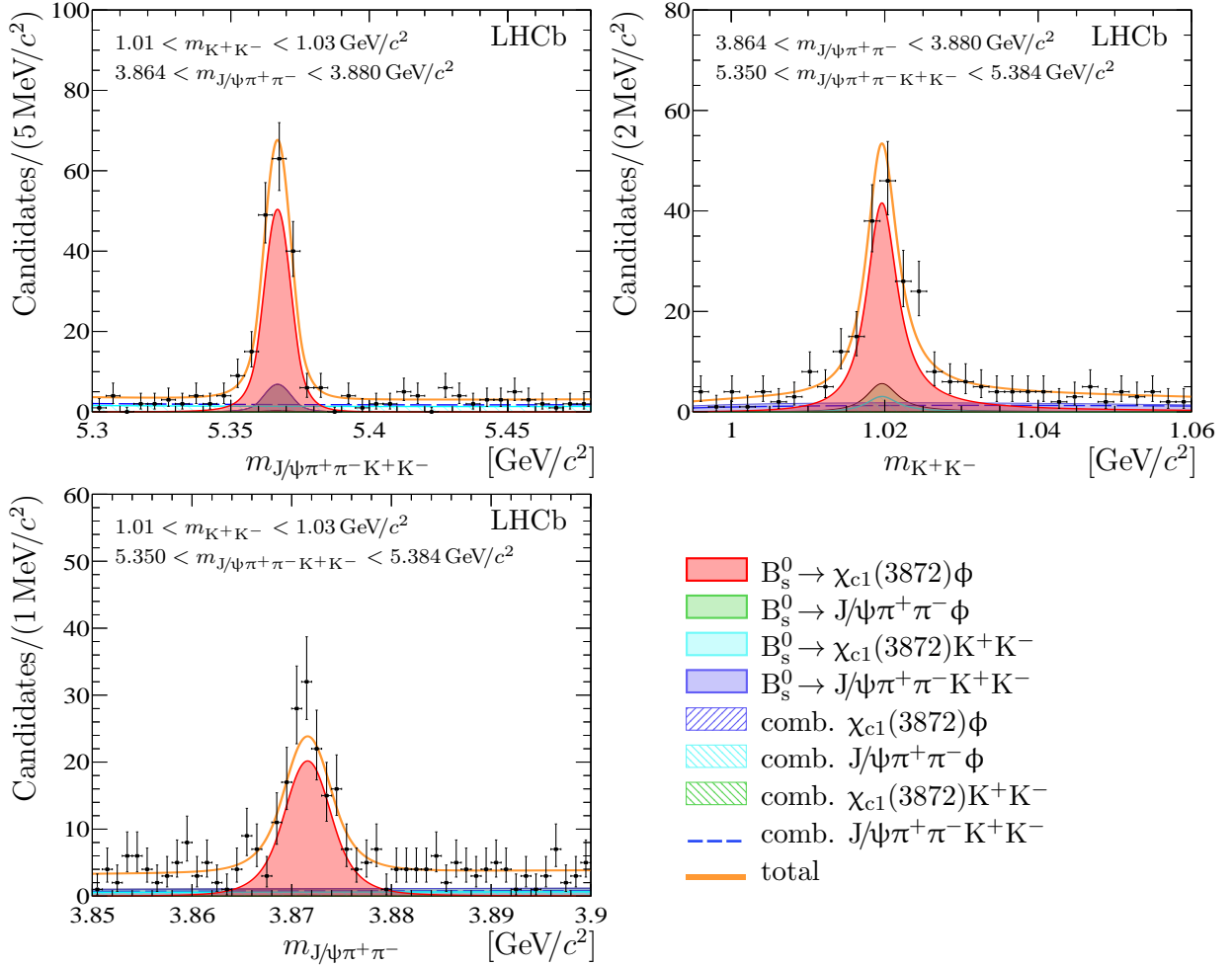

\begin{figure}[t]
  \setlength{\unitlength}{1mm}
  \centering
  \begin{picture}(160,130)
    %
    \put(  0, 65){ 
      \includegraphics*[width=80mm,height=65mm,%
      ]{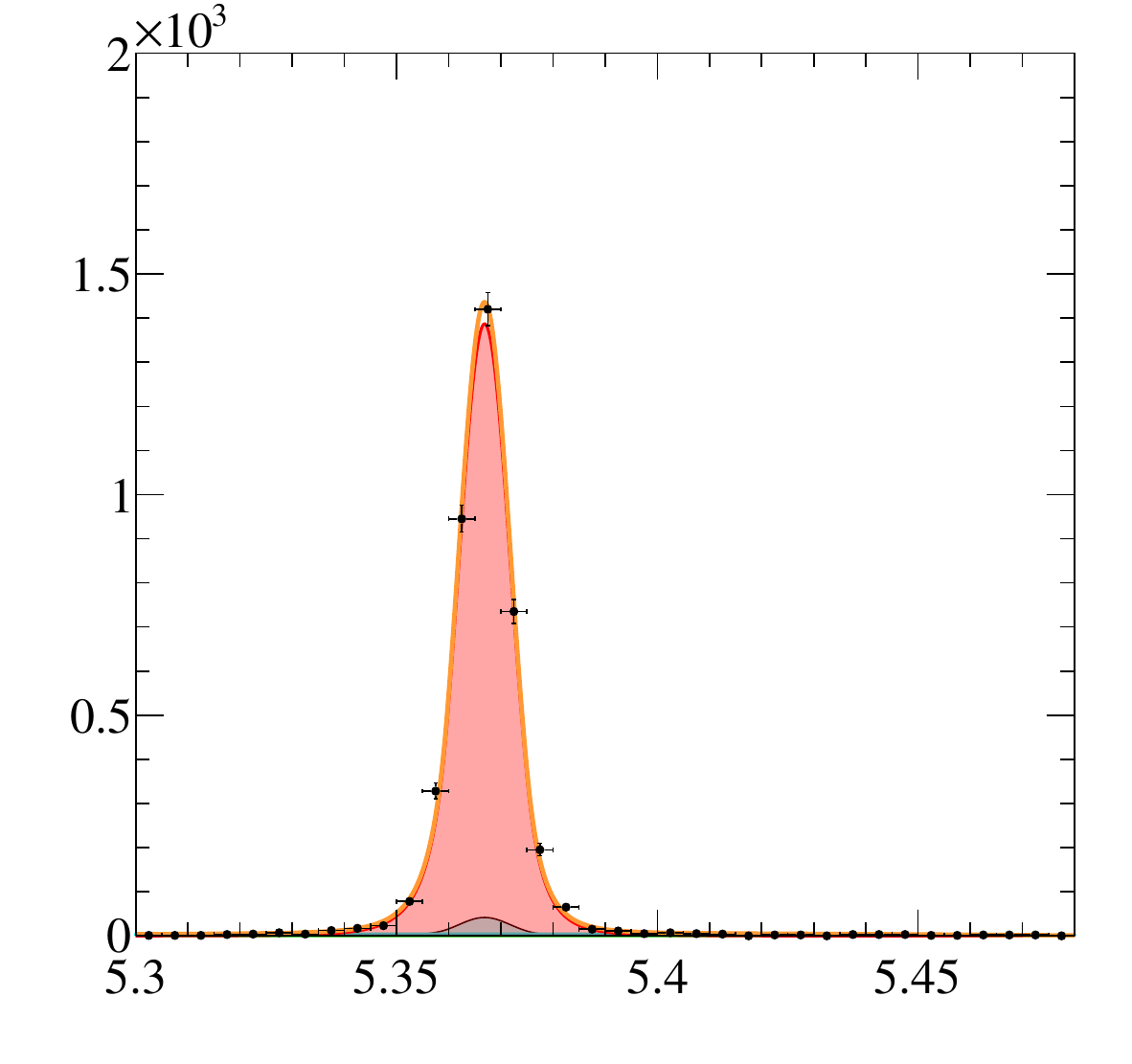}
    }
     \put( 80, 65){ 
      \includegraphics*[width=80mm,height=65mm,%
      ]{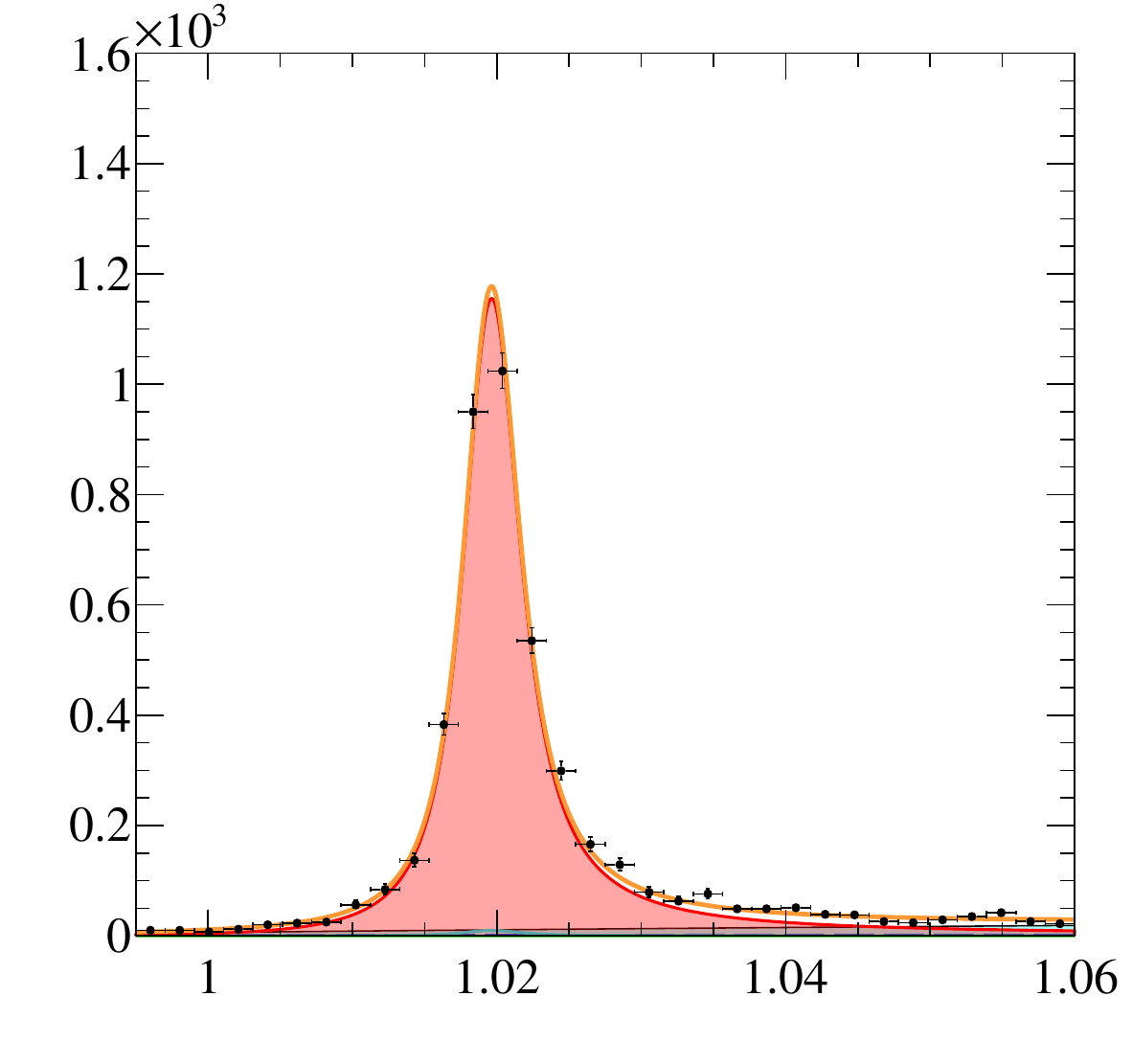}
    }
    \put(  0,  0){ 
      \includegraphics*[width=80mm,height=65mm,%
       ]{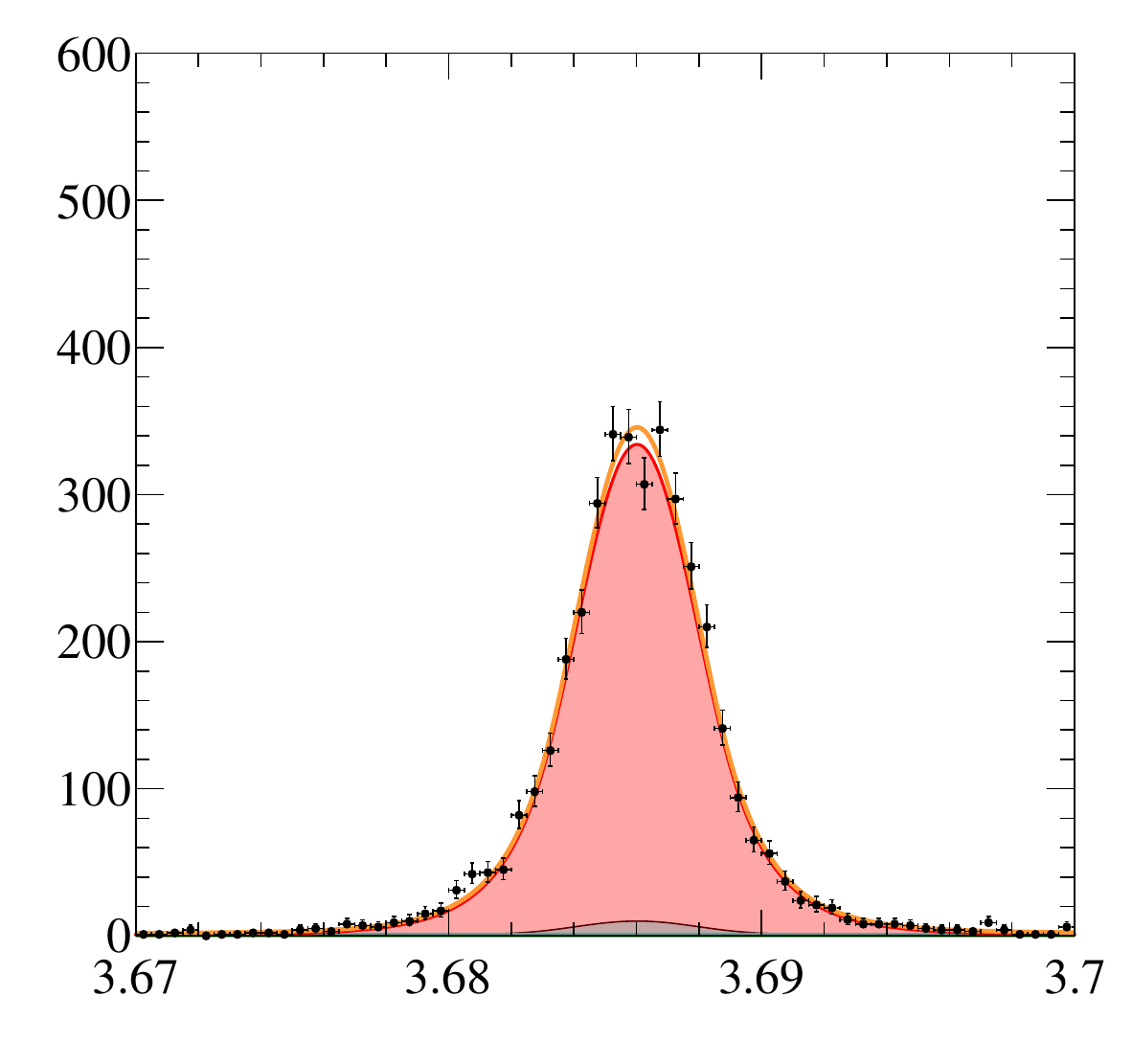}
    }
    \put(  0, 105){\rotatebox[]{90} {Candidates$/(5\mevcc)$} }
    \put( 80, 105) {\rotatebox[]{90}{Candidates$/(2\mevcc)$} }
    \put(  0,  40) {\rotatebox[]{90}{Candidates$/(0.5\mevcc)$} }
    \put( 30, 65) { $m_{\jpsi\pip\pim\Kp\Km}$} 
    \put( 62, 65) { $\left[\!\gevcc\right]$} 
    \put(115, 65) { $m_{\Kp\Km}$} 
    \put(142,65) { $\left[\!\gevcc\right]$} 
    \put( 35, 0) { $m_{\jpsi\pip\pim}$} 
    \put( 62, 0) { $\left[\!\gevcc\right]$} 
    \put( 14,52.5){\scriptsize$5.350<m_{\jpsi\pip\pim\Kp\Km}<5.384\gevcc$}
	\put( 14,56.5){\scriptsize$1.01<m_{\Kp\Km}<1.03\gevcc$}

	\put( 14,117.5){\scriptsize$3.679<m_{\jpsi\pip\pim}<3.693\gevcc$}
	\put( 14,121.5){\scriptsize$1.01<m_{\Kp\Km}<1.03\gevcc$}

	\put( 94,117.5){\scriptsize$5.350<m_{\jpsi\pip\pim\Kp\Km}<5.384\gevcc$}
	\put( 94,121.5){\scriptsize$3.679<m_{\jpsi\pip\pim}<3.693\gevcc$}
  
  	\put( 64, 57){\small\lhcb}
	\put( 64,122){\small\lhcb}
	\put(144,122){\small\lhcb}
	
	\definecolor{gr}{rgb}{0.35, 0.83, 0.33}
	\definecolor{br}{rgb}{0.43, 0.98, 0.98}
	\definecolor{vi}{rgb}{0.39, 0.37, 0.96}

	\definecolor{db}{rgb}{0.1, 0.08, 0.41}
	 \put(90,50) {\begin{tikzpicture}[x=1mm,y=1mm]\filldraw[fill=red!35!white,draw=red,thick]  (0,0) rectangle (8,3);\end{tikzpicture} }
    \put(90,45){\begin{tikzpicture}[x=1mm,y=1mm]\filldraw[fill=gr!35!white,draw=gr,thick]  (0,0) rectangle (8,3);\end{tikzpicture} }
    \put(90,40){\begin{tikzpicture}[x=1mm,y=1mm]\filldraw[fill=br!35!white,draw=br,thick]    (0,0) rectangle (8,3);\end{tikzpicture} }
    \put(90,35) {\begin{tikzpicture}[x=1mm,y=1mm]    \filldraw[fill=vi!35!white,draw=vi,thick]  (0,0) rectangle (8,3);\end{tikzpicture} }
	\put(90,30) {\begin{tikzpicture}[x=1mm,y=1mm] \draw[thin,vi,pattern=north east lines, pattern color=vi] (0,0) rectangle (8,3);\end{tikzpicture} }
	\put(90,25)
	{\begin{tikzpicture}[x=1mm,y=1mm]	\draw[thin,br,pattern=north west lines, pattern color=br]    (0,0) rectangle (8,3);\end{tikzpicture} }
	\put(90,20) {\begin{tikzpicture}[x=1mm,y=1mm]	\draw[thin,gr,pattern=north west lines, pattern color=gr]    (0,0) rectangle (8,3);\end{tikzpicture} }
	\put(90,15) {\color[RGB]{36,70,246}     {\hdashrule[0.0ex][x]{8mm}{1.0pt}{2.0mm 0.3mm} } }
	\put(90,10){\color[RGB]{255,153,51} {\rule{8mm}{2.0pt}}}
    \put( 100,50){\small{\decay{\Bs}{\psitwos\Pphi}}}
	\put( 100,45){\small{\decay{\Bs}{\jpsi\pip\pim\Pphi}}}
	\put( 100,40){\small{\decay{\Bs}{\psitwos\Kp\Km}}}

	\put( 100,35){\small{\decay{\Bs}{ \jpsi\pip\pim\Kp\Km }}}

	\put( 100,30){\small{comb. $\psitwos\Pphi$}}
	\put( 100,25){\small{comb. $\jpsi\pip\pim\Pphi$}}
	\put( 100,20){\small{comb. $\psitwos\Kp\Km$}}

	\put( 100,15){\small{comb. $\jpsi\pip\pim\Kp\Km$ }}
	\put( 100,10){\small{total}}
  \end{picture}
  \caption { \small
     Distributions 
   of the~(top left)~\mbox{$\jpsi\pip\pim\Kp\Km$},
   (top right)~\mbox{$\Kp\Km$} and 
   (bottom left)~\mbox{$\jpsi\pip\pim$}~mass
   of selected \mbox{$\decay{\Bs}{\psitwos\Pphi}$}~candidates
   shown as points with error bars. 
   A~fit, described in the~text, is overlaid.
  }
  \label{fig:fit_sim_psi}
\end{figure}
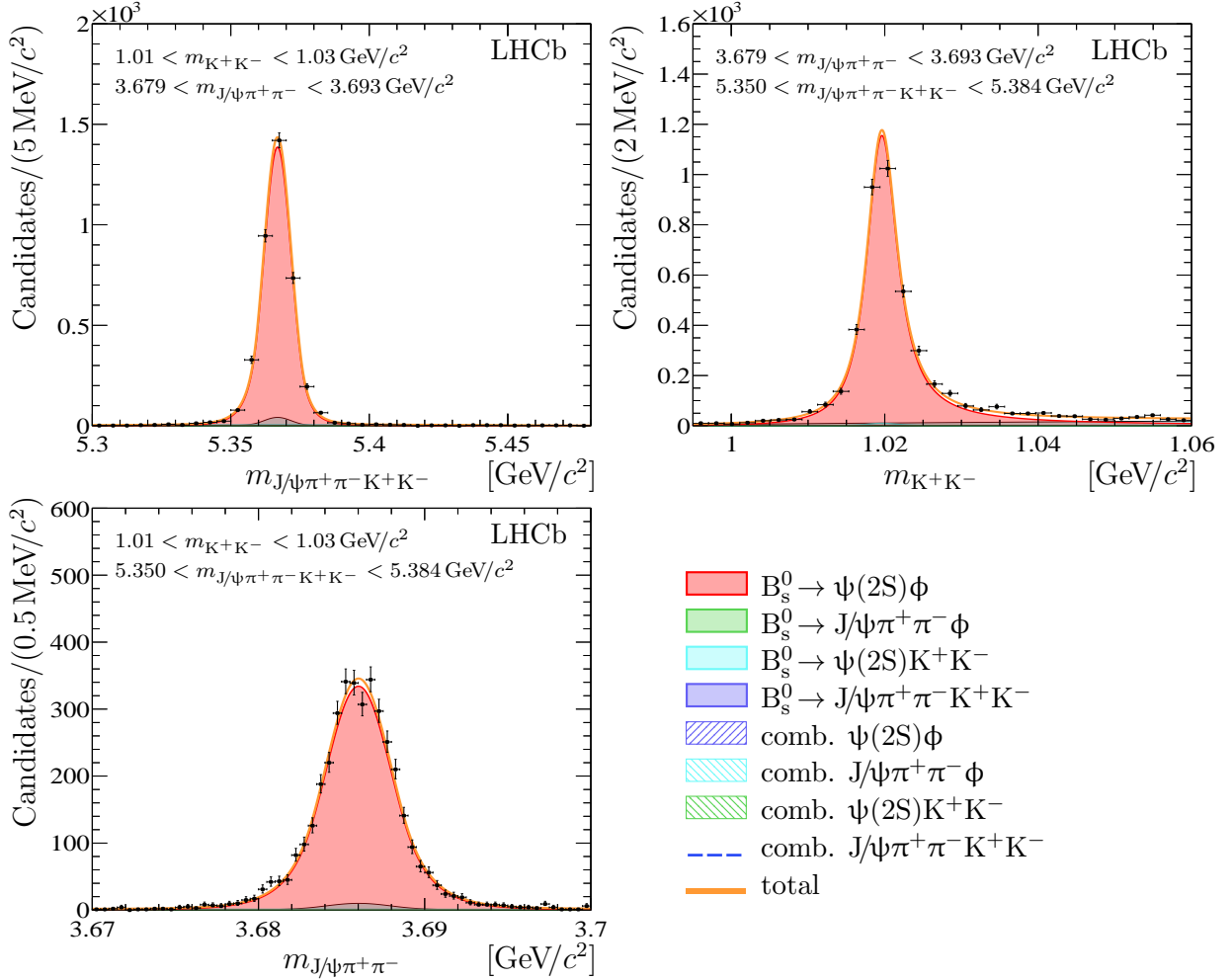

The~fit is performed
simultaneously to the~two \mbox{$\jpsi\pip\pim$} mass regions, 
with the~\Bs~and $\PX_{\ccbar}$~masses and the~resolution scale factors, 
$s_{\Bs}$ and $s_{\PX_{\ccbar}}$, as shared parameters. 
The~\mbox{$\jpsi\pip\pim\Kp\Km$},
\mbox{$\jpsi\pip\pim$}
and \mbox{$\Kp\Km$}~mass distributions together with 
projections of the simultaneous fit 
are shown in Figs.~\ref{fig:fit_sim_chi}
and~\ref{fig:fit_sim_psi}. 
The~fit procedure is tested using a~large sample of  
pseudoexperiments, generated using the~nominal model
with parameters extracted from  data. 
Biases of \mbox{${\mathcal{O}}(1\%)$}  
on the~yields of different 
fit components are observed and
the~results are corrected    for  these biases.
The~corrected 
yields of 
the~\mbox{$\decay{\Bs}{\chicone(3872)\Pphi}$}
and \mbox{$\decay{\Bs}{\psitwos\Pphi}$}~decays
and the~resolution scale factors 
are listed in Table~\ref{tab:sim_res}.
The~statistical significance 
for the~\mbox{$\decay{\Bs}{\chicone(3872)\Pphi}$}~signal is 
calculated to be 
in excess of 10~standard deviations
using Wilks' theorem~\cite{Wilks:1938dza}. 
Apart from the~signal \mbox{$\decay{\Bs}{\psitwos\Pphi} $}~component,  
only 
the~\mbox{$\decay{\Bs}{\psitwos\Kp\Km}$},
and the~combinatorial 
\mbox{$\jpsi\pip\pim\Pphi$} and 
\mbox{$\jpsi\pip\pim\Kp\Km$}~components are  found to contribute 
in a~non\nobreakdash-negligible way to the~\psitwos~mass region. 
In~the~\mbox{$\chicone(3872)$}~region, 
the~contribution 
from 
the~\mbox{$\decay{\Bs}{\chicone(3872)\Kp\Km}$}
component is found to be small and 
the~combinatorial components 
\mbox{$\chicone(3872)\Pphi$} and 
\mbox{$\chicone(3872)\Kp\Km$} 
are negligible.
The~resolution scale factors,  
$s_{\Bs}$ and $s_{\PX_{\ccbar}}$, 
are similar to those 
obtained in Refs.~\cite{LHCb-PAPER-2020-008,LHCb-PAPER-2020-009}.

\begin{table}[tb]
	\centering
	\caption{\small 
	Signal yields,
	 $N_{\decay{\Bs}{\PX_{\ccbar}\Pphi}}$,
	mass of the \Bs~meson, $m_{\Bs}$,
	and detector resolution scale 
	factors, $s_{\Bs}$ and $s_{\PX_{\ccbar}}$, 
	from the~fit described in the~text.
	The~uncertainties are statistical only. 
	}	
	\label{tab:sim_res}
	\vspace{2mm}
	\begin{tabular*}{0.75\textwidth}{@{\hspace{3mm}}l@{\extracolsep{\fill}}lcc@{\hspace{2mm}}}
	Parameter & \BsTopsitwophi & \BsTochiophi 
   \\[1mm]
  \hline 
  \\[-2mm]
   $N_{\decay{\Bs}{\PX_{\ccbar}\Pphi}}$  & $4180 \pm 66$  &  $154 \pm 15 $ \\
   $m_{\Bs}~\left[\!\mevcc\right]$ &  
  \multicolumn{2}{c}{$5366.89 \pm 0.08\phantom{000}$}
  \\
  $s_{\Bs}$ & \multicolumn{2}{c}{ $1.04 \pm   0.02$   }  
  \\
  $s_{\PX_{\ccbar}}$ & \multicolumn{2}{c}{ $1.06 \pm      0.02$ } 

	\end{tabular*}
	\vspace{3mm}
\end{table}

The~results are cross\nobreakdash-checked
using a
two\nobreakdash-dimensional unbinned extended
maximum\nobreakdash-likelihood fit to 
the~background\nobreakdash-subtracted
\mbox{$\jpsi\pip\pim$} and \mbox{$\Kp\Km$}~mass
distributions, where 
the~\sPlot~technique~\cite{Pivk:2004ty} 
is used with the~\mbox{$\jpsi\pip\pim\Kp\Km$}~mass  
as a~discriminating variable.  
The~results of this fit are found to be in 
very good agreement 
with the results listed in Table~\ref{tab:sim_res}.

The ratio of branching fractions defined in Eq.~\eqref{eq:r_chi} 
is calculated from 
\begin{equation}\label{eq:rcalcone}
\Rchi 
 = 
 \dfrac { N_{\decay{\Bs}{\chicone(3872) \Pphi}}}
        { N_{\decay{\Bs}{\psitwos \Pphi}}}
  \times 
  \dfrac { \varepsilon_{\decay{\Bs}{\psitwos \Pphi}}}
         { \varepsilon_{\decay{\Bs}{\chicone(3872) \Pphi}}}\,,
\end{equation}
where the~signal yields, 
$N_{\decay{\Bs}{\chicone(3872) \Pphi}}$
and $N_{\decay{\Bs}{\psitwos \Pphi}}$,
are taken from 
Table~\ref{tab:sim_res}
and 
$\varepsilon_{\decay{\Bs}{\chicone(3872) \Pphi}}$ and 
$\varepsilon_{\decay{\Bs}{\psitwos \Pphi}}$ are 
the~efficiencies to reconstruct and select 
the~\mbox{$\decay{\Bs}{\chicone(3872) \Pphi}$} and 
\mbox{$\decay{\Bs}{\psitwos \Pphi}$}~decays, respectively. 
The~efficiencies are defined as the~product of 
the~detector geometric acceptance and the  
reconstruction,  selection, 
hadron identification and trigger efficiencies. 
All~of the~efficiency contributions, except the~hadron\nobreakdash-identification efficiency, 
are determined using simulated samples.
The~hadron\nobreakdash-identification efficiency is determined
using large calibration samples
of  
\mbox{$\decay{\Dstarp}{ \left(\decay{\Dz}{\Km\pip}\right)\pip}$}, 
\mbox{$\decay{\KS}{\pip\pim}$}
and \mbox{$\decay{\Ds}{\left(\decay{\Pphi}{\Kp\Km}\right)\pip}$}
decays selected in data~\cite{LHCb-DP-2012-003, LHCb-DP-2018-001}.
 The~efficiency ratio is found to be 
  \mbox{$0.66\pm0.01$}, 
 where the~uncertainty is only that  due to the~size of the simulated samples. 
 The~efficiency ratio differs from unity due 
to the~harder \pt~spectrum of pions
 in the~\mbox{$\decay{\Bs}{\chicone(3872)\Pphi}$}~decays.
 The~resulting value of \Rchi~is
 \begin{equation}
     \Rchi = \left( 2.42 \pm 0.23\right)\times 10^{-2}\,,
 \end{equation}
 where the~uncertainty is statistical.
     Systematic uncertanties are discussed in Sec.~\ref{sec:sys}.

%% file: XKK.tex
\section{$\decay{\Bs}{\chicone(3872)\Kp\Km}$~decays}

The decay~\mbox{$\decay{\Bs}{\chicone(3872) \Kp\Km}$}, 
where 
the $\Kp\Km$~pair does not originate from a~\Pphi~meson, is studied using 
a~sample of 
selected \mbox{$\decay{\Bs}{\jpsi\pip\pim\Kp\Km}$}~candidates
with the~$\jpsi\pip\pim$
and \mbox{$\jpsi\pip\pim\Kp\Km$}~masses in the~ranges
\mbox{$3.85<m_{\jpsi\pip\pim}<3.90\gevcc$} and 
\mbox{$5.30<m_{\jpsi\pip\pim\Kp\Km}<5.48\gevcc$}.
A~two\nobreakdash-dimensional unbinned extended
maximum\nobreakdash-likelihood fit 
is performed to the~\mbox{$\jpsi\Kp\Km\pip\pim$} and 
\mbox{$\jpsi\pip\pim$}~mass distributions. 
The fit function comprises the~sum of four components: 
\begin{enumerate}
    \item a~component corresponding to~\mbox{$\decay{\Bs}{\chicone(3872)\Kp\Km}$}~decays,
    parameterised as a~product of the~\Bs and $\chicone(3872)$~signal templates described in Sec.~\ref{sec:xcc_phi};
    \item a~component corresponding 
    to~\mbox{$\decay{\Bs}{\jpsi\pip\pim\Kp\Km}$}~decays, 
    parameterised as a~product of the~\Bs~signal template and the~non\nobreakdash-resonant
    $\jpsi\pip\pim$~function;
    \item a~component corresponding to random combinations of~$\chicone(3872)$~particles 
    with a $\Kp\Km$~pair, parameterised as 
    a~product of the~$\chicone(3872)$~signal template 
    and the~$\mathcal{F}_{\Bs}$~function;
   \item a component corresponding to random  $\jpsi\pip\pip\Kp\Km$~combinations,
   parameterised as a~product of 
   the~three\nobreakdash-body phase\nobreakdash-space  function
   $\Phi_{3,5}\left(m_{\jpsi\pip\pim}\right)$
   and a~two\nobreakdash-dimensional non\nobreakdash-factorisable 
   bilinear function. 
\end{enumerate}

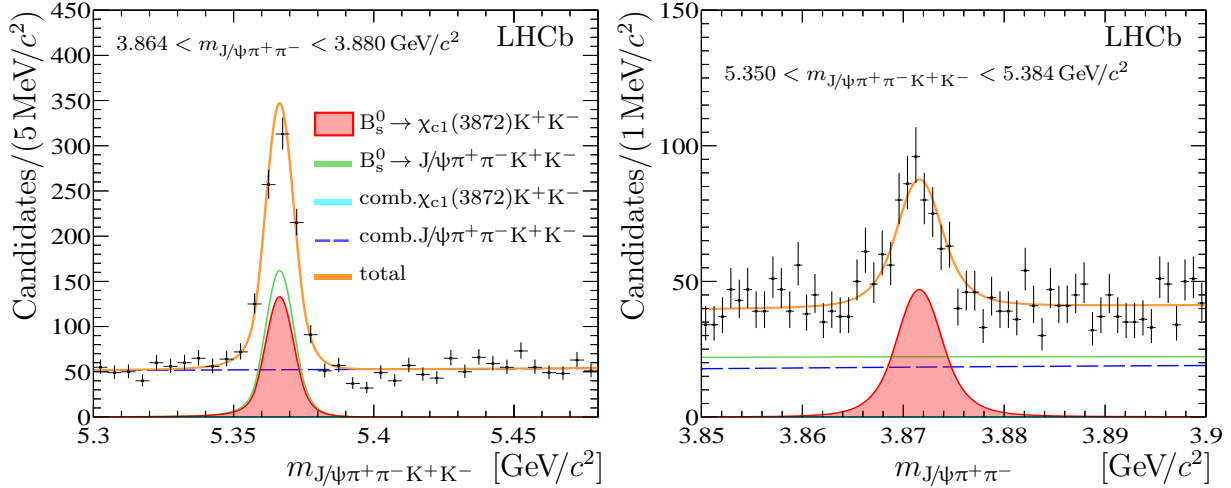
\begin{figure}[t]
  \setlength{\unitlength}{1mm}
  \centering
  \begin{picture}(160,65)
    \put(  0, 0){ 
      \includegraphics*[width=80mm,height=65mm,%
      ]{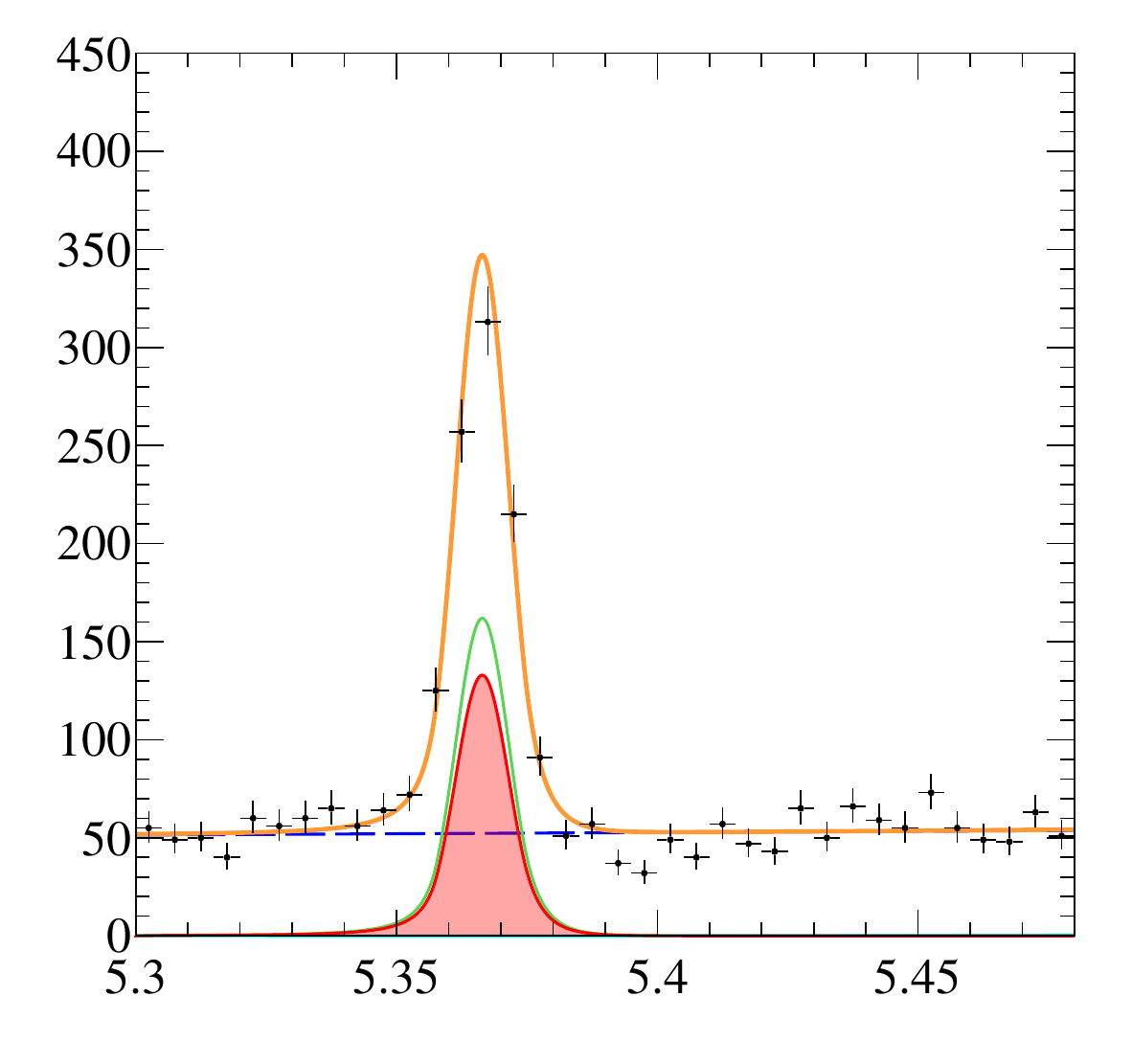}
    }
     \put( 80, 00){ 
      \includegraphics*[width=80mm,height=65mm,%
      ]{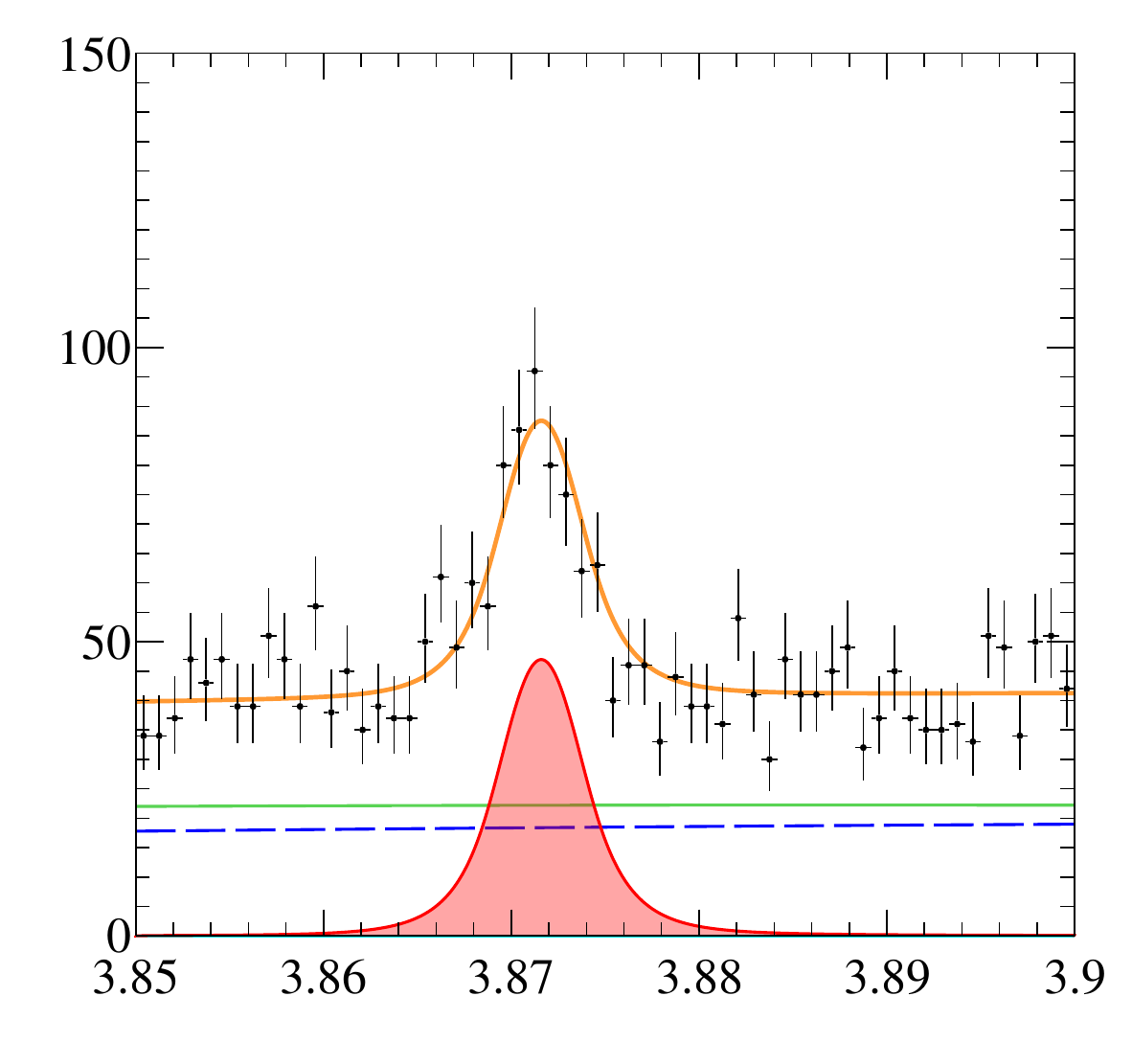}
    }
    \put( 80,40) {\rotatebox[]{90}{Candidates$/(1\mevcc)$} }
    \put(  0,40) {\rotatebox[]{90}{Candidates$/(5\mevcc)$} }
    \put(115, 0) { $m_{\jpsi\pip\pim}$} 
    \put(142,0) { $\left[\!\gevcc\right]$} 
    \put( 35,  0) { $m_{\jpsi\pip\pim\Kp\Km}$} 
    \put( 62, 0) { $\left[\!\gevcc\right]$} 
    \put( 94,52.5){\scriptsize$5.350<m_{\jpsi\pip\pim\Kp\Km}<5.384\gevcc$}
	\put( 14,56.5){\scriptsize$3.864<m_{\jpsi\pip\pim}<3.880\gevcc$}

	\put( 64, 57){\small\lhcb}
	\put(144, 57){\small\lhcb}

		\definecolor{vi}{rgb}{0.39, 0.37, 0.96}
    \put(40,45) {\begin{tikzpicture}[x=1mm,y=1mm]\filldraw[fill=red!35!white,draw=red,thick]  (0,0) rectangle (5,3);\end{tikzpicture} }
	\put(40,41) {\color[RGB] {121,220,117} {\rule{5mm}{2.0pt}}}
	\put(40,36) {\color[RGB] {110,251,251} {\rule{5mm}{2.0pt}}}
	\put(40,31){\color[RGB]{85,83,246}     {\hdashrule[0.0ex][x]{5mm}{1.0pt}{2.0mm 0.3mm} } }
	\put(40,26){\color[RGB]{255,153,51} {\rule{5mm}{2.0pt}}}

	\put( 46,46){{\scriptsize{\decay{\Bs}{\chicone(3872) \Kp\Km}}}}
	\put( 46,41){{\scriptsize{\decay{\Bs}{\jpsi \pip \pim \Kp \Km}}}}
	\put( 46,36){{\scriptsize{comb.$\chicone(3872) \Kp\Km$}}}
	\put( 46,31){{\scriptsize{comb.$\jpsi\pip\pim\Kp\Km$}}}
	\put( 46,26){{\scriptsize{total}}}

  \end{picture}
	\caption{\small 
	Distributions of 
	the~(left)~$\jpsi\pip\pim\Kp\Km$ and 
	(right)~$\jpsi\pip\pim$~mass of 
	selected $\decay{\Bs}{\chicone(3872)\Kp\Km}$ candidates shown as points with error bars.
	A~fit, described in the~text, is overlaid.}
	\label{fig:fit_XKK}
\end{figure}

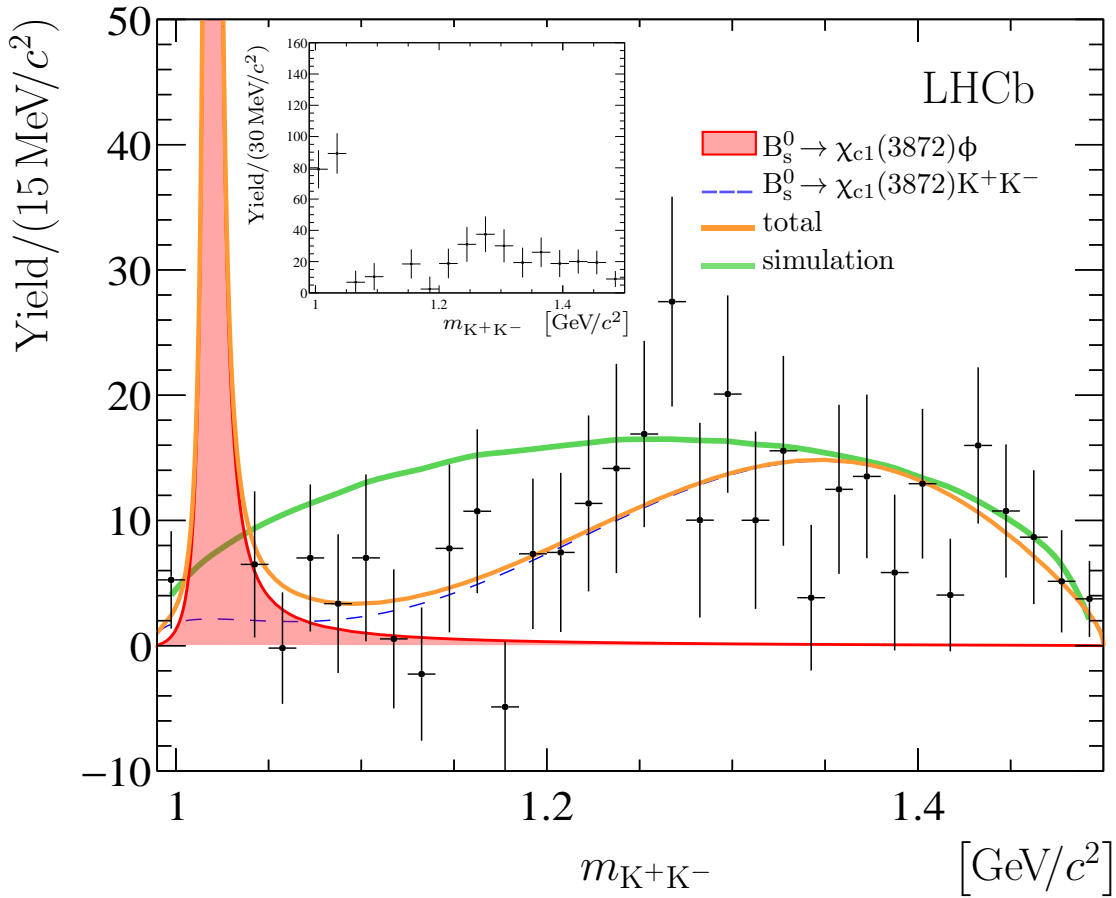
\begin{figure}[t]
  \setlength{\unitlength}{1mm}
  \centering
    \begin{picture}(150,120)
     \put( 0, 0){ 
      \includegraphics*[width=150mm,height=120mm,%
      ]{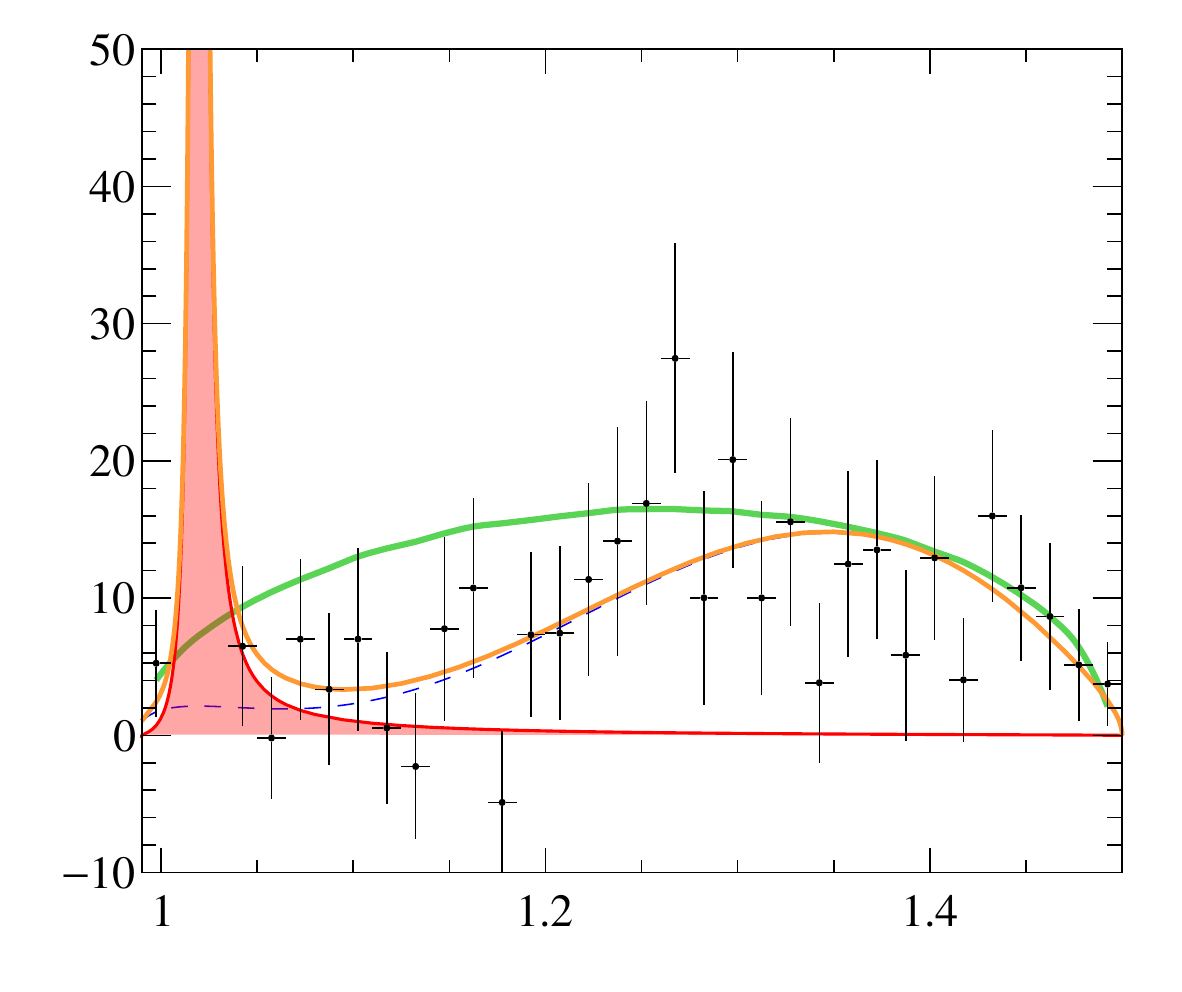}
    }
     \put(32,73){ 
      \includegraphics*[width=50mm,height=40mm,%
     ]{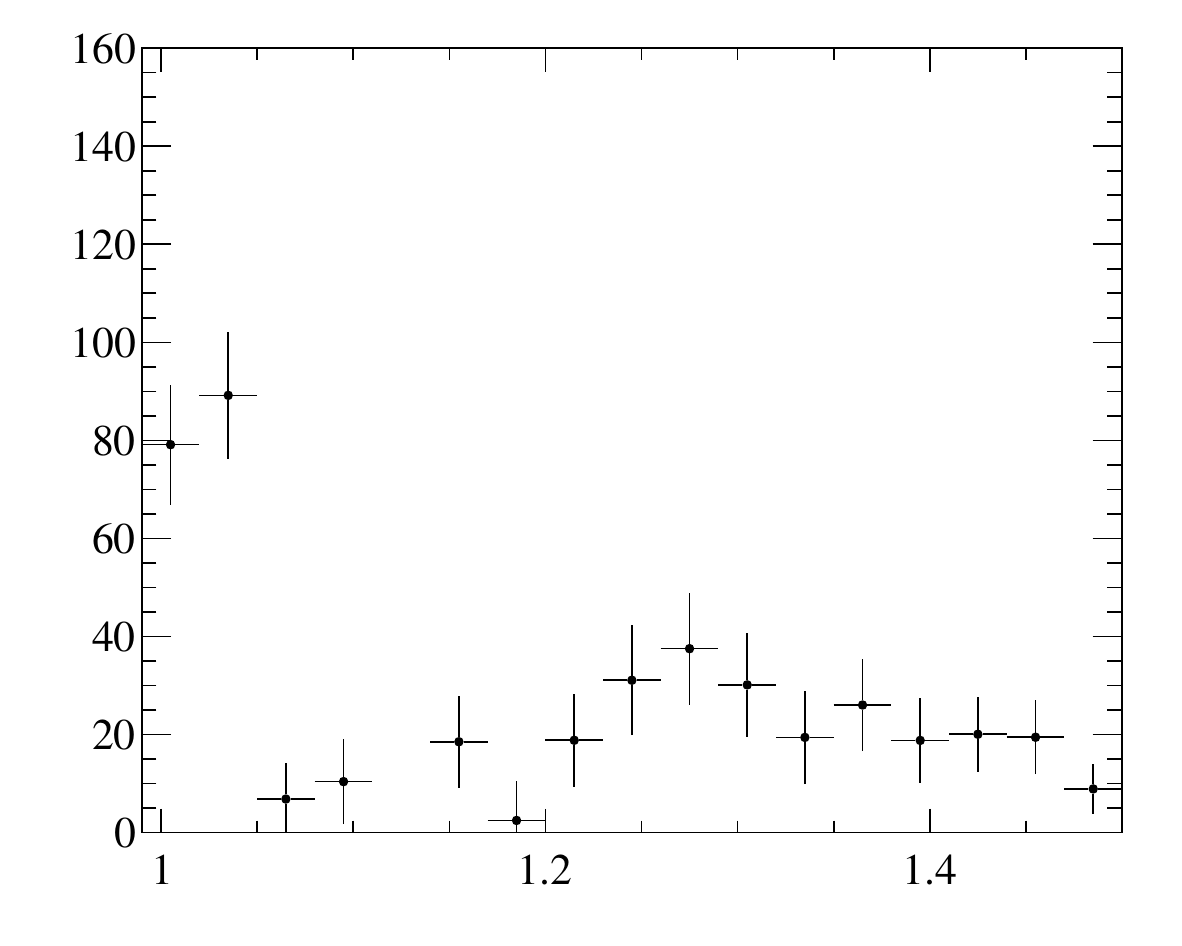}
    }
    \put(0,  70) {\Large\begin{sideways}Yield/(15\mevcc)\end{sideways}}
    \put(31, 98) {\scriptsize\rotatebox[]{90}{Yield/(30\mevcc)} }

    \put(57,73) {\scriptsize$m_{\Kp\Km}$} 
    \put(70,73) {\scriptsize$\left[\!\gevcc\right]$} 
    
    \put(75, 0) {\Large$m_{\Kp\Km}$} 
    \put(124, 0) {\Large$\left[\!\gevcc\right]$} 

\put(120,103){\Large\lhcb}

   \put( 91,96) {\begin{tikzpicture}[x=1mm,y=1mm]\filldraw[fill=red!35!white,draw=red,thick]  (0,0) rectangle (7,3);\end{tikzpicture} }

	\put(91,91){\color[RGB]{85,83,246}     {\hdashrule[0.0ex][x]{7mm}{1.0pt}{2.0mm 0.3mm} } }
	\put(91,86){\color[RGB]{255,153,51} {\rule{7mm}{2.0pt}}}
   \put( 91,81){\color[RGB]{121,220,117} {\rule{7mm}{2.0pt}}}

	\put( 99,96){{\small{\decay{\Bs}{\chicone(3872) \Pphi}}}}
	\put( 99,91){{\small{\decay{\Bs}{\chicone(3872) \Kp\Km}}}} 
	\put( 99,86){{\small{total}}}
	\put( 99,81){{\small{simulation}}}

  \end{picture}
 	\caption{\small 
 	Background-subtracted and 
 	efficiency-corrected 
 	$\Kp\Km$~mass distribution (points with error bars) 
    of 
    the~$\mbox{$\decay{\Bs}{\chicone(3872)\Kp\Km}$}$~decays.
    For~a~better 
 	visualisation, 
 	the~high\nobreakdash-mass region
 	the~plot is shown with a~reduced vertical scale.
 	A~fit, described in the~text, is overlaid.
 	The~expectation for phase\nobreakdash-space
 	simulated decays
    is shown as a~green solid line. 
    A~distribution with extended vertical scale is shown inset. 
    }
	\label{fig:KK_mss}
\end{figure}

The~$\jpsi\pip\pim\Kp\Km$ and $\jpsi\pip\pim$~mass distributions
together with projections of the~fit are shown in Fig.~\ref{fig:fit_XKK}. 
The~yield of~\mbox{$\decay{\Bs}{\chicone(3872)\Kp\Km}$}~signal decays 
is
\begin{equation}
N_{\decay{\Bs}{\chicone(3872)\Kp\Km}}  = 378 \pm 33\,,
\end{equation} 
which significantly exceeds 
the~yield of~$N_{\decay{\Bs}{\chicone(3782)\Pphi}}$ shown in
Table~\ref{tab:sim_res},
pointing to a~sizeable 
contribution from 
the~\mbox{$\decay{\Bs}{\chicone(3872)\Kp\Km}$}~decays,
where the~$\Kp\Km$~pair does not originate from a~\Pphi~meson.

The~fraction of 
\mbox{$\decay{\Bs}{\chicone(3872)\left(\decay{\Pphi}{\Kp\Km}\right)}$}~decays 
is estimated using an~unbinned 
maximum\nobreakdash-likelihood fit
to the~background\nobreakdash-subtracted 
$\Kp\Km$~mass distribution
from~signal \mbox{$\decay{\Bs}{\chicone(3872)\Kp\Km}$}~decays. 
The~background\nobreakdash-subtracted $\Kp\Km$~mass
distribution is obtained by 
applying the~\sPlot technique~\cite{Pivk:2004ty} 
to the~results of the~two\nobreakdash-dimensional 
fit to 
the~\mbox{$\decay{\Bs}{\chicone(3872)\Kp\Km}$}~decays
described above. 
The~background\nobreakdash-subtracted $\Kp\Km$~mass 
distribution is 
further corrected for 
the~$\Kp\Km$~mass\nobreakdash-dependent efficiency by applying
a~weight, 
\begin{equation}\label{eq:rcalctwo}
       w_{\varepsilon} \left(m_{\Kp\Km}\right) \equiv 
    \dfrac 
    { \varepsilon_{\decay{\Bs}{\chicone(3872)\Pphi}}}  
    { \varepsilon_{\decay{\Bs}{\chicone(3872)\Kp\Km}} \left(m_{\Kp\Km}\right)}\,, 
\end{equation}
 to each candidate.
The~efficiencies 
 $\varepsilon_{\decay{\Bs}{\chicone(3872)\Pphi}}$ 
 and 
$\varepsilon_{\decay{\Bs}{\chicone(3872)\Kp\Km}}$ 
are calculated using simulated samples,  
where a~phase\nobreakdash-space decay model is 
used for 
the~three\nobreakdash-body~\mbox{$\decay{\Bs}{\chicone(3872)}\Kp\Km$}~decays.
The~background\nobreakdash-subtracted 
and efficiency\nobreakdash-corrected  
 	$\Kp\Km$~mass distribution  
 	of the~$\mbox{$\decay{\Bs}{\chicone(3872)\Kp\Km}$}$ ~candidates
 	is  shown in Fig.~\ref{fig:KK_mss}. 
 In~addition to a~clear narrow structure, corresponding 
 to \mbox{$ \decay{\Bs}{\chicone(3872) \left(  \decay{\Pphi}{\Kp\Km}\right)  }$}
 decays, a~sizeable number of 
 \mbox{$\decay{\Bs}{\chicone(3872) \Kp\Km}$}~decays,
 where the $\Kp\Km$~pair does not originate from the~\Pphi~meson is visible.  
 The~$\Kp\Km$~mass distribution for 
 \mbox{$m_{\Kp\Km}>1.1\gevcc$} 
 cannot be described by phase\nobreakdash-space, 
 and possibly contains contributions from 
 the~\mbox{${\mathrm{f}}_0(980)$},
 \mbox{${\mathrm{f}}_2(1270)$}, 
 \mbox{${\mathrm{f}}_0(1370)$} 
 and 
 \mbox{${\mathrm{f}}^{\prime}_2(1525)$}~resonances 
 decaying to a~pair of kaons, 
 as has been observed 
 in~\mbox{$\decay{\Bs}{\jpsi\Kp\Km}$}~decays~\cite{LHCb-PAPER-2012-040,LHCb-PAPER-2017-008}.
 An~amplitude analysis of a larger data sample 
 would be required to properly disentangle 
 individual contributions.   However,   
 a~narrow \Pphi~component can be separated from the 
 non-\Pphi~components using 
 an~unbinned maximum\nobreakdash-likelihood fit 
 to the~background\nobreakdash-subtracted and 
 efficiency\nobreakdash-corrected $\Kp\Km$~mass 
 distribution.   
 The~fit function comprises two components 
 \begin{enumerate}
     \item a~component corresponding 
     to~\mbox{$\decay{\Bs}{\chicone(3872) \left( \decay{\Pphi}{\Kp\Km} \right) }$}~decays,
     modelled  by the~\Pphi~signal template (see Sec.~\ref{sec:xcc_phi})  multiplied by 
     the~phase\nobreakdash-space function
     $\Phi_{2,3}(m_{\Kp\Km})$ for~the three\nobreakdash-body
     \mbox{$\decay{\Bs}{\chicone(3872)\Kp\Km}$}~decay;
     \item a component that accounts for 
     non\nobreakdash-resonant
     \mbox{$\decay{\Bs}{\chicone(3872)\Kp\Km}$}~decays  
     and decays via broad~high\nobreakdash-mass
     $\Kp\Km$~intermediate states, modelled 
     by a~product of a~phase\nobreakdash-space function
     $\Phi_{2,3}(m_{\Kp\Km})$ for three\nobreakdash-body
     \mbox{$\decay{\Bs}{\chicone(3872)\Kp\Km}$}~decays and 
     a~third\nobreakdash-order  polynomial function.
     
 \end{enumerate}
 The~shape of the~second component is flexible enough to 
     accommodate 
     contributions
     from wide \mbox{$\Kp\Km$}~resonances.
 The~projection of the~fit is overlaid 
 in Fig.~\ref{fig:KK_mss}.
 The~fraction of the~\mbox{$\decay{\Pphi}{\Kp\Km}$}~signal component is found  
 to be 
 \begin{equation}
    {f_{\Pphi}=\left(38.9\pm 4.9\right)\%\,.}
 \end{equation} 
 This fraction is converted into 
 the~ratio of branching fractions
\Rnonphi,
defined in Eq.~\eqref{eq:rnonphi},
 \begin{equation}
 \Rnonphi
 = 
 \dfrac{1}{f_{\Pphi}} -1 
 = 1.57 \pm 0.32 \,, 
 \end{equation} 
 where the~uncertainty is statistical.
     Systematic uncertanties are discussed in Sec.~\ref{sec:sys}.
 This~is the~first observation  of
the~decay \mbox{$\decay{\Bs}{\chicone(3872)\Kp\Km}$}, 
where $\Kp\Km$~pair does not originate from a~\Pphi~meson.

%% file: kstar.tex
\section{
$\decay{\Bs}{\jpsi\Kstarz\Kstarzb}$~decays}
\label{sec:kstar}

The yield of~$\decay{\Bs}{\jpsi\Kstarz\Kstarzb}$~decays is 
determined using a~three-dimensional unbinned extended 
maximum\nobreakdash-likelihood
fit to the~\mbox{$\jpsi\pip\pim\Kp\Km$},
\mbox{$\Kp\pim$} and \mbox{$\Km\pip$}~mass distributions in 
the region defined by 
$m_{\Kp\pim}<1.2\gevcc$ 
and $m_{\Km\pip}<1.2\gevcc$. 
To~eliminate overlap with 
the~samples used in Sec.~\ref{sec:xcc_phi}, 
only $\jpsi\pip\pim\Kp\Km$~combinations 
with $m_{\Kp\Km}>1.06\gevcc$ 
 that do not fall into 
the~narrow regions around the $\psitwos$ and $\chicone(3872)$~masses, 
$3.679<m_{\jpsi\pip\pim}<3.694\gevcc$ and 
$3.864<m_{\jpsi\pip\pim}<3.881\gevcc$, are used here. 

\begin{figure}[tb]
  \setlength{\unitlength}{1mm}
  \centering
  \begin{picture}(160,130)
    \put(  0, 65){ 
      \includegraphics*[width=80mm,height=65mm,%
      ]{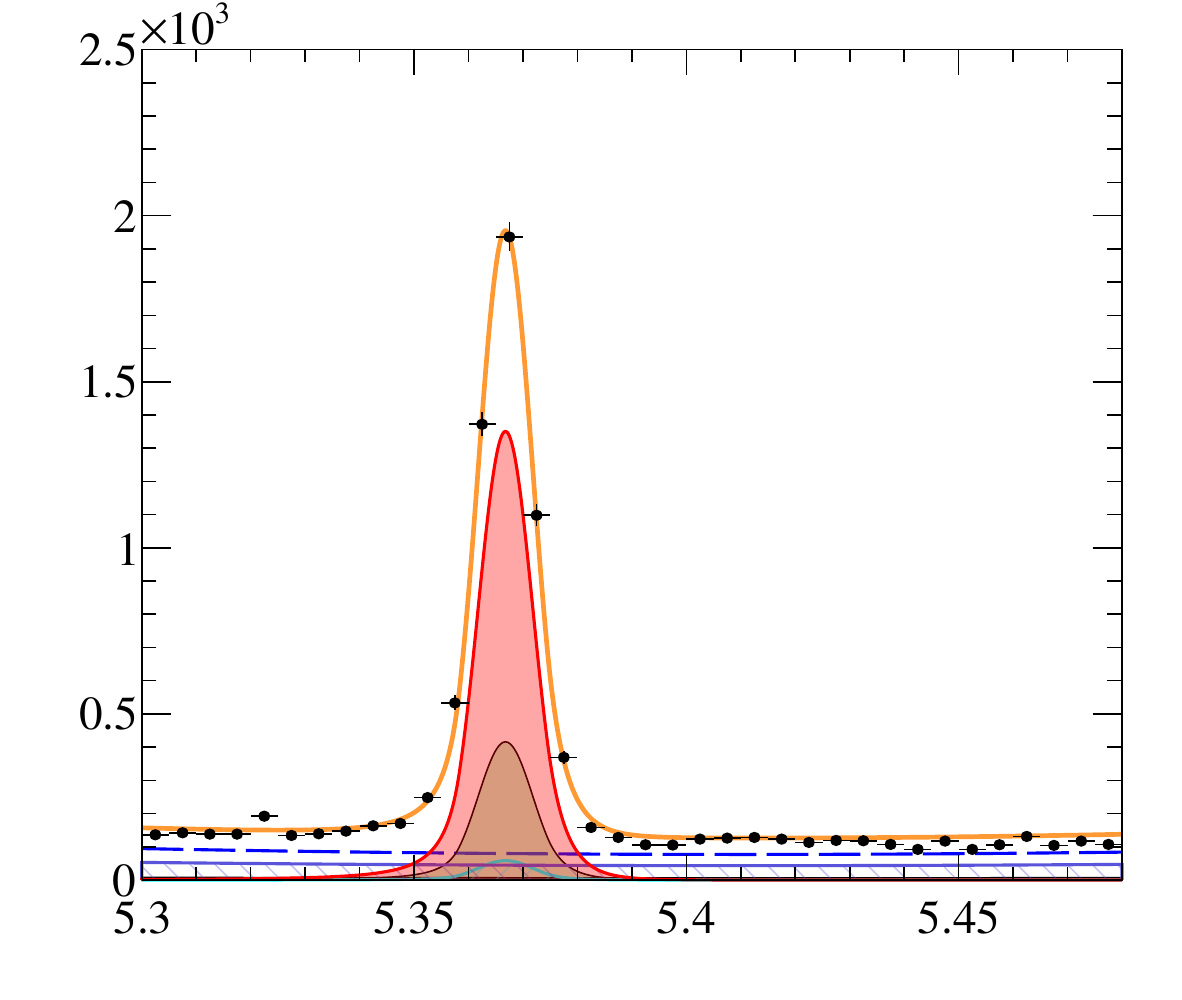}
    }
     \put( 80, 65){ 
      \includegraphics*[width=80mm,height=65mm,%
      ]{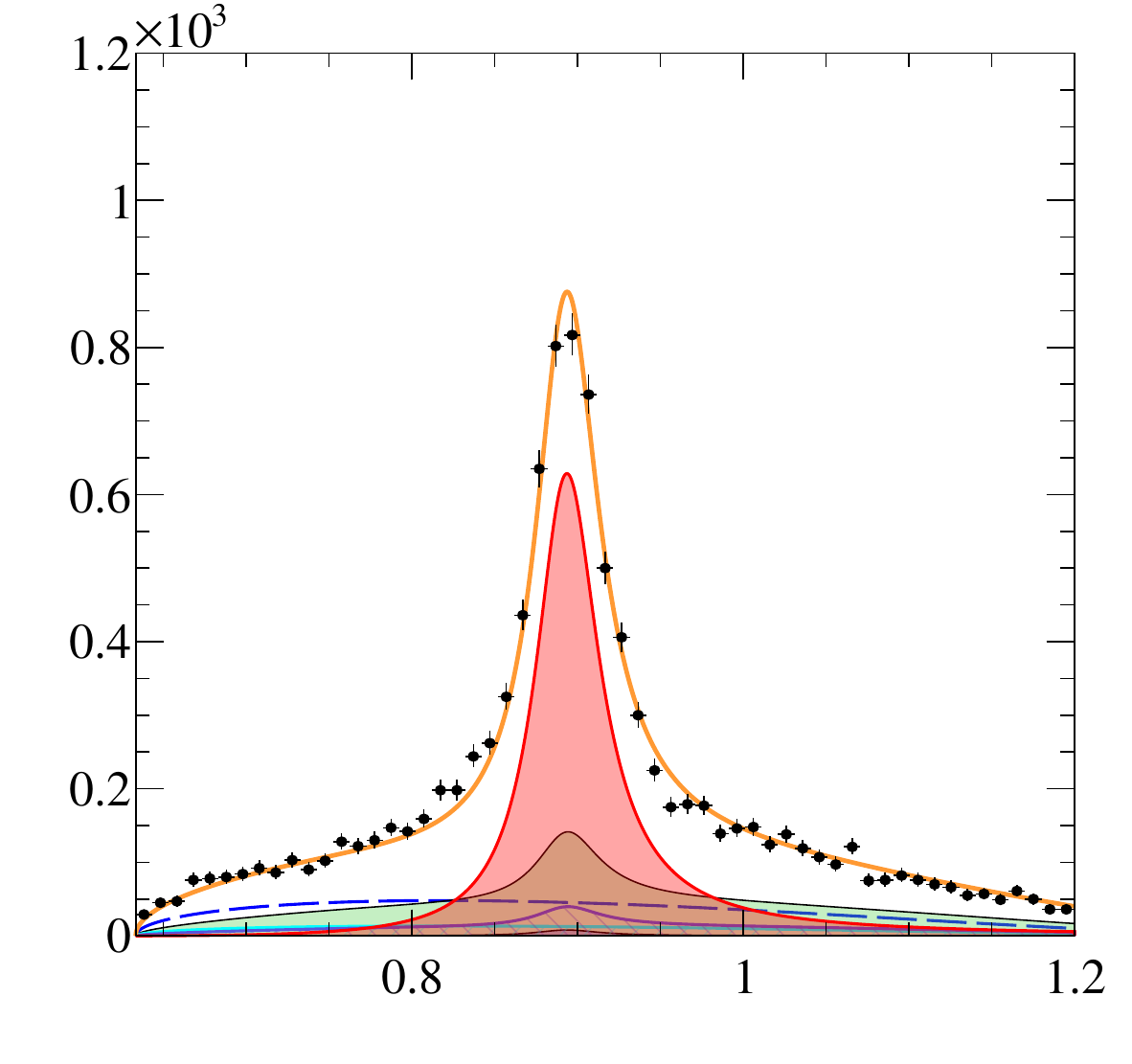}
    }
    \put(  0,  0){ 
      \includegraphics*[width=80mm,height=65mm,%
      ]{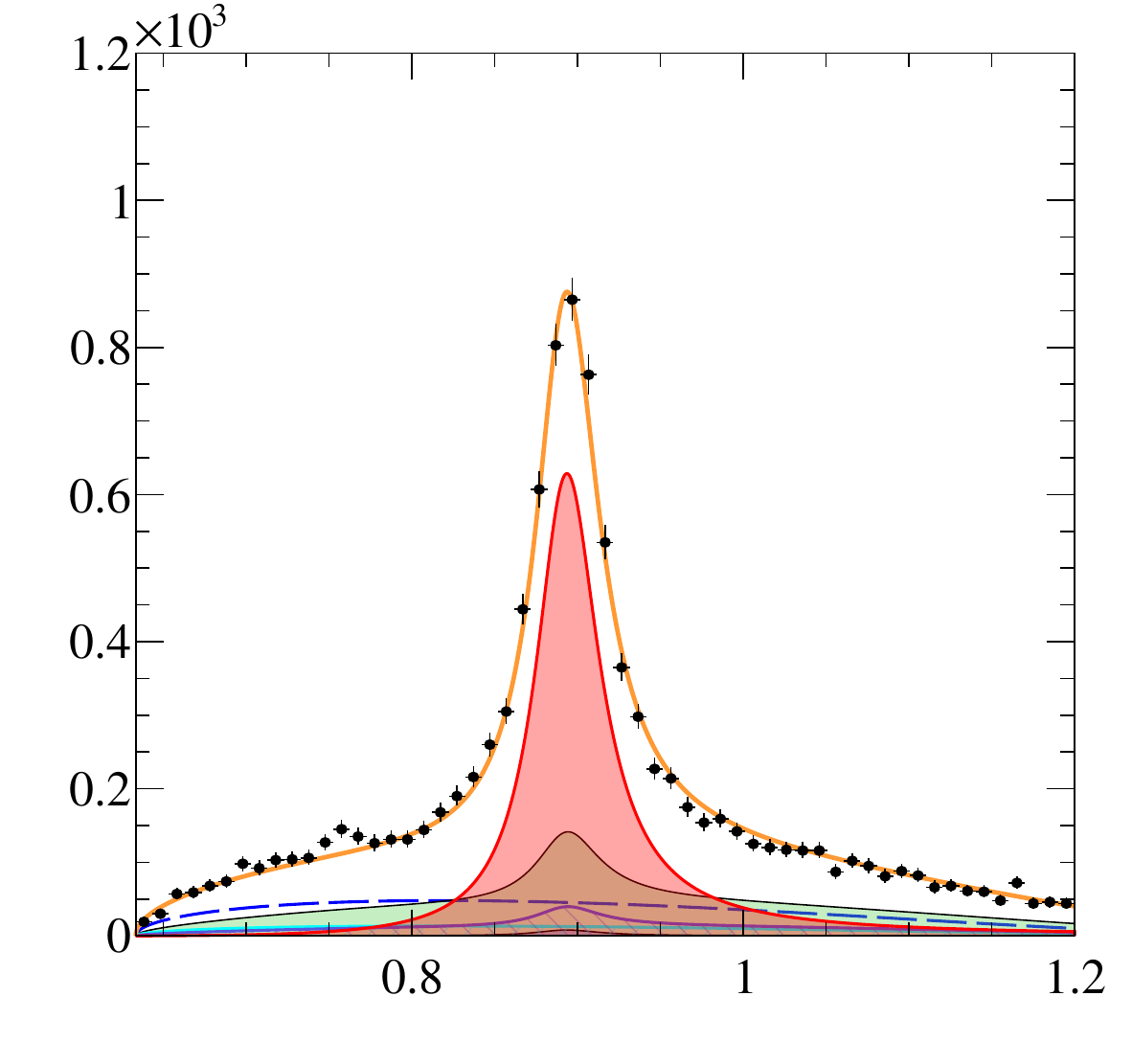}
    }
    \put(  0, 105){\rotatebox[]{90}{Candidates$/(5\mevcc)$} }
    \put( 81, 105){\rotatebox[]{90}{Candidates$/(10\mevcc)$} }
    \put(  0,  40){\rotatebox[]{90}{Candidates$/(10\mevcc)$} }
    \put( 30, 65) { $m_{\jpsi\pip\pim\Kp\Km}$} 
    \put( 62, 65) { $\left[\!\gevcc\right]$} 
    \put(115, 65) { $m_{\Kp \pim}$} 
    \put(142, 65) { $\left[\!\gevcc\right]$} 
    \put( 35,  0) { $m_{\Km \pip}$} 
    \put( 62,  0) { $\left[\!\gevcc\right]$} 
    \put( 14,52.5){\scriptsize$5.350<m_{\jpsi\pip\pim\Kp\Km}<5.384\gevcc$}
	\put( 14,56.5){\scriptsize$0.835<m_{\Kp \pim}<0.955\gevcc$}

	\put( 14,117.5){\scriptsize$0.835<m_{\Kp \pim}<0.955\gevcc$}
	\put( 14,121.5){\scriptsize$0.835<m_{\Km \pip}<0.955\gevcc$}

	\put( 94,117.5){\scriptsize$5.350<m_{\jpsi\pip\pim\Kp\Km}<5.384\gevcc$}
	\put( 94,121.5){\scriptsize$0.835<m_{\Km \pip}<0.955\gevcc$}

	\definecolor{gr}{rgb}{0.35, 0.83, 0.33}
	\definecolor{br}{rgb}{0.43, 0.98, 0.98}
	\definecolor{vi}{rgb}{0.39, 0.37, 0.96}
	\definecolor{db}{rgb}{0.1, 0.08, 0.41}
	\put( 64, 57){\small\lhcb}
	\put( 64,122){\small\lhcb}
	\put(144,122){\small\lhcb}

	 \put(90,50) {\begin{tikzpicture}[x=1mm,y=1mm]\filldraw[fill=red!35!white,draw=red,thick]  (0,0) rectangle (8,3);\end{tikzpicture} }
    \put(90,45) {\begin{tikzpicture}[x=1mm,y=1mm]\filldraw[fill=gr!35!white,draw=gr,thick]  (0,0) rectangle (8,3);\end{tikzpicture} }
    \put(90,40){\begin{tikzpicture}[x=1mm,y=1mm]\draw[thin,br,pattern=north east lines, pattern color=br]  (0,0) rectangle (8,3);\end{tikzpicture} }
    \put(90,35) {\begin{tikzpicture}[x=1mm,y=1mm]\filldraw[fill=vi!35!white,draw=vi,thick]  (0,0) rectangle (8,3);\end{tikzpicture} }
    \put(90,30){\begin{tikzpicture}[x=1mm,y=1mm]\draw[thin,vi,pattern=north west lines, pattern color=vi]  (0,0) rectangle (8,3);\end{tikzpicture} }
    \put(90,25){\color[RGB]{85,83,246}     {\hdashrule[0.0ex][x]{8mm}{1.0pt}{2.0mm 0.3mm} } }
	\put(90,20){\color[RGB]{255,153,51} {\rule{8mm}{2.0pt}}}
	\put( 100,51){\small{\decay{\Bs}{\jpsi \Kstarz \Kstarzb}}}
	\put( 100,46){\small{\decay{\Bs}{\jpsi\Kstar  \kaon \pion}}}
   
    \put( 100,41){\small{\decay{\Bs}{\jpsi\pip\pim\Kp\Km}}}
	\put( 100,35){\small comb. $\jpsi \Kstarz \Kstarzb$}
    \put( 100,30){\small comb. $\jpsi\Kstar\kaon \pion$}

	\put( 100,25){\small{comb. bkg.}}
	\put( 100,20.05){\small{total}}
  \end{picture}
  \caption { \small
   Distributions of 
   the~(top left)~\mbox{$\jpsi\pip\pim\Kp\Km$},
   (top right)~\mbox{$\Kp\pim$}
   and 
   (bottom left)~\mbox{$\Km\pip$}
   mass of selected $\BsTokstkst$ candidates 
   shown as points with error bars.   
   A~fit, described in the~text, is overlaid.
  }
  \label{fig:fit_kstar}
\end{figure}

The~fit model is similar to that used in Sec.~\ref{sec:xcc_phi}
but with some modifications.
First, the~model is symmetric with respect to 
an~interchange of ~$\Kp\pim$ and $\Km\pip$~pairs. 
Second, for components that account  for~\mbox{$\Kstarz\Kstarzb$},  
\mbox{$\Kstarz\Km\pip$} 
or \mbox{$\Kp\pim\Kstarzb$}~combinations, 
corrections 
are applied due to  the~limited phase
space available in the decays.
These shapes are  
derived from fits to simulated samples and comprise~symmetric  
products of phase\nobreakdash-space functions 
and linear polynomials.
The~$\Kstarz(\Kstarzb)$~signal is parameterised by 
a~relativistic P\nobreakdash-wave Breit\nobreakdash--Wigner 
function.  
The~width of the~\Kstarz~meson, 
\mbox{$47.3\pm0.5\mev$}, 
is not small~\cite{PDG2020} and 
the~fit ranges are wide, hence the~correct determination 
of all  components
would require a~full amplitude analysis that properly accounts for 
interference effects. Such an analysis is beyond 
the~scope of this paper.   
However, fits to simulated samples 
of \mbox{$\decay{\Bs}{\jpsi\pip\pim\Kp\Km}$}
decays with different compositions 
of intermediate states  
show that 
the~simple model described here  
allows for a~reliable determination of
the~\mbox{$\decay{\Bs}{\jpsi\Kstarz\Kstarzb}$} component.  
The~\mbox{$\jpsi\pip\pim\Kp\Km$},
\mbox{$\Kp\pim$}
and \mbox{$\Km\pip$}~mass distributions together with 
projections of the~fit 
are shown in Fig.~\ref{fig:fit_kstar}
and the~parameters of interest are summarized in 
Table~\ref{tab:fit_kstar}.
A~study of a~large sample of pseudoexperiments 
generated and fitted with the~nominal~model,  
indicates a~small bias of ${\mathcal{O}}(1\%)$~on 
the~signal yield. 
The~quoted yield is corrected for 
this bias.

\begin{table}[bt]
	\centering
	\caption{\small 
	Signal yield, $N_{\decay{\Bs}{\jpsi\Kstarz\Kstarzb}}$, 
	and mass of the~\Bs~meson,   $m_{\Bs}$,
	from the~fit   described in the~text. 
	The~uncertainties are statistical only. 
	}
	\label{tab:fit_kstar}
	\vspace{2mm}
	\begin{tabular*}{0.45\textwidth}{@{\hspace{3mm}}l@{\extracolsep{\fill}}lc@{\hspace{2mm}}}
	\multicolumn{2}{l}{Parameter} &  $\decay{\Bs}{\jpsi\Kstarz\Kstarzb}$  
   \\[1mm]
  \hline 
  \\[-2mm]
   \multicolumn{2}{l}{$N_{\decay{\Bs}{\jpsi\Kstarz\Kstarzb}}$}   
   & $\phantom{0.0}5447\pm 125\phantom{.}$   
   \\ 
   $m_{\Bs}$  &  $\left[\!\mevcc\right]$ 
   & $5366.79 \pm 0.06$ 
	\end{tabular*}
	\vspace{3mm}
\end{table}

The~ratio of branching fractions 
\Rkst,
defined in Eq.~\eqref{eq:r_kstar},  
is calculated as
\begin{equation}\label{eq:rcalcthree}
 \Rkst
 = 
 \dfrac { N_{\decay{\Bs}{\jpsi\Kstarz\Kstarzb}}}
        { N_{\decay{\Bs}{\psitwos \Pphi}}}
  \times 
  \dfrac { \varepsilon_{\decay{\Bs}{\psitwos \Pphi}}}
         { \varepsilon_{\decay{\Bs}{\jpsi\Kstarz\Kstarzb}}}
         = 1.22 \pm 0.03\,,
\end{equation}
where  
$\varepsilon_{\decay{\Bs}{\jpsi\Kstarz\Kstarzb}}$ and 
$\varepsilon_{\decay{\Bs}{\psitwos \Pphi}}$ are the
efficiencies for~\mbox{$\decay{\Bs}{\jpsi\Kstarz\Kstarzb}$} and 
\mbox{$\decay{\Bs}{\psitwos \Pphi}$}~decays, respectively,  
and the~signal yields
$N_{\decay{\Bs}{\psitwos\Pphi}}$ and 
$\decay{\Bs}{\jpsi\Kstarz\Kstarzb}$ 
are taken from 
Tables~\ref{tab:sim_res} and~\ref{tab:fit_kstar}, 
respectively. 
 The~efficiency ratio is found to be 
\mbox{$0.93\pm0.01$}, 
 where the~uncertainty is only that 
 due to the~size of the~simulated samples. 
     Systematic uncertanties are discussed in Sec.~\ref{sec:sys}.

%% file: mass.tex
\section{$\Bs$ mass measurement}
\label{sec:mmass}

The~precision  on the~\Bs~mass value, 
reported in Table~\ref{tab:sim_res}, 
is improved by  
imposing a~constraint 
on the reconstructed mass of the  $\psitwos$~state~\cite{Hulsbergen:2005pu}.
Applying this constraint improves 
the~\Bs~mass resolution and significantly 
decreases systematic uncertainties 
on the~mass measurement, since the~mass of the~$\psitwos$~meson
is known with high precision~\cite{Anashin:2015rca}. 
The~mass of the~\Bs~meson 
is determined from
an~unbinned extended 
maximum\nobreakdash-likelihood fit to 
the~\mbox{$\psitwos\Kp\Km$}~mass 
distribution for a~sample of~\mbox{$\decay{\Bs}{\jpsi\Kp\Km\pip\pim}$}~decays 
with \mbox{$m_{\Kp\Km}<1.06\gevcc$}   
and with 
the~\mbox{$\jpsi\pip\pim$}~mass
within a~narrow region 
around the~known mass of the~\psitwos~meson,
\mbox{$3.679<m_{\jpsi\pip\pim}<3.694\gevcc$}.  

\begin{figure}[t]
  \setlength{\unitlength}{1mm}
  \centering
  \begin{picture}(150,120)
    %
    \put(0,  0){ 
      \includegraphics*[width=150mm,height=120mm,%
      ]{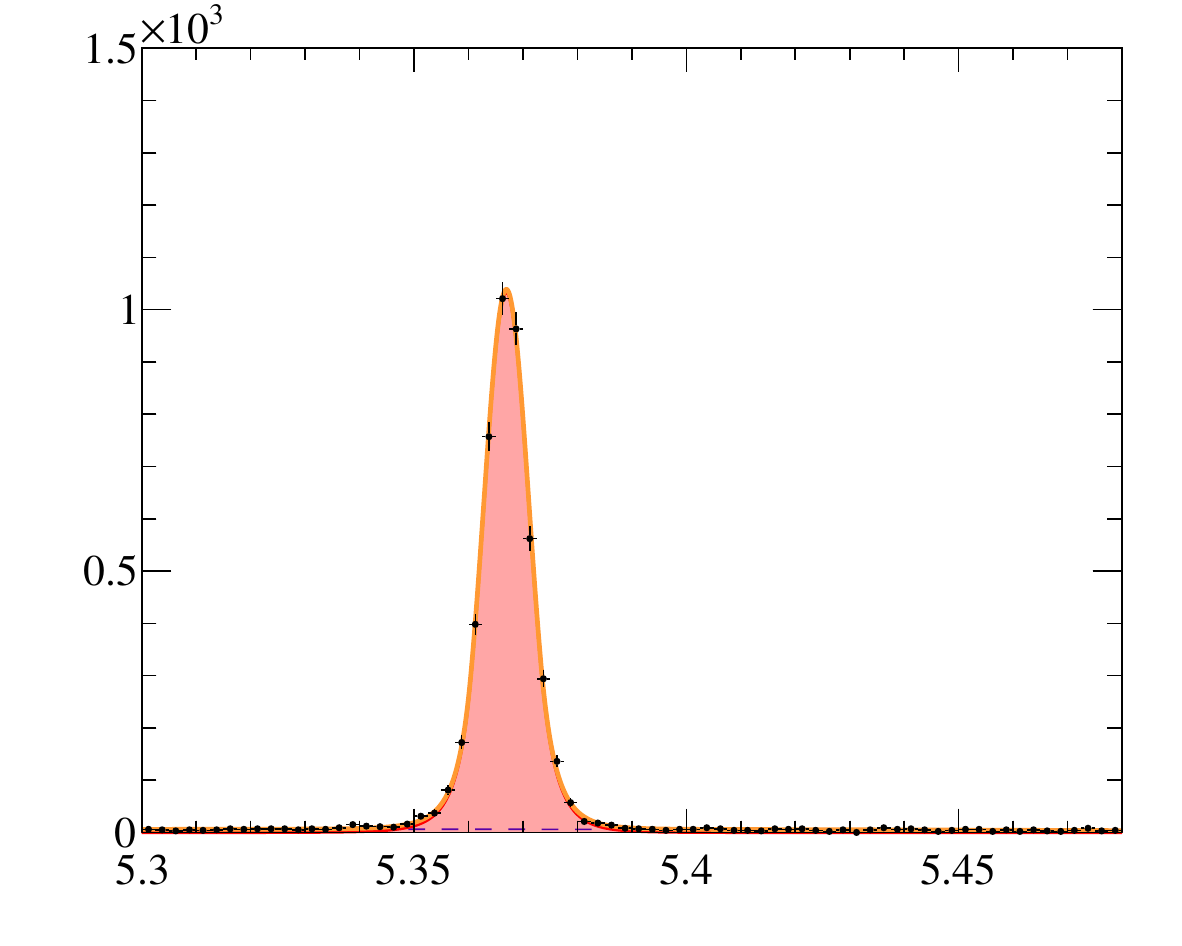}
    }
    \put( 3, 53) {\Large\begin{sideways}Candidates$/(2.5\mevcc)$\end{sideways} }
    \put(60, 2)  {\Large$m_{\psitwos\Kp\Km}$}
    \put(123,2)  {\Large$\left[\!\gevcc\right]$} 
    \put(125,105){\large\lhcb}
	\put(74,92) {\begin{tikzpicture}[x=1mm,y=1mm]\filldraw[fill=red!35!white,draw=red,thick]  (0,0) rectangle (8,3);\end{tikzpicture} }
	\put(74,85){\color[RGB]{85,83,246}     {\hdashrule[0.0ex][x]{8mm}{1.5pt}{2.0mm 0.3mm} } }
	\put(74,78){\color[RGB]{255,153,51} {\rule{8mm}{4.0pt}}}
	\put( 83,92.5){{$\decay{\Bs}{\left(\decay{\psitwos}{\jpsi\pip\pim}\right)\Kp\Km}$ }}
	\put( 83,85.0){{background}}
	\put( 83,78){{total}}
  \end{picture}
	\caption {\small
	Distribution of 
	the~\mbox{$\psitwos\Kp\Km$}~mass
	for~selected~\mbox{$\decay{\Bs}{\jpsi\Kp\Km\pip\pim}$}~candidates, 
	enriched in~\mbox{$\decay{\Bs}{\left(\decay{\psitwos}{\jpsi\pip\pim}\right) \left( \decay{\Pphi}{\Kp\Km}\right) }$}~decays~(points with error bars).
	A~fit, described in the text, is overlaid. 
	}
	\label{fig:bs_mass}
\end{figure}

\begin{table}[t]
	\centering
	\caption{\small 
	Signal yield, $N_{\Bs}$
	and mass of the~\Bs~meson, $m_{\Bs}$,
	from the~fit
	described in the~text
	to the~sample enriched in 
	the~\mbox{$\decay{\Bs}{\psitwos\Pphi}$}~decays.
	The~uncertainties are statistical only. 
	}	\label{tab:fit_mass}
	\vspace{2mm}
	\begin{tabular*}{0.45\textwidth}{@{\hspace{3mm}}l@{\extracolsep{\fill}}lc@{\hspace{2mm}}}
	\multicolumn{2}{l}{Parameter} &  
   \\[1mm]
  \hline 
  \\[-2mm]
   $N_{\Bs}$ &  
   & $\phantom{0.0}4505\pm 69\phantom{.0}$   \\
   $m_{\Bs}$  &  $\left[\!\mevcc\right]$ 
   & $5366.95 \pm 0.07$
	\end{tabular*}
\vspace{3mm}
\end{table}

The~\mbox{$\psitwos\Kp\Km$}~mass 
distribution is fitted with a~two-component function
comprising a signal component modelled with the~\Bs~signal 
template  and a~background component modelled 
with a~second\nobreakdash-order polynomial function. 
The~\mbox{$\psitwos\Kp\Km$}~mass 
distribution 
together with the~fit 
results is shown in Fig.~\ref{fig:bs_mass}. 
The~fit results are summarized in Table~\ref{tab:fit_mass}. 
Studies of simulated samples show 
that the~selection requirements
introduce a~small bias in the~measured 
mass of long\nobreakdash-lived 
heavy\nobreakdash-flavour 
hadrons~\cite{LHCb-PAPER-2014-002,LHCb-PAPER-2019-037,LHcb-PAPER-2020-003}. 
The~corrected value for the~\Bs~mass 
is found to be 
\begin{equation}\label{eq:bs_mass}
     m^{\mathrm{corr}}_{\Bs} =  5366.98\pm 0.07\mevcc\,,
\end{equation}
where the~uncertainty is statistical only.
    Systematic uncertanties are discussed in Sec.~\ref{sec:sys}.

%% file: Xpeak.tex
\section{$\jpsi \Pphi$ mass spectrum}
\label{sec:peak}

\begin{figure}[t]
  \setlength{\unitlength}{1mm}
  \centering
  \begin{picture}(160,65)
    %
    \put(  0, 0){ 
      \includegraphics*[width=80mm,height=65mm,%
      ]{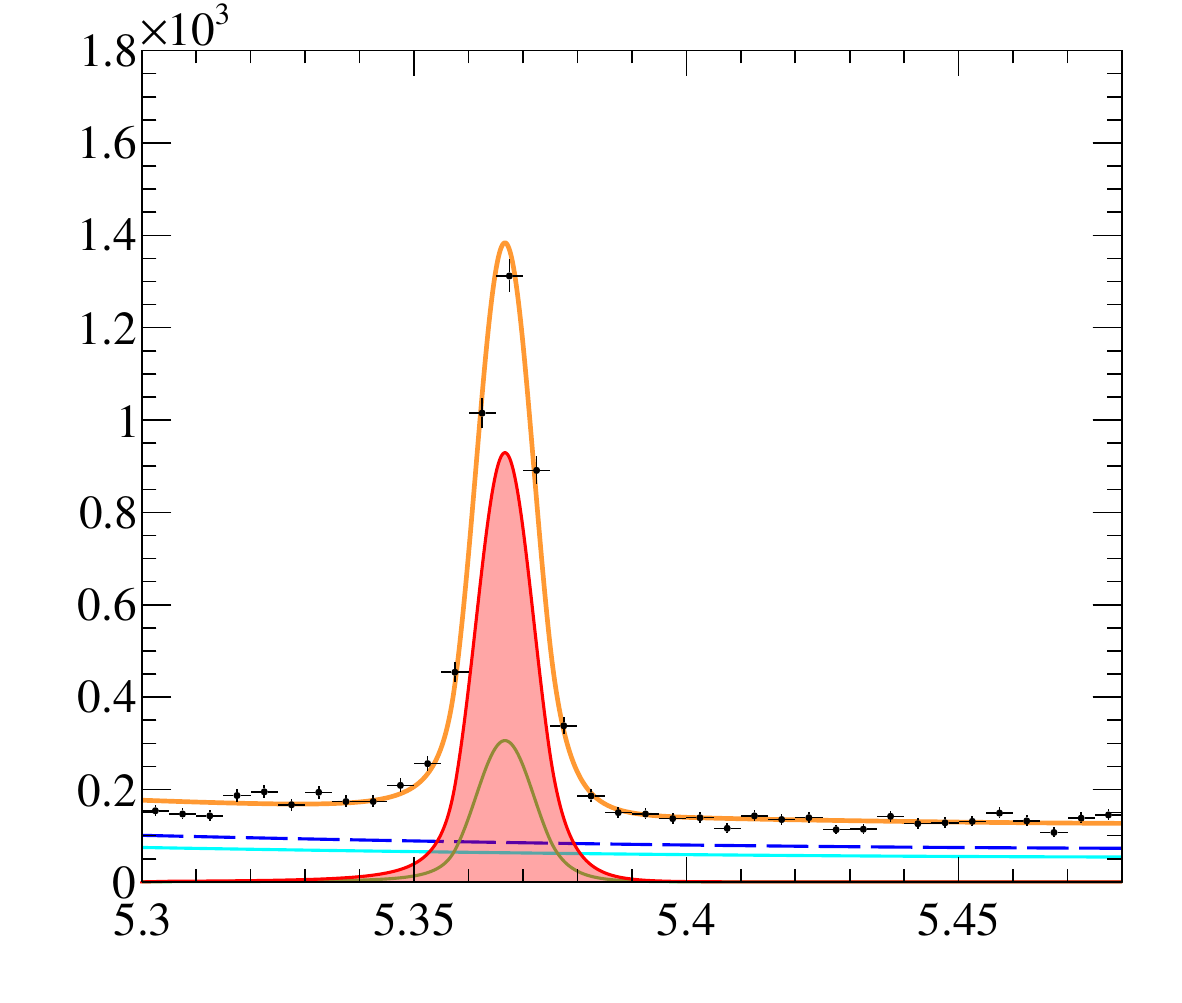}
    }
     \put( 80, 00){ 
      \includegraphics*[width=80mm,height=65mm,%
       ]{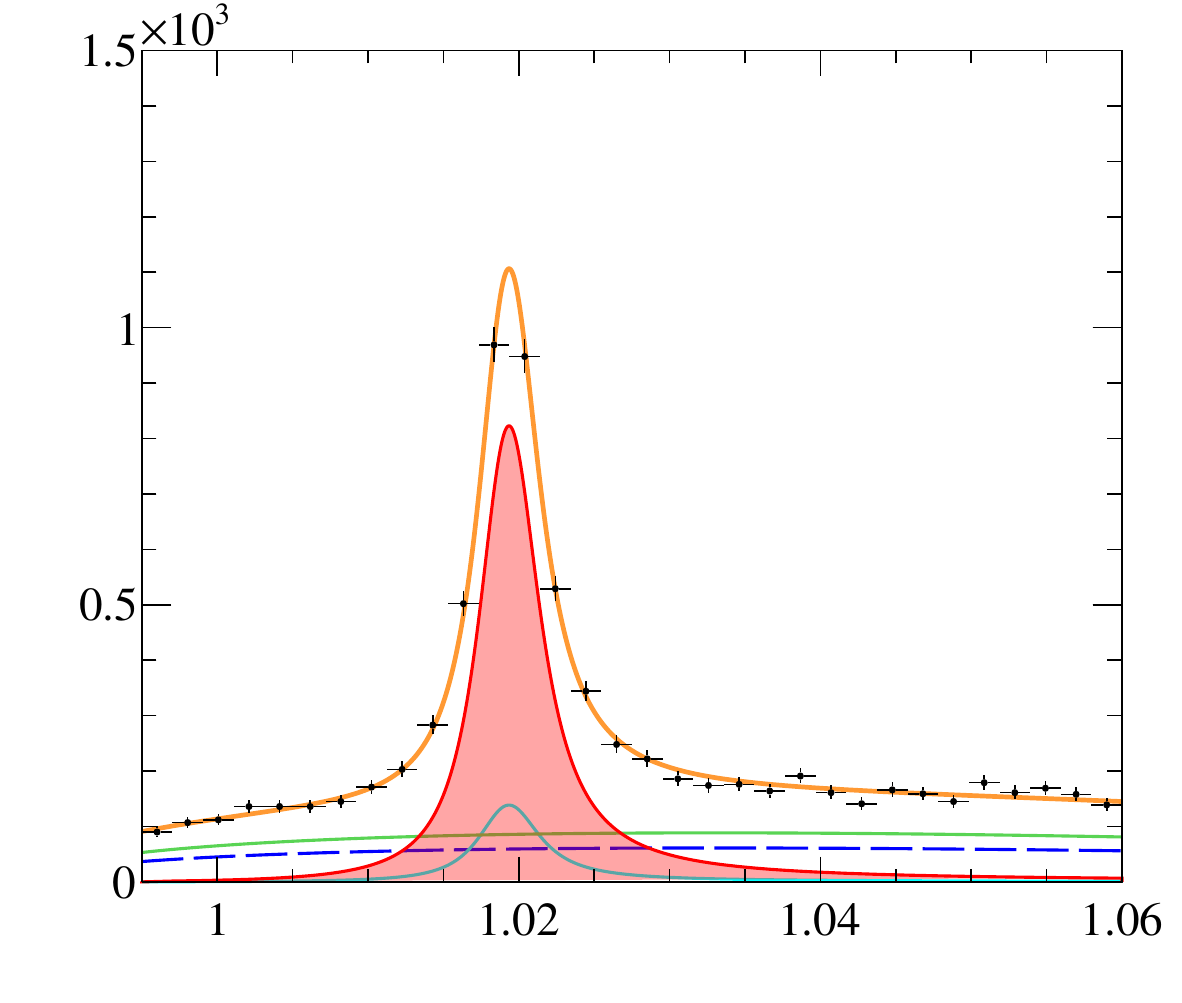}
    }
    \put( 80, 40) {\rotatebox[]{90}{Candidates$/(2\mevcc)$} }
    \put(  0,40) {\rotatebox[]{90}{Candidates$/(5\mevcc)$} }
    \put(115, 0) { $m_{\Kp\Km}$} 
    \put(142,0) { $\left[\!\gevcc\right]$} 
    \put( 30,  0) { $m_{\jpsi\pip\pim\Kp\Km}$} 
    \put( 62, 0) { $\left[\!\gevcc\right]$} 

    \put( 94,52.5){\scriptsize$5.35<m_{\jpsi\pip\pim\Kp\Km}<5.384\gevcc$}
	\put( 14,56.5){\scriptsize$1.01<m_{\Kp\Km}<1.03\gevcc$}
	
	\put( 64, 57){\small\lhcb}
	\put(144, 57){\small\lhcb}

	
	\put(40,45) {\begin{tikzpicture}[x=1mm,y=1mm]\filldraw[fill=red!35!white,draw=red,thick]  (0,0) rectangle (5,3);\end{tikzpicture} }
	\put(40,41) {\color[RGB]  {121,220,117}    {\rule{5mm}{2.0pt}}}
	\put(40,36) {\color[RGB]  {110,251,251}  {\rule{5mm}{2.0pt}}}
	\put(40,31){\color[RGB]{85,83,246}     {\hdashrule[0.0ex][x]{5mm}{1.0pt}{2.0mm 0.3mm} } }
	\put(40,26){\color[RGB]{255,153,51} {\rule{5mm}{2.0pt}}}

	\put( 47,46){{\scriptsize{\decay{\Bs}{\jpsi \pip \pim\Pphi}}}}
	\put( 47,41){{\scriptsize{\decay{\Bs}{\jpsi \pip \pim \Kp \Km  }}}}
	\put( 47,36){{\scriptsize{comb. $\jpsi \pip \pim \Pphi$}}}
	\put( 47,31){{\scriptsize{comb.$\jpsi\pip\pim\Kp\Km$}}}
	\put( 47,26){{\scriptsize{total}}}

  \end{picture}
	\caption {\small
	Distribution of  the~(left)~\mbox{$\jpsi\pip\pim\Kp\Km$} 
	and (right)~\mbox{$\Kp\Km$}~mass for selected
	\mbox{$\decay{\Bs}{\jpsi\pip\pim\Pphi}$}~candidates.
	A~fit, described in the text, is overlaid.}
	\label{fig:phi_BS}
\end{figure}

\begin{figure}[t]
  \setlength{\unitlength}{1mm}
  \centering
  \begin{picture}(150,60)

    \put(  0, 0){ 
      \includegraphics*[width=75mm,height=60mm,%
      ]{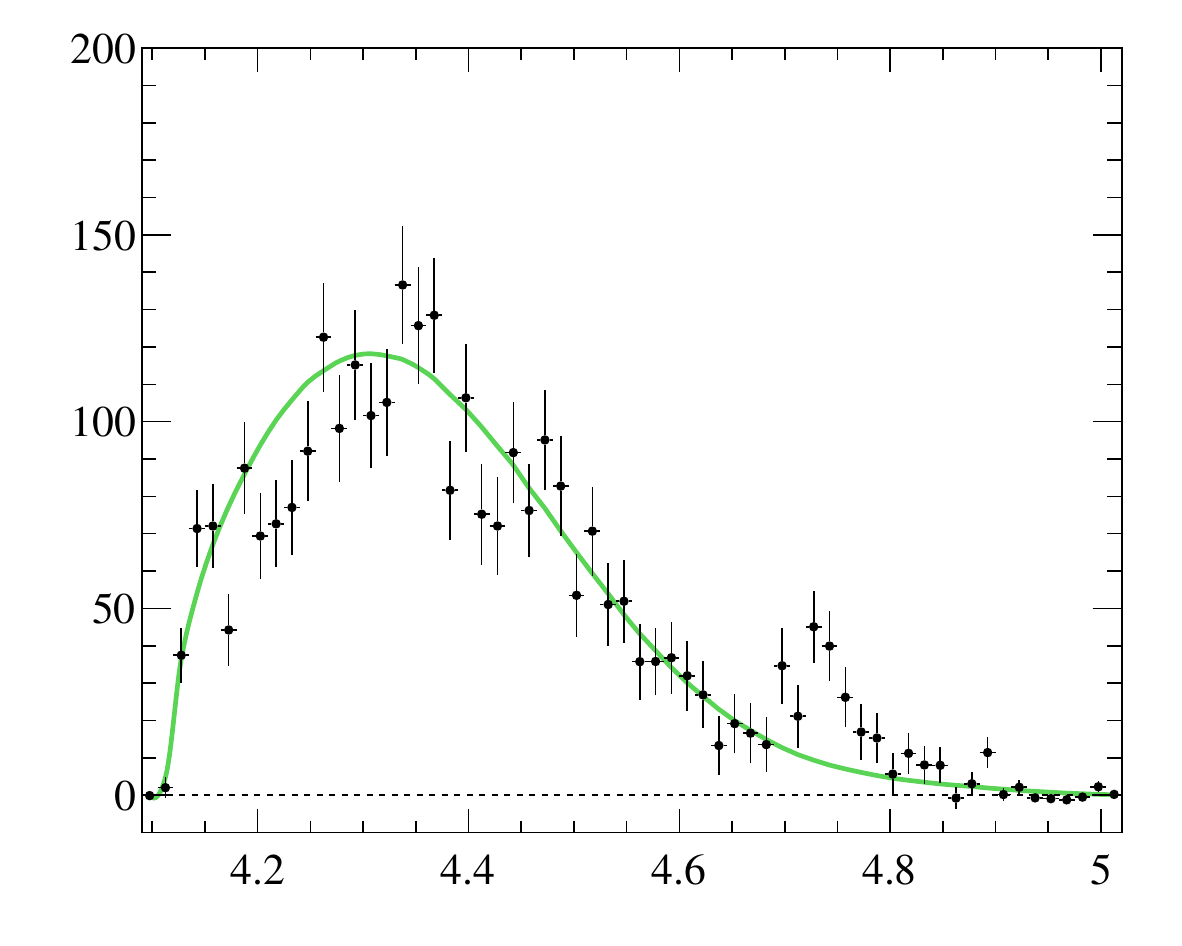}
    }
    \put( 75, 0){ 
      \includegraphics*[width=75mm,height=60mm,%
       ]{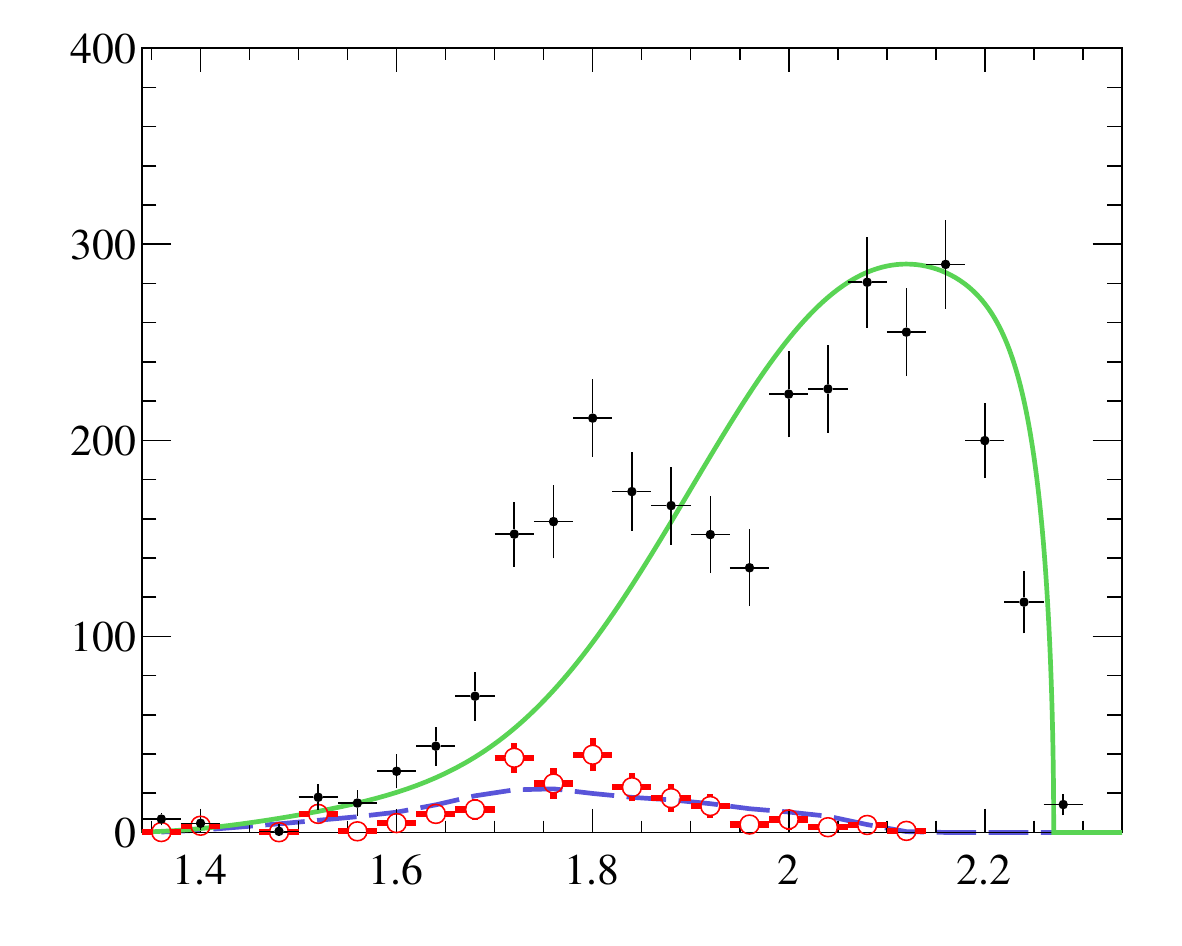}
    }
 
    \put( 76,40) {\rotatebox[]{90}{Yield/(40\mevcc)} }
    \put(  0,40) {\rotatebox[]{90}{Yield/(15\mevcc)} }

    \put(110, 0) { $m_{\Pphi\pip\pim}$} \put(132,0) { $\left[\!\gevcc\right]$} 
    \put( 30,  0) { $m_{\jpsi\Pphi}$} \put( 57, 0) { $\left[\!\gevcc\right]$} 
    \put(133,51){\small\lhcb}
    \put(58,51){\small\lhcb}

    \put(86.5,47){\color[RGB]{89,84,217} {\hdashrule[0.5ex][x]{6mm}{1.2pt}{2.2mm 0.2mm}}}
    \put(86.5,52){\color[RGB]{121,220,117}  {\rule{6mm}{1.2pt}}}
    \put(93,52){{\scriptsize{simulation}}}
    
    \put(93,47){\scriptsize{simulation, $4.68 < m_{\jpsi\Pphi} < 4.78\gevcc$}}

  \end{picture}
	\caption{\small 
	Background-subtracted 
	(left)~\mbox{$\jpsi\Pphi$}
	and (right)~\mbox{$\Pphi\pip\pim$}~mass 
	distributions from 
	\mbox{$\BsTopsiphtwopi$}~decays\,(points with error bars).
	The expectation from  
	simulated \mbox{$\decay{\Bs}{\jpsi\pip\pim\Pphi}$}~decays
	is overlaid (green solid line).
	In~the~right figure,
    the~background-subtracted 
    \mbox{$\Pphi\pip\pim$}~mass distribution 
    in the region~\mbox{$4.68<m_{\jpsi\Pphi}<4.78\gevcc$}
    is shown\,(red open circles  with error bars)
    together with the~corresponding expectation 
    from simulated \mbox{$\decay{\Bs}{\jpsi\pip\pim\Pphi}$}~decays\,(blue dashed line).
    }
	\label{fig:xxx}

\end{figure}

The $\jpsi\Pphi$~mass spectrum in~\mbox{$\decay{\Bs}{\jpsi\pip\pim\Pphi}$}~decays is 
studied using a~sample of selected \mbox{$\decay{\Bs}{\jpsi\pip\pim\Kp\Km}$}~candidates
with the~$\Kp\Km$~mass in the~range \mbox{$m_{\Kp\Km}<1.06\gevcc$}
and excluding the~\mbox{$\jpsi\pip\pim$}~mass regions 
around the~narrow \psitwos and $\chicone(3872)$~states, \ie
\mbox{$3.672<m_{\jpsi\pip\pim}<3.700\gevcc$} and 
\mbox{$3.864<m_{\jpsi\pip\pim}<3.880\gevcc$}.
A~two\nobreakdash-dimensional  
unbinned extended maximum\nobreakdash-likelihood 
is performed to the~\mbox{$\jpsi\pip\pim\Kp\Km$} and \mbox{$\Kp\Km$}~mass distributions.
The~fit function comprises  a~sum of four components:  
\begin{enumerate}
    \item a~component corresponding to~\mbox{$\decay{\Bs}{\jpsi\pip\pim\Pphi}$}~decays,
    parameterised by the~product of the~\Bs~and \Pphi~signal templates described in Sec.~\ref{sec:xcc_phi};
    \item a~component corresponding 
     to~\mbox{$\decay{\Bs}{\jpsi\pip\pim \Kp\Km  } $}~decays, 
     parameterised by the~product of the~\Bs~signal template and the~non\nobreakdash-resonant
    $\Kp\Km$~function;
    \item a~component corresponding to random 
    $\jpsi\pip\pim\Pphi$~combinations,  
    parameterised by the product of the~\Pphi~signal template and the~$\mathcal{F}_{\Bs}$~function; 
    \item  a~component describing random 
    \mbox{$\jpsi\pip\pim\Kp\Km$}~combinations, parameterised by the~product of the phase-space function
   $\Phi_{2,5}\left(m_{\Kp\Km}\right)$
   and the two\nobreakdash-dimensional non\nobreakdash-factorisable bilinear function described in Sec.~\ref{sec:xcc_phi}.
\end{enumerate}
The~$\jpsi\pip\pim\Kp\Km$ and $\Kp\Km$~mass spectra  
together with the~projections of the~fit are shown in 
Fig.~\ref{fig:phi_BS}. 
The~\sPlot technique is applied  to obtain
a~background\nobreakdash-subtracted $\jpsi\Pphi$~mass distribution
of~\mbox{$\decay{\Bs}{\jpsi\pip\pim\Pphi}$}~decays.
The~resulting distribution is shown in Fig.~\ref{fig:xxx}\,(left). 
It~shows a~prominent structure at a~mass 
around~\mbox{$4.74\gevcc$}.
No~such structure 
is seen if the~$\Kp\Km$~mass 
is restricted 
to the~region of~\mbox{$1.06<m_{\Kp\Km}<1.15\gevcc$}.  
This~structure cannot be explained 
by~\mbox{$\decay{\Bs}{\PX_{\ccbar}\Pphi}$}~decays 
via a~narrow intermediate $\PX_{\ccbar}$~resonance 
since 
contributions from \mbox{$\decay{\Bs}{\psitwos\Pphi}$}
and \mbox{$\decay{\Bs}{\chicone(3872)\Pphi}$}~decays are explicitly vetoed.
If no veto is applied, 
\mbox{$ \decay{\Bs}{\psitwos\Pphi} $}~decays would produce 
a~broad structure in 
the~$\jpsi\Pphi$~mass spectrum
that peaks around 4.76\gevcc 
and has a~width that is approximately twice 
that of the~observed structure.
Studies with simulated samples
indicate that   after the veto  is applied
the~remaining 
contributions from 
these decays 
are totally negligible. 
No sizeable contributions from 
decays via other narrow charmonium states 
are observed in 
the~background\nobreakdash-subtracted 
\mbox{$\jpsi\pip\pim$}~mass 
spectrum. 
The~background\nobreakdash-subtracted  $\pip\pim$~mass 
distribution of candidates in the~mass   range 
\mbox{$4.68<m_{\jpsi\Pphi}<4.78\gevcc$}
 is found to have no structure.
The~background\nobreakdash-subtracted 
$\Pphi\pip\pim$~mass spectrum 
of the~\mbox{$\decay{\Bs}{\jpsi\pip\pim\Pphi}$}~decays
is shown in Fig.~\ref{fig:xxx}\,(right). 
The~spectrum exhibits 
significant deviations from 
the~phase\nobreakdash-space distribution, 
indicating possible
presence of excited \Pphi~states,
referred to as $\Pphi^{\ast}$~states hereafter.
The~decays 
\mbox{$\decay{\Bs}{\jpsi\Pphi^{\ast}}$}
 via 
intermediate  
\mbox{$\Pphi(1680)$},
\mbox{$\Pphi(1850)$} or 
\mbox{$\Pphi(2170)$}~states~\cite{PDG2020}
are studied using simulated samples. 
It~is found that 
the~$\jpsi\Pphi$~mass spectra
from \mbox{$\decay{\Bs}{\jpsi\Pphi^{\ast}}$}~decays 
exhibit no structure 
and  for 
the~$\jpsi\Pphi$~mass
exceeding 4.4\gevcc 
can be described by 
a~monotonically decreasing function.
If the~intervals used to reject 
the~\mbox{$\decay{\Bs}{\psitwos\Pphi}$} and 
\mbox{$\decay{\Bs}{\chicone(3872)\Pphi}$} decays	
are significantly increased, in excess of 60\mevcc, 
it is possible to generate
two wide regions with decreased yields around 4.65\gevcc and 4.82\gevcc
in the~$\jpsi\Pphi$~mass spectrum. 
The~positions and shapes of these~dips 
depend on the~assumed mass and width of 
the~$\Pphi^{\ast}$~state and for 
certain choices 
of the~$\Pphi^{\ast}$~states, 
two dips in the~monotonically 
decreased spectrum could sculpt 
a~bump.
The~complicated interference between 
several decay chains, including 
different intermediate $\Pphi^{\ast}$~states, 
 could result in a~distorted
$\jpsi\Pphi$~mass spectrum. 
In~order to ascertain if
the~structure at 4.74\gevcc, 
seen in Fig.~\ref{fig:xxx}\,(left),
is resonant and not due to 
 interference
an~amplitude analysis, 
similar to that in  Refs.~\cite{LHCb-PAPER-2014-014,
LHCb-PAPER-2015-029,
LHCb-PAPER-2016-018,
LHCb-PAPER-2016-019} would be required.
Such an analysis is beyond the scope of this paper.

\begin{figure}[t]
  \setlength{\unitlength}{1mm}
  \centering
  \begin{picture}(150,120)
    \put(0, 0){ 
      \includegraphics*[width=150mm,height=120mm,%
      ]{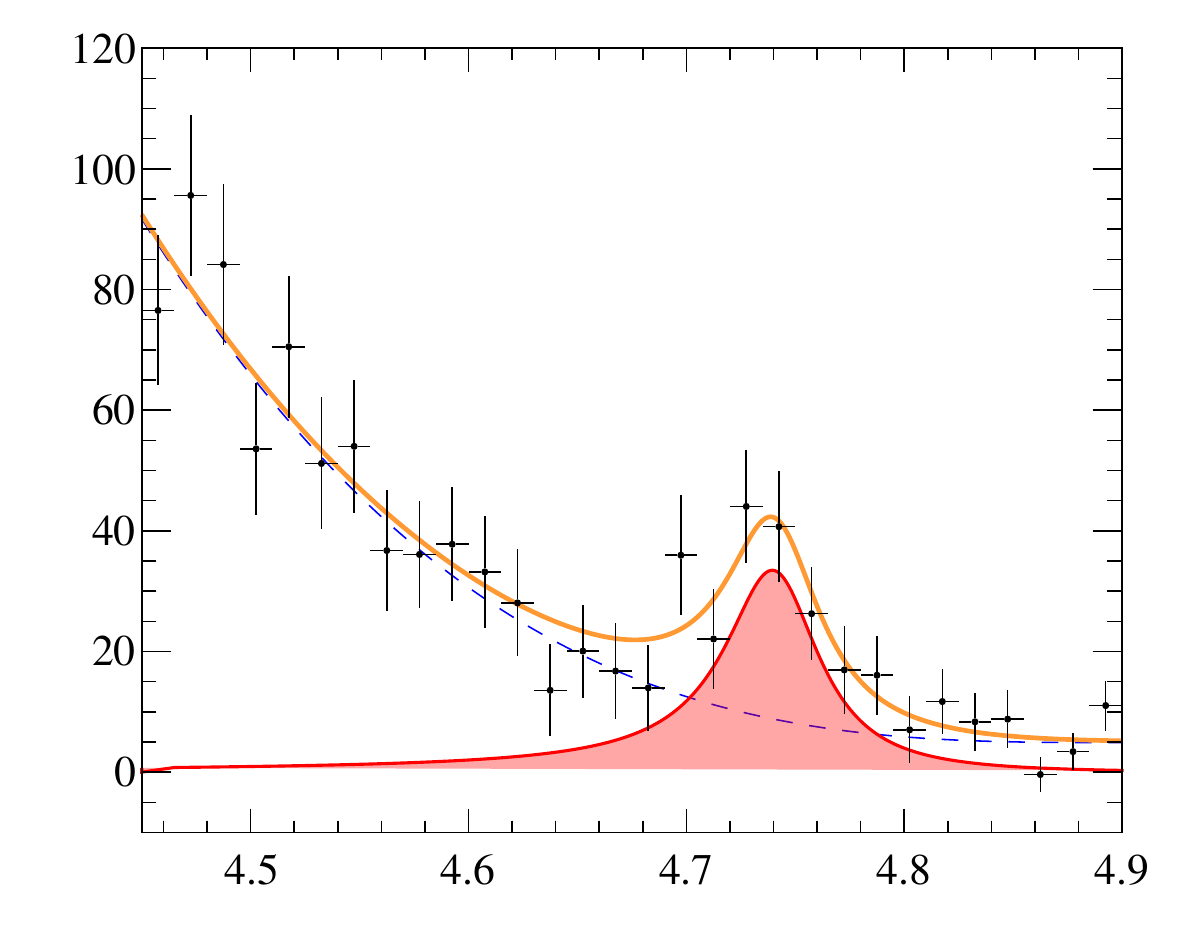}
    }
     \put(2,75) {\large\begin{sideways}Yield/(15\mevcc)\end{sideways}}
  
    \put(120,102){\Large\lhcb}
   \put(40,100) {\begin{tikzpicture}[x=1mm,y=1mm]\filldraw[fill=red!35!white,draw=red,thick]  (0,0) rectangle (13,3);\end{tikzpicture} }
	\put(40,94){\color[RGB]{85,83,246}     {\hdashrule[0.0ex][x]{13mm}{1.0pt}{2.0mm 0.5mm} } }
	\put(40,88){\color[RGB]{255,153,51} {\rule{13mm}{2.0pt}}}
	\put(55,100){$\decay{\Bs}{\PX(4740)\pip\pim}$}
	\put( 55,94){\decay{\Bs}{\jpsi \pip \pim \Pphi}}
	\put( 55,88){total}
    \put( 65,2){\Large{$m_{\jpsi\Pphi}$}}
    \put(125,2){\Large{$\left[\!\gevcc\right]$}}
  \end{picture}
	\caption{\small 
	Background-subtracted $\jpsi\Pphi$~mass 
	distribution from 
	\mbox{$\BsTopsiphtwopi$}~decays\,(points with error bars).
	A~fit, described in 
	the~text, is
	overlaid.}
	\label{fig:xxx_fit}

\end{figure}
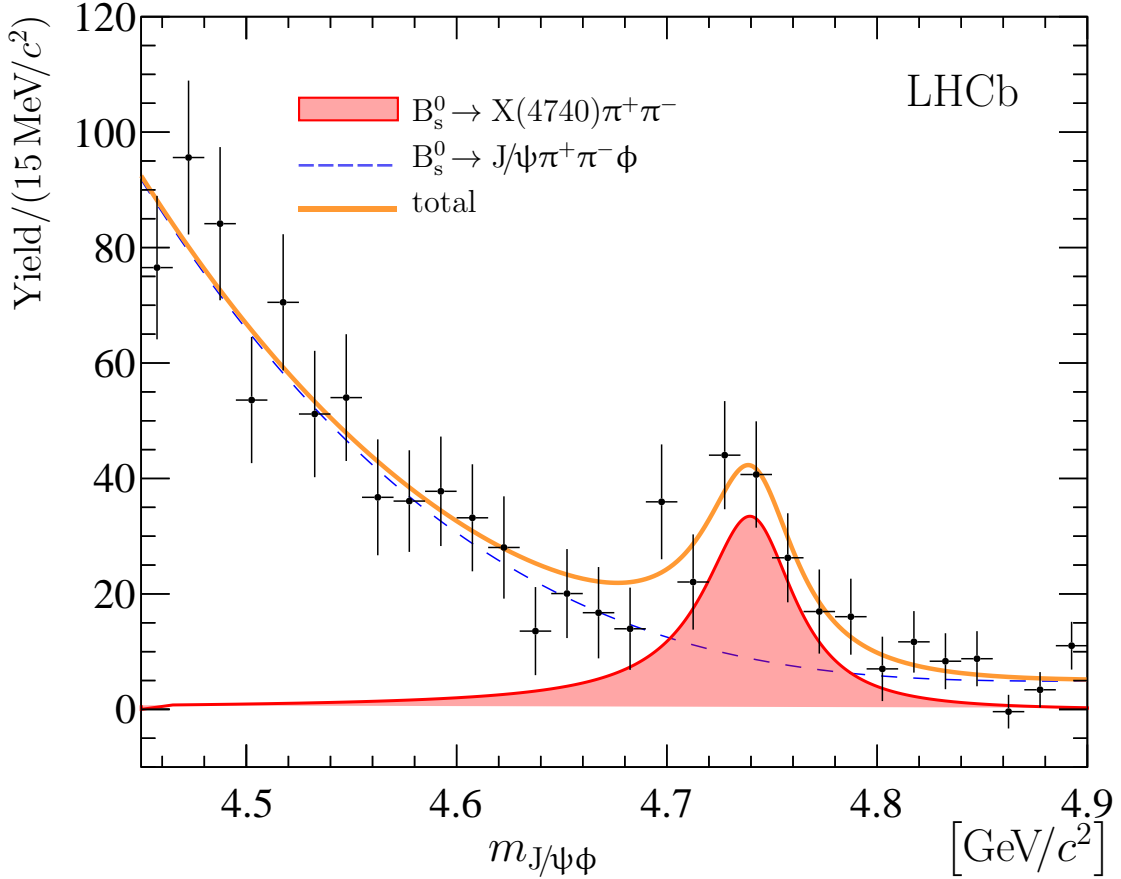

Under the~assumption that this~structure, 
referred to as $\PX(4740)$ hereafter, 
has a~resonant  nature, 
its~mass and width  are determined through 
an~unbinned extended maximum\nobreakdash-likelihood fit 
to the~background\nobreakdash-subtracted 
\mbox{$\jpsi\Pphi$}~mass distribution 
in the~range~\mbox{$4.45<m_{\jpsi\Pphi}<4.90\gevcc$}. 
The~fit function comprises
two components:
\begin{enumerate}
    \item a~signal component, parameterised by the~product 
    of the~squared absolute value of 
    a~relativistic 
    S\nobreakdash-wave Breit\nobreakdash--Wigner amplitude and a~two\nobreakdash-body
    phase\nobreakdash-space distribution  
    from four\nobreakdash-body \mbox{$\decay{\Bs}{\jpsi\pip\pim\Pphi}$}~decays, \mbox{$\Phi_{2,4}\left(m_{\jpsi\Pphi}\right)$};
    \item a~component, corresponding 
    to \mbox{$\decay{\Bs}{\jpsi\pip\pim\Pphi}$}~decays, 
    parameterised by the~product 
    of the~\mbox{$\Phi_{2,4}\left(m_{\jpsi\Pphi}\right)$}~function 
    and a~third\nobreakdash-order polynomial function.
\end{enumerate}
The~background\nobreakdash-subtracted 
\mbox{$\jpsi\Pphi$}~mass spectrum 
with superimposed results of the~fit 
is  shown in 
Fig.~\ref{fig:xxx_fit} and 
the~results are listed in Table~\ref{tab:xxx}.

The~statistical significance of the~observed structure
is estimated using Wilks' theorem~\cite{Wilks:1938dza}
and found   
to be 5.5~standard deviations.
The~significance estimate is 
validated using 
a~large number of pseudoexperiments
comprising 
 no~$\PX(4740)$~signal component.
The~mass and width of the~$\PX(4740)$~structure   
qualitatively agree with 
those of 
the~$\chiczero(4700)$~state~observed by the~LHCb collaboration 
in an~amplitude analysis  of~\mbox{$\decay{\Bu}{\jpsi\Pphi\Kp}$}~decays
of 
\mbox{$m_{\Pchi_{\cquark0}(4700)}= 4704\pm10^{\,+\,14}_{\,-\,24}\mevcc$}
and 
\mbox{$\Gamma_{\Pchi_{\cquark0}(4700)}=120\pm31^{\,+\,42}_{\,-\,33}\mev$}~\cite{LHCb-PAPER-2016-018,LHCb-PAPER-2016-019}.
Interpreting the~observed  structure as the~$\chiczero(4700)$~state 
and repeating the~fit 
using the~measurements from Refs.~\cite{LHCb-PAPER-2016-018,LHCb-PAPER-2016-019} 
as~Gaussian constraints, 
the~resulting mass and width  
differ only slightly from those listed in Table~\ref{tab:xxx}.
    A~$p$\nobreakdash-value of
    the~hypothesis that the~$\PX(4740)$~state is
    the~$\chiczero(4700)$~state
    is estimated neglecting correlations
    for the~systematic uncertainties,
    discussed in Sec.~\ref{sec:sys},  
    and it corresponds to 6\%. 
        The~measured mass is close to the~value expected
     for
     a~$\cquark\squark\cquarkbar\squarkbar$~tetraquark with
     quantum numbers $\mathrm{J}^{\mathrm{PC}}=2^{++}$~\cite{Ebert:2008kb}.

\begin{table}[tb]
	\centering
	\caption{\small 
    Signal yield $N_{\PX(4740)}$, 
    mass $m_{\PX(4740)}$ and 
    width $\Gamma_{\PX(4740)}$ of 
    the~$\PX(4740)$~structure, 
    obtained from the~fit to the~background\nobreakdash-subtracted 
    $\fsX$~mass distribution. 
    The~uncertainties are statistical only.
	}
	\vspace{2mm}
	\begin{tabular*}{0.50\textwidth}{@{\hspace{3mm}}l@{\extracolsep{\fill}}lc@{\hspace{3mm}}}
	\multicolumn{2}{l}{$\PX(4740)$~structure } & 
    \\[1.5mm]
    \hline 
    \\[-3mm]
    $N_{\PX(4740)}$ &  
    & $\phantom{0.0} 175 \pm 39 \phantom{.0}$                     

    \\
    $m_{\PX(4740)}$ & $\left[\!\mevcc\right]$ 
       & $ 4740.6 \pm 6.0\phantom{0}$ 
   \\ 
    $\Gamma_{\PX(4740)}$ &  $\left[\!\mev\right]$ 
    & $\phantom{00} 52.8 \pm 15.1$ 
	\end{tabular*}
	\label{tab:xxx}
\end{table}

%% file: systematics.tex
\section{Systematic uncertainties}
\label{sec:sys}

Due to the~similar decay topologies, 
systematic uncertainties largely cancel 
in the~ratios~$\mathcal{R}$. 
The~remaining contributions to systematic uncertainties
are summarized in Table~\ref{tab:sys_ratios}  
and discussed below.

The largest source of systematic uncertainty on the ratios arise from imperfect knowledge of the~shapes 
of signal and background components used in the~fits. 
To~estimate this uncertainty, several alternative models for 
the~signal, non\nobreakdash-resonant signal and  
background components are tested.
For the~\Bs~signal shape
and the~detector resolution functions
in the~$\PX_{\ccbar}$~signal templates, 
the~bifurcated Student's $t$\nobreakdash-distribution 
is tested as an~alternative model.
For~the~\mbox{Breit}\nobreakdash-Wigner functions 
describing 
the~\Pphi and \Kstarz~signal
shapes, the~meson radii in the~Blatt\nobreakdash-Weisskopf 
barrier factors~\cite{Blatt:1952ije} are varied 
between 1.5 and $5\gev^{-1}$.
The~mass and width of the~\Kstarz~meson
are varied  within their~uncertainties~\cite{PDG2020}. 
The~degree of the polynomials
used in
the~non\nobreakdash-resonant 
\mbox{$\jpsi\pip\pim$}~and 
\mbox{$\Kp\Km$}~functions,
the~$\mathcal{F}_{\Bs} $ and $P_{\mathrm{bkg}}$~functions  
and all other polynomial functions 
used in the~fits are increased by one.  
The~largest systematic uncertainty
for the~ratio 
\Rnonphi
is associated with the~parameterisation of 
the~fit component for 
the~\mbox{$\decay{\Bs}{\chicone(3872)\Kp\Km}$}~decays,
where the~$\Kp\Km$~pair does not originate from a~\Pphi~meson.
The~explicit inclusion  of 
a~\mbox{$\decay{\Bs}{\chicone(3872)\mathrm{f}_0(980)}$}~component
 is considered, 
where the~$\mathrm{f}_0(980)$~state
decays into a~\Kp\Km~pair. 
The~$\mathrm{f}_0(980)$~line shape is modelled 
by a~Flatt\'e function~\cite{Flatte:1976xu},
with parameters taken from Refs.~\cite{Ablikim:2004wn,LHCb-PAPER-2011-002}.
The~systematic uncertainty on the~ratio 
\Rkst
due to the~fit range for~$\kaon^{\pm}\Ppi^{\mp}$~masses
is studied by increasing this range to \mbox{$0.63< m_{\kaon^{\pm}\Ppi^{\mp}}<1.25\gevcc$}. 
For~each alternative model  
the~ratio of event yields is remeasured, 
and the~maximum deviation
with respect to the~nominal model, 
1.8\%,
2.6\% and 7.3\%
for 
the~\Rchi, \Rkst and \Rnonphi~ratios, respectively,
is assigned as a~systematic uncertainty.

\begin{table}[bt]
	\centering
	\caption{Relative systematic uncertainties (in \%) for 
the~ratios of branching fractions. 
The~sources are described in the~text. } 
		\label{tab:sys_ratios}
	\vspace*{2mm}
    \begin{tabular*}{0.90\textwidth}{@{\hspace{3mm}}l@{\extracolsep{\fill}}ccc@{\hspace{2mm}}}
    Source  
    & \Rchi 
    & \Rkst
    & \Rnonphi 
     \\[1mm]
    \hline 
    \\[-2mm]
  Fit model   
  & 1.8
  & 2.6
  & 7.3
  \\
  Efficiency corrections
  & 0.3
  & 0.1
  & 0.3
  \\
  Trigger efficiency
  &  1.1 
  &  1.1 
  &  1.1 
  \\
  Data\nobreakdash-simulation difference
  & 2.0 
  & 2.0 
  & 2.0
  \\
  Simulated sample size 
  & 1.0
  & 0.9
  & 1.3
  \\[1mm] 
  \hline 
  \\[-2mm]
    Sum in quadrature & 3.1 & 3.6 & 7.8
\end{tabular*}	
	\vspace*{3mm}
\end{table}

An~additional systematic uncertainty on 
the~ratios arises due to
differences between data and  simulation. 
In particular, there are differences in 
the~reconstruction efficiency 
of charged\nobreakdash-particle tracks that 
do not cancel completely in the~ratio due 
to the~different kinematic distributions 
of the~final\nobreakdash-state particles.
The~track\nobreakdash-finding efficiencies 
obtained from 
simulated samples 
are corrected \mbox{using}
data\nobreakdash-driven techniques~\cite{LHCb-DP-2013-002}.
The~uncertainties related to~the~efficiency 
correction factors, 
\mbox{together}
with the~uncertainty on the~hadron\nobreakdash-identification efficiency 
due to the~finite size of 
the~calibration samples~\mbox{\cite{LHCb-DP-2012-003, LHCb-DP-2018-001}},
are propagated to the~ratios of 
the~total efficiencies using pseudoexperiments
and account for
0.3\%,
0.1\% and 0.3\%
for 
the~\Rchi, \Rkst and \Rnonphi~ratios, respectively.

A~systematic uncertainty on the~ratios related to the knowledge of the~trigger efficiencies is estimated 
by comparing the~ratios of trigger
efficiencies in data and simulation for large samples of~$\decay{\Bp}{\jpsi\Kp}$ 
and 
$\decay{\Bp}{\psitwos\Kp}$ decays~\cite{LHCb-PAPER-2012-010}
and is taken to be 1.1\% for all three ratios 
of branching fractions.
Other data\nobreakdash-simulation differences 
are investigated by varying 
selection criteria in data.
The~resulting variations in the~efficiency 
ratios do not exceed 2\%, which  
is taken as the corresponding systematic 
uncertainty.
The~final systematic uncertainty 
considered on the~ratios of branching fractions is 
due to the~knowledge of  
the~ratios of efficiencies in 
Eqs.~\eqref{eq:rcalcone},  
\eqref{eq:rcalctwo}~and~\eqref{eq:rcalcthree}
due to the~finite size of the~simulated samples.
It~is determined to be 
1.0\%,
0.9\% and 1.3\% 
for 
the~\Rchi, \Rkst and \Rnonphi~ratios, respectively.
No~systematic uncertainty is
included for the~admixture of 
the~$\CP$\nobreakdash-odd and 
$\CP$\nobreakdash-even 
\Bs~eigenstates in the~decays,
which is assumed to be 
the~same for all four 
involved channels~\cite{DeBruyn:2012wj}.
In~the~extreme case that one decay
is only from 
the~short\nobreakdash-lifetime
eigenstate and the~other 
from the~long\nobreakdash-lifetime
eigenstate, the~corresponding ratio of 
branching fractions 
would change by 3.8\%.

The~systematic uncertainties on the~\Bs~mass measurement 
are summarised in Table~\ref{tab:sys_bmass}.
The~most important source of systematic 
uncertainty 
is related to the~momentum scale calibration 
in~data. 
This~effect 
is evaluated by varying the~scale 
within its known uncertainty~\cite{LHCb-PAPER-2013-011}.
The resulting change in 
the~mass  of 122\kevcc 
is assigned 
as a~systematic uncertainty. 
Other sources of uncertainty  are related to energy loss corrections
and the~imprecise knowledge of 
the~$\kaon^{\pm}$~and $\psitwos$~meson masses~\cite{PDG2020}. 
The~amount of material traversed in the~tracking system 
by a~particle is known to 10\% accuracy, 
which leads to an~uncertainty on the~estimated 
energy loss of particles in the detector. 
This~systematic uncertainty
is calculated in Ref.~\cite{LHCb-PAPER-2013-011}
to be 15\kevcc. 
The~uncertainties on the~known kaon and \psitwos~masses~\cite{PDG2020, Anashin:2015rca} 
are propagated to the~uncertainty in the~\Bs~mass using simulated samples
and are found to be 27 and 10\kevcc, respectively.
Using the~\psitwos~mass constraint
significantly reduces the~systematic uncertainties 
associated with the~momentum scale and energy loss correction.
The~\mbox{$\decay{\Bs}{\jpsi\Kstarz\Kstarzb}$}~signal sample has a
smaller~statistical uncertainty, see Table~\ref{tab:fit_kstar}, 
however,  
the~systematic uncertainties 
due to momentum scaling and energy loss are 
twice as large, making this sample  
non\nobreakdash-competitive for 
a~precise measurement of the~\Bs~mass.

\begin{table}[tb]
  \centering
  \caption{\small  
  Systematic uncertainties on the \Bs~mass measurement
  using the~sample enriched in the~\mbox{$\decay{\Bs}{\psitwos\Pphi}$}~decays. 
  The~sources are described in the~text.
  }\label{tab:sys_bmass}
  \vspace*{2mm}
  \begin{tabular*}{0.50\textwidth}{@{\hspace{2mm}}l@{\extracolsep{\fill}}c@{\hspace{2mm}}}
  Source  
  &   $\sigma_{m_{\Bs}}~\left[\!\kevcc\right]$ 
  \\[1mm]
  \hline 
  \\[-2mm]
  Fit model      &   \phantom{0}51 \\ 
  Momentum scale &   122           \\ 
  Energy loss    &  \phantom{0}15  \\ 
  Kaon mass      &  \phantom{0}27  \\ 
  \psitwos mass  &  \phantom{0}10     
   \\[1mm]
  \hline 
  \\[-2mm]
  Sum in quadrature  &  133 
  \end{tabular*}	
  \vspace*{3mm}
 \end{table}

\begin{table}[tb]
  \centering
  \caption{\small  
  Systematic uncertainties
  for the~measurements of 
  the~mass and width of the~$\PX(4740)$~structure. 
  The~sources are described in the~text. 
  }\label{tab:sys_x}
  \vspace*{2mm}
  \begin{tabular*}{0.75\textwidth}{@{\hspace{2mm}}l@{\extracolsep{\fill}}cc@{\hspace{2mm}}}
  Source  
  &   $\sigma_{m_{\PX(4740)}}~\left[\!\mevcc\right]$
  &   $\sigma_{\Gamma_{\PX(4740)}}~\left[\!\mev\right]$
  \\[1mm]
  \hline 
  \\[-2mm]
  Fit model                       
  & 2.8 
  & $\phantom{0}8.4$ 
  \\
  \psitwos, $\chicone(3872)$~veto  
  & 4.6 
  & $\phantom{0}5.1$ 
  \\
   Interference  
   &  1.2 
   &  $\phantom{0}5.1$ 
   \\[1mm]
  \hline 
  \\[-2mm]
  Sum in quadrature  
  &  5.5 
  & 11.1
  \end{tabular*}	
  \vspace*{3mm}
 \end{table}

Systematic uncertainties on 
the~mass and width of 
the~$\PX(4740)$~structure 
are summarised in Table~\ref{tab:sys_x}.
The~uncertainty related to 
the~imperfect knowledge of 
the~signal and background shapes  
is estimated using alternative  
fit models.  
Relativistic P-~and D\nobreakdash-wave
Breit\nobreakdash--Wigner functions are 
used  as~alternative shapes for signal
with the~value 
of the~\mbox{Blatt}\nobreakdash--Weisskopf
barrier factor meson radius
varied between 1.5~and~$5\gev^{-1}$.
A~model comprising a~product 
of an~S\nobreakdash-wave
Breit\nobreakdash--Wigner function
with a~phase\nobreakdash-space function
that accounts for the~proximity of 
the~upper edge for the~$\jpsi\Pphi$~mass spectrum 
from~\mbox{$\decay{\Bs}{\jpsi\pip\pim\Pphi}$}~decays
is also used. 
For~the~background component, 
a~convex 
monotonically decreasing 
third\nobreakdash-order 
polynomial function and 
a~product of 
the~$\Phi_{2,4}\left(m_{\jpsi\Pphi}\right)$~function
with a~third\nobreakdash-order polynomial function are tested
as alternative models. 
The~maximal deviations
with respect to the~baseline model 
of $2.8\mevcc$ and 
$8.4\mev$ for the~mass and 
width of the~$\PX(4740)$~state, respectively,  
are taken as corresponding systematic uncertainties.
The~contributions from~\mbox{$\decay{\Bs}{\psitwos\Kp\Km}$} and 
\mbox{$\decay{\Bs}{\chicone(3872)\Kp\Km}$}~decays
are explicitly suppressed 
in the~analysis by excluding
the~mass regions~\mbox{$3.672<m_{\jpsi\pip\pim}<3.700\gevcc$}  
and  \mbox{$3.864<m_{\jpsi\pip\pim}<3.880\gevcc$} 
around the~known masses of 
the~\psitwos  and 
\mbox{$\chicone(3872)$}~states~\cite{Anashin:2015rca,
LHCb-PAPER-2020-008,
LHCb-PAPER-2020-009,PDG2020}.
Repeating the~analysis 
using wider exclusion ranges, 
causes changes of~$4.6\mevcc$ and~$5.1\mev$
in the~mass and width 
of the~$\PX(4740)$~structure, respectively. 
These~changes are taken as systematic 
uncertainties due to possible 
remaining contributions
from~\mbox{$\decay{\Bs}{\psitwos\Kp\Km}$}
and \mbox{ $\decay{\Bs}{\chicone(3872)\Kp\Km}$}~decays.
Large interference effects 
between the~signal and coherent 
part of the background 
can also distort 
the~visible shape of the~resonance. 
To~probe the~importance of this effect, 
the~signal fit component $\mathcal{F}_{\mathrm{S}}$ 
is modelled with 
a~coherent sum of an~S\nobreakdash-wave  
Breit\nobreakdash-Wigner
amplitude $\mathcal{A}\left(m_{\jpsi\Pphi}\right)$ and 
a~coherent background 
\begin{equation}
    \mathcal{F}_{\mathrm{S}}
    \left(m_{\jpsi\Pphi}\right)
    \propto
    \left|  \mathcal{A} \left(m_{\jpsi\Pphi}\right)
    +  b\left( m_{\jpsi\Pphi}\right) \mathrm{e}^{i\varphi}
    \right|^2 \Phi_{2,4}\left(m_{\jpsi\Pphi}\right)\,, 
\end{equation}
where the~positive linear 
polynomial $b(m_{\jpsi\Pphi})$~stands for
the~magnitude of the~coherent 
background amplitude 
and $\varphi$~denotes 
the~phase of the coherent background, 
chosen to be independent of the~$\jpsi\Pphi$~mass.
The~deviations of  
the~mass and width of the~$\PX(4740)$~structure 
obtained from this fit are
taken as systematic uncertainties 
related to neglecting 
possible interference 
effects between the~signal and 
the~coherent part of the~background. 
The~complicated interference pattern 
for the~\mbox{$\decay{\Bs}{\jpsi\Pphi^{\ast}}$}~decays
via different $\Pphi^{\ast}$~states 
also can distort the~$\jpsi\Pphi$~mass spectrum.
However, 
to quantify this
effect a~full amplitude analysis, 
similar to Refs.~\cite{LHCb-PAPER-2014-014,
LHCb-PAPER-2015-029,
LHCb-PAPER-2016-018,
LHCb-PAPER-2016-019} is needed, 
that is   
beyond the~scope of this paper,
and no systematic uncertainty is assigned. 
Other sources of systematic uncertainties on the~mass
and width of the~$\PX(4740)$~structure,
namely the~momentum scale and the~background\nobreakdash-subtraction 
procedure
are found to be negligible with respect to the~leading systematic 
uncertainties related to the~fit~model. 
For~each choice of the~fit model, 
the~statistical significance of 
the~observed $\PX(4740)$~structure 
is calculated from data 
using Wilks' theorem~\cite{Wilks:1938dza}.
The~smallest significance
found is 5.3~standard deviations, 
taken as its~significance 
including~systematic uncertainties.

%% file: results.tex
\section{Summary} 
\label{sec:results}

A~study of 
\mbox{$\decay{\Bs}{\jpsi\pip\pim\Kp\Km}$}~decays 
is made using $\proton\proton$~collision data 
corresponding to an~integrated luminosity of 1, 2 and 6\invfb, 
collected with the~LHCb detector at centre\nobreakdash-of\nobreakdash-mass  
energies of 7, 8 and 13\tev, respectively. 
The~ratios of the~branching fractions via intermediate resonances, 
defined via Eqs.~\eqref{eq:r}, are measured to be
\begin{eqnarray*}
\Rchi 
& = &    \left( 2.42 \pm 0.23 \pm 0.07 \right) \times 10^{-2} \,, 
\\
\Rnonphi 
& = &
\phantom{(} 1.57 \pm 0.32 \pm 0.12 \,, 
\\ 
\Rkst
& = &   
\phantom{(}1.22 \pm 0.03  \pm 0.04 \,, 
\end{eqnarray*}
where the~first uncertainty is statistical 
and the~second systematic. 
The~ratio 
\Rchi~is consistent with but more precise 
than the~value of \mbox{$\left(2.21\pm0.29\pm0.17\right)\times10^{-2}$}
recently reported by the~CMS~collaboration~\cite{Sirunyan:2020qir}.
The~decays
\mbox{$\decay{\Bs}{\jpsi\Kstarz\Kstarzb}$}
and 
\mbox{$\decay{\Bs}{\chicone(3872)\Kp\Km}$},
where the~$\Kp\Km$~pair does not originate from 
a~\Pphi~meson,  are observed for the~first time. 
A~full amplitude analysis, 
similar to Refs.~\cite{LHCb-PAPER-2012-040,LHCb-PAPER-2017-008}, 
is needed to resolve 
possible contributions from 
two\nobreakdash-body decays  
via $\Kp\Km$~resonances,
like~\mbox{$\decay{\Bs}{\chicone(3872)\mathrm{f}_0(980)}$}
and  \mbox{$\decay{\Bs}{\chicone(3872)\mathrm{f}^{\prime}_2(1525)}$}, 
that in turn could be useful for a~better understanding of 
the~nature of the~$\chicone(3872)$~state.

A~precise measurement of the~\Bs~mass is performed 
using a~sample of selected \mbox{$\decay{\Bs}{\jpsi\pip\pim\Kp\Km}$}~candidates
enriched in~\mbox{$\decay{\Bs}{\psitwos\Pphi}$}~decays. The~mass of the~\Bs~meson is determined to be 
\begin{equation*} 
m_{\Bs} = 5366. 98 \pm 0.07 \pm 0.13 \mevcc\,,
\end{equation*} 
which is the~most precise single measurement of 
this observable.
This result is combined with other 
precise measurements 
by the~LHCb collaboration using~\mbox{$\decay{\Bs}{\jpsi\Pphi}$}~\cite{LHCb-PAPER-2011-035},
\mbox{$\decay{\Bs}{\jpsi\Pphi\Pphi}$}~\cite{LHCb-PAPER-2015-033}, 
\mbox{$\decay{\Bs}{\chictwo\Kp\Km}$}~\cite{LHCb-PAPER-2018-018} and 
\mbox{$\decay{\Bs}{\jpsi\proton\antiproton}$}~\cite{LHCb-PAPER-2018-046}~decays. 
The~combined mass is calculated using the~best linear 
unbiased estimator~\cite{Lyons:1988rp}, 
accounting for correlations of systematic uncertainties between 
the~measurements. 
The~LHCb average for the~mass of the~\Bs~meson is found to be 
\begin{equation*} 
m_{\Bs}^{\mathrm{LHCb}} = 5366.94 \pm 0.08 \pm 0.09 \mevcc\,.
\end{equation*}

A~structure with significance exceeding
5.3~standard deviations, denoted as $\PX(4740)$, 
is also seen in 
the~$\jpsi\Pphi$~mass 
spectrum of~\mbox{$\decay{\Bs}{\jpsi\pip\pim\left( \decay{\Pphi}{\Kp\Km} \right)}$}~decays.
The~mass and width of the~structure  are determined to be 
\begin{eqnarray*}
 m_{\PX(4740)}      & = 
 & 4741 \pm 6  \phantom{0}\pm 6  \phantom{0}\mevcc \,, \\ 
 \Gamma_{\PX(4740)} & = 
 & \phantom{00}53 \pm 15 \pm 11\mev\,. 
\end{eqnarray*}
A~dedicated analysis using 
a~larger data set 
is needed to resolve 
if this state is different from 
the~$\chiczero(4700)$~state, observed in 
the~\mbox{$\decay{\Bu}{\jpsi\Pphi\Kp}$}~decays~\cite{LHCb-PAPER-2016-018,
LHCb-PAPER-2016-019}.

%% file: acknowledgements.tex
\section*{Acknowledgements}
%
%
\noindent We express our gratitude to our colleagues in 
the~CERN
accelerator departments for the~excellent performance of the~LHC. 
We~thank the technical and administrative staff at the LHCb
institutes.
We acknowledge support from CERN and from the national agencies:
CAPES, CNPq, FAPERJ and FINEP\,(Brazil); 
MOST and NSFC\,(China); 
CNRS/IN2P3\,(France); 
BMBF, DFG and MPG\,(Germany); 
INFN\,(Italy); 
NWO\,(Netherlands); 
MNiSW and NCN\,(Poland); 
MEN/IFA\,(Romania); 
MSHE\,(Russia); 
MICINN\,(Spain); 
SNSF and SER\,(Switzerland); 
NASU\,(Ukraine); 
STFC\,(United Kingdom); 
DOE NP and NSF\,(USA).
We acknowledge the computing resources that are provided
by CERN, 
IN2P3\,(France), KIT and DESY\,(Germany), 
INFN\,(Italy), SURF\,(Netherlands),
PIC\,(Spain), GridPP\,(United Kingdom), 
RRCKI and Yandex
LLC\,(Russia), CSCS\,(Switzerland),
IFIN\nobreakdash-HH\,(Romania), 
CBPF\,(Brazil),
PL\nobreakdash-GRID\,(Poland) and OSC\,(USA).
We are indebted to the communities behind the multiple open-source
software packages on which we depend.
Individual groups or members have received support from
AvH Foundation\,(Germany);
EPLANET, Marie Sk\l{}odowska\nobreakdash-Curie Actions 
and ERC\,(European Union);
A*MIDEX, ANR, Labex P2IO and OCEVU, 
and 
R\'{e}gion Auvergne\nobreakdash-Rh\^{o}ne\nobreakdash-Alpes\,(France);
Key Research Program of Frontier Sciences of CAS, CAS PIFI,
Thousand Talents Program, and Sci. \& Tech. Program of 
Guangzhou\,(China);
RFBR, RSF and Yandex LLC\,(Russia);
GVA, XuntaGal and GENCAT\,(Spain);
the Royal Society
and the Leverhulme Trust\,(United Kingdom).

%% file: LHCb_Authorship_07-Sep-2020.tex
\centerline
{\large\bf LHCb collaboration}
\begin
{flushleft}
\small
R.~Aaij$^{31}$,
C.~Abell{\'a}n~Beteta$^{49}$,
T.~Ackernley$^{59}$,
B.~Adeva$^{45}$,
M.~Adinolfi$^{53}$,
H.~Afsharnia$^{9}$,
C.A.~Aidala$^{84}$,
S.~Aiola$^{25}$,
Z.~Ajaltouni$^{9}$,
S.~Akar$^{64}$
% ...

{\footnotesize \it

% ...
$ ^{9}$Universit{\'e} Clermont Auvergne, CNRS/IN2P3, LPC, Clermont-Ferrand, France\\
% ...
$ ^{25}$INFN Sezione di Milano, Milano, Italy\\
% ...
$ ^{31}$Nikhef National Institute for Subatomic Physics, Amsterdam, Netherlands\\
% ...
$ ^{45}$Instituto Galego de F{\'\i}sica de Altas Enerx{\'\i}as (IGFAE), Universidade de Santiago de Compostela, Santiago de Compostela, Spain\\
% ...
$ ^{49}$Physik-Institut, Universit{\"a}t Z{\"u}rich, Z{\"u}rich, Switzerland\\
% ...
$ ^{53}$H.H. Wills Physics Laboratory, University of Bristol, Bristol, United Kingdom\\
% ...
$ ^{59}$Oliver Lodge Laboratory, University of Liverpool, Liverpool, United Kingdom\\
% ...
$ ^{64}$University of Cincinnati, Cincinnati, OH, United States\\
% ...
$ ^{67}$Syracuse University, Syracuse, NY, United States\\
% ...
$ ^{84}$University of Michigan, Ann Arbor, United States, associated to $^{67}$\\
\bigskip
% ...
\medskip
}
\end{flushleft}

%% file: main.bbl
\ifx\mcitethebibliography\mciteundefinedmacro
\PackageError{LHCb.bst}{mciteplus.sty has not been loaded}
{This bibstyle requires the use of the mciteplus package.}\fi
\providecommand{\href}[2]{#2}
\begin{mcitethebibliography}{10}
\mciteSetBstSublistMode{n}
\mciteSetBstMaxWidthForm{subitem}{\alph{mcitesubitemcount})}
\mciteSetBstSublistLabelBeginEnd{\mcitemaxwidthsubitemform\space}
{\relax}{\relax}

\bibitem{Choi:2003ue}
Belle collaboration, S.~K. Choi {\em et~al.},
  \ifthenelse{\boolean{articletitles}}{\emph{{Observation of a~narrow
  charmoniumlike state in exclusive \mbox{$\decay{\B^{\pm}}{\kaon^{\pm} \pip
  \pim \jpsi}$}~decays}},
  }{}\href{https://doi.org/10.1103/PhysRevLett.91.262001}{Phys.\ Rev.\ Lett.\
  \textbf{91} (2003) 262001},
  \href{http://arxiv.org/abs/hep-ex/0309032}{{\normalfont\ttfamily
  arXiv:hep-ex/0309032}}\relax
\mciteBstWouldAddEndPuncttrue
\mciteSetBstMidEndSepPunct{\mcitedefaultmidpunct}
{\mcitedefaultendpunct}{\mcitedefaultseppunct}\relax
\EndOfBibitem
\bibitem{LHCb-PAPER-2015-029}
LHCb collaboration, R.~Aaij {\em et~al.},
  \ifthenelse{\boolean{articletitles}}{\emph{{Observation of $\jpsi\proton$
  resonances consistent with pentaquark states in
  \mbox{\decay{\Lb}{\jpsi\proton\Km}} decays}},
  }{}\href{https://doi.org/10.1103/PhysRevLett.115.072001}{Phys.\ Rev.\ Lett.\
  \textbf{115} (2015) 072001},
  \href{http://arxiv.org/abs/1507.03414}{{\normalfont\ttfamily
  arXiv:1507.03414}}\relax
\mciteBstWouldAddEndPuncttrue
\mciteSetBstMidEndSepPunct{\mcitedefaultmidpunct}
{\mcitedefaultendpunct}{\mcitedefaultseppunct}\relax
\EndOfBibitem
\bibitem{LHCb-PAPER-2016-009}
LHCb collaboration, R.~Aaij {\em et~al.},
  \ifthenelse{\boolean{articletitles}}{\emph{{Model-independent evidence for
  $\jpsi\proton$ contributions to \mbox{\decay{\Lb}{\jpsi\proton\Km}} decays}},
  }{}\href{https://doi.org/10.1103/PhysRevLett.117.082002}{Phys.\ Rev.\ Lett.\
  \textbf{117} (2016) 082002},
  \href{http://arxiv.org/abs/1604.05708}{{\normalfont\ttfamily
  arXiv:1604.05708}}\relax
\mciteBstWouldAddEndPuncttrue
\mciteSetBstMidEndSepPunct{\mcitedefaultmidpunct}
{\mcitedefaultendpunct}{\mcitedefaultseppunct}\relax
\EndOfBibitem
\bibitem{LHCb-PAPER-2019-014}
LHCb collaboration, R.~Aaij {\em et~al.},
  \ifthenelse{\boolean{articletitles}}{\emph{{Observation of a narrow
  $\PP_\cquark(4312)^+$ state, and of two-peak structure of
  the~$\PP_\cquark(4450)^+$}},
  }{}\href{https://doi.org/10.1103/PhysRevLett.122.222001}{Phys.\ Rev.\ Lett.\
  \textbf{122} (2019) 222001},
  \href{http://arxiv.org/abs/1904.03947}{{\normalfont\ttfamily
  arXiv:1904.03947}}\relax
\mciteBstWouldAddEndPuncttrue
\mciteSetBstMidEndSepPunct{\mcitedefaultmidpunct}
{\mcitedefaultendpunct}{\mcitedefaultseppunct}\relax
\EndOfBibitem
\bibitem{LHCb-PAPER-2016-015}
LHCb collaboration, R.~Aaij {\em et~al.},
  \ifthenelse{\boolean{articletitles}}{\emph{{Evidence for exotic hadron
  contributions to \mbox{\decay{\Lb}{\jpsi\proton\pim}} decays}},
  }{}\href{https://doi.org/10.1103/PhysRevLett.117.082003}{Phys.\ Rev.\ Lett.\
  \textbf{117} (2016) 082003},
  \href{http://arxiv.org/abs/1606.06999}{{\normalfont\ttfamily
  arXiv:1606.06999}}\relax
\mciteBstWouldAddEndPuncttrue
\mciteSetBstMidEndSepPunct{\mcitedefaultmidpunct}
{\mcitedefaultendpunct}{\mcitedefaultseppunct}\relax
\EndOfBibitem
\bibitem{Choi:2007wga}
Belle collaborationn, S.~K. Choi {\em et~al.},
  \ifthenelse{\boolean{articletitles}}{\emph{{Observation of a~resonance-like
  structure in the~\mbox{$\pion^\pm \Ppsi^\prime$}~mass distribution in
  exclusive \mbox{$\decay{\B}{\kaon \pion^{\pm}\Ppsi^\prime}$}~decays}},
  }{}\href{https://doi.org/10.1103/PhysRevLett.100.142001}{Phys.\ Rev.\ Lett.\
  \textbf{100} (2008) 142001},
  \href{http://arxiv.org/abs/0708.1790}{{\normalfont\ttfamily
  arXiv:0708.1790}}\relax
\mciteBstWouldAddEndPuncttrue
\mciteSetBstMidEndSepPunct{\mcitedefaultmidpunct}
{\mcitedefaultendpunct}{\mcitedefaultseppunct}\relax
\EndOfBibitem
\bibitem{Mizuk:2009da}
Belle collaboration, R.~Mizuk {\em et~al.},
  \ifthenelse{\boolean{articletitles}}{\emph{{Dalitz analysis of
  \mbox{$\decay{\B}{\kaon \pip \Ppsi^{\prime}}$}~decays and
  the~\mbox{$\PZ(4430)^+$}}},
  }{}\href{https://doi.org/10.1103/PhysRevD.80.031104}{Phys.\ Rev.\
  \textbf{D80} (2009) 031104},
  \href{http://arxiv.org/abs/0905.2869}{{\normalfont\ttfamily
  arXiv:0905.2869}}\relax
\mciteBstWouldAddEndPuncttrue
\mciteSetBstMidEndSepPunct{\mcitedefaultmidpunct}
{\mcitedefaultendpunct}{\mcitedefaultseppunct}\relax
\EndOfBibitem
\bibitem{Chilikin:2013tch}
Belle collaboration, K.~Chilikin {\em et~al.},
  \ifthenelse{\boolean{articletitles}}{\emph{{Experimental constraints on
  the~spin and parity of the~\mbox{$\PZ(4430)^+$}}},
  }{}\href{https://doi.org/10.1103/PhysRevD.88.074026}{Phys.\ Rev.\
  \textbf{D88} (2013) 074026},
  \href{http://arxiv.org/abs/1306.4894}{{\normalfont\ttfamily
  arXiv:1306.4894}}\relax
\mciteBstWouldAddEndPuncttrue
\mciteSetBstMidEndSepPunct{\mcitedefaultmidpunct}
{\mcitedefaultendpunct}{\mcitedefaultseppunct}\relax
\EndOfBibitem
\bibitem{LHCb-PAPER-2014-014}
LHCb collaboration, R.~Aaij {\em et~al.},
  \ifthenelse{\boolean{articletitles}}{\emph{{Observation of the resonant
  character of the~$\PZ(4430)^-$ state}},
  }{}\href{https://doi.org/10.1103/PhysRevLett.112.222002}{Phys.\ Rev.\ Lett.\
  \textbf{112} (2014) 222002},
  \href{http://arxiv.org/abs/1404.1903}{{\normalfont\ttfamily
  arXiv:1404.1903}}\relax
\mciteBstWouldAddEndPuncttrue
\mciteSetBstMidEndSepPunct{\mcitedefaultmidpunct}
{\mcitedefaultendpunct}{\mcitedefaultseppunct}\relax
\EndOfBibitem
\bibitem{LHCb-PAPER-2015-038}
LHCb collaboration, R.~Aaij {\em et~al.},
  \ifthenelse{\boolean{articletitles}}{\emph{{Model-independent confirmation of
  the~$\PZ(4430)^-$ state}},
  }{}\href{https://doi.org/10.1103/PhysRevD.92.112009}{Phys.\ Rev.\
  \textbf{D92} (2015) 112009},
  \href{http://arxiv.org/abs/1510.01951}{{\normalfont\ttfamily
  arXiv:1510.01951}}\relax
\mciteBstWouldAddEndPuncttrue
\mciteSetBstMidEndSepPunct{\mcitedefaultmidpunct}
{\mcitedefaultendpunct}{\mcitedefaultseppunct}\relax
\EndOfBibitem
\bibitem{LHCb-PAPER-2016-018}
LHCb collaboration, R.~Aaij {\em et~al.},
  \ifthenelse{\boolean{articletitles}}{\emph{{Observation of exotic
  $\jpsi\Pphi$ structures from amplitude analysis of
  \mbox{\decay{\Bp}{\jpsi\Pphi\Kp}} decays}},
  }{}\href{https://doi.org/10.1103/PhysRevLett.118.022003}{Phys.\ Rev.\ Lett.\
  \textbf{118} (2017) 022003},
  \href{http://arxiv.org/abs/1606.07895}{{\normalfont\ttfamily
  arXiv:1606.07895}}\relax
\mciteBstWouldAddEndPuncttrue
\mciteSetBstMidEndSepPunct{\mcitedefaultmidpunct}
{\mcitedefaultendpunct}{\mcitedefaultseppunct}\relax
\EndOfBibitem
\bibitem{LHCb-PAPER-2016-019}
LHCb collaboration, R.~Aaij {\em et~al.},
  \ifthenelse{\boolean{articletitles}}{\emph{{Amplitude analysis of
  \mbox{\decay{\Bp}{\jpsi\Pphi\Kp}} decays}},
  }{}\href{https://doi.org/10.1103/PhysRevD.95.012002}{Phys.\ Rev.\
  \textbf{D95} (2017) 012002},
  \href{http://arxiv.org/abs/1606.07898}{{\normalfont\ttfamily
  arXiv:1606.07898}}\relax
\mciteBstWouldAddEndPuncttrue
\mciteSetBstMidEndSepPunct{\mcitedefaultmidpunct}
{\mcitedefaultendpunct}{\mcitedefaultseppunct}\relax
\EndOfBibitem
\bibitem{LHCb-PAPER-2018-034}
LHCb collaboration, R.~Aaij {\em et~al.},
  \ifthenelse{\boolean{articletitles}}{\emph{{Evidence for a
  $\Peta_\cquark(1\PS) \pim$ resonance in \mbox{\decay{\Bz}{\Peta_\cquark(1\PS)
  \Kp\pim}} decays}},
  }{}\href{https://doi.org/10.1140/epjc/s10052-018-6447-z}{Eur.\ Phys.\ J.\
  \textbf{C78} (2018) 1019},
  \href{http://arxiv.org/abs/1809.07416}{{\normalfont\ttfamily
  arXiv:1809.07416}}\relax
\mciteBstWouldAddEndPuncttrue
\mciteSetBstMidEndSepPunct{\mcitedefaultmidpunct}
{\mcitedefaultendpunct}{\mcitedefaultseppunct}\relax
\EndOfBibitem
\bibitem{LHCb-PAPER-2018-043}
LHCb collaboration, R.~Aaij {\em et~al.},
  \ifthenelse{\boolean{articletitles}}{\emph{{Model-independent observation of
  exotic contributions to \mbox{\decay{\Bz}{\jpsi\Kp\pim}} decays}},
  }{}\href{https://doi.org/10.1103/PhysRevLett.122.152002}{Phys.\ Rev.\ Lett.\
  \textbf{122} (2019) 152002},
  \href{http://arxiv.org/abs/1901.05745}{{\normalfont\ttfamily
  arXiv:1901.05745}}\relax
\mciteBstWouldAddEndPuncttrue
\mciteSetBstMidEndSepPunct{\mcitedefaultmidpunct}
{\mcitedefaultendpunct}{\mcitedefaultseppunct}\relax
\EndOfBibitem
\bibitem{Bhardwaj:2013rmw}
Belle collaboration, V.~Bhardwaj {\em et~al.},
  \ifthenelse{\boolean{articletitles}}{\emph{{Evidence of a~new narrow
  resonance decaying to~\mbox{$\chicone\g$} in
  \mbox{$\decay{\B}{\chicone\g\kaon}$}}},
  }{}\href{https://doi.org/10.1103/PhysRevLett.111.032001}{Phys.\ Rev.\ Lett.\
  \textbf{111} (2013) 032001},
  \href{http://arxiv.org/abs/1304.3975}{{\normalfont\ttfamily
  arXiv:1304.3975}}\relax
\mciteBstWouldAddEndPuncttrue
\mciteSetBstMidEndSepPunct{\mcitedefaultmidpunct}
{\mcitedefaultendpunct}{\mcitedefaultseppunct}\relax
\EndOfBibitem
\bibitem{LHCb-PAPER-2020-009}
LHCb collaboration, R.~Aaij {\em et~al.},
  \ifthenelse{\boolean{articletitles}}{\emph{{Study of the $\Ppsi_2(3823)$ and
  $\chicone(3872)$ states in $\decay{\Bu}{\left(\jpsi \pip\pim\right)\Kp}$
  decays}}, }{}\href{https://doi.org/10.1007/JHEP08(2020)123}{JHEP \textbf{08}
  (2020) 123}, \href{http://arxiv.org/abs/2005.13422}{{\normalfont\ttfamily
  arXiv:2005.13422}}\relax
\mciteBstWouldAddEndPuncttrue
\mciteSetBstMidEndSepPunct{\mcitedefaultmidpunct}
{\mcitedefaultendpunct}{\mcitedefaultseppunct}\relax
\EndOfBibitem
\bibitem{Maiani:2017kyi}
L.~Maiani, A.~D. Polosa, and V.~Riquer,
  \ifthenelse{\boolean{articletitles}}{\emph{{A~theory of \PX~and \Z~multiquark
  resonances}}, }{}\href{https://doi.org/10.1016/j.physletb.2018.01.039}{Phys.\
  Lett.\  \textbf{B778} (2018) 247},
  \href{http://arxiv.org/abs/1712.05296}{{\normalfont\ttfamily
  arXiv:1712.05296}}\relax
\mciteBstWouldAddEndPuncttrue
\mciteSetBstMidEndSepPunct{\mcitedefaultmidpunct}
{\mcitedefaultendpunct}{\mcitedefaultseppunct}\relax
\EndOfBibitem
\bibitem{Artoisenet:2010va}
P.~Artoisenet, E.~Braaten, and D.~Kang,
  \ifthenelse{\boolean{articletitles}}{\emph{{Using line shapes to discriminate
  between binding mechanisms for the~\mbox{$\PX(3872)$}}},
  }{}\href{https://doi.org/10.1103/PhysRevD.82.014013}{Phys.\ Rev.\
  \textbf{D82} (2010) 014013},
  \href{http://arxiv.org/abs/1005.2167}{{\normalfont\ttfamily
  arXiv:1005.2167}}\relax
\mciteBstWouldAddEndPuncttrue
\mciteSetBstMidEndSepPunct{\mcitedefaultmidpunct}
{\mcitedefaultendpunct}{\mcitedefaultseppunct}\relax
\EndOfBibitem
\bibitem{Braaten:2019yua}
E.~Braaten, L.-P. He, and K.~Ingles,
  \ifthenelse{\boolean{articletitles}}{\emph{{Production of \mbox{$\PX(3872)$}
  accompanied by a~pion in $\B$~meson decay}},
  }{}\href{https://doi.org/10.1103/PhysRevD.100.074028}{Phys.\ Rev.\
  \textbf{D100} (2019) 074028},
  \href{http://arxiv.org/abs/1902.03259}{{\normalfont\ttfamily
  arXiv:1902.03259}}\relax
\mciteBstWouldAddEndPuncttrue
\mciteSetBstMidEndSepPunct{\mcitedefaultmidpunct}
{\mcitedefaultendpunct}{\mcitedefaultseppunct}\relax
\EndOfBibitem
\bibitem{Maiani:2020zhr}
L.~Maiani, A.~D. Polosa, and V.~Riquer,
  \ifthenelse{\boolean{articletitles}}{\emph{{The~
  $\PX\mathrm{(3872)}$~tetraquarks in \B and $\B_\squark$~decays}},
  }{}\href{https://doi.org/10.1103/PhysRevD.102.034017}{Phys.\ Rev.\
  \textbf{D102} (2020) 034017},
  \href{http://arxiv.org/abs/2005.08764}{{\normalfont\ttfamily
  arXiv:2005.08764}}\relax
\mciteBstWouldAddEndPuncttrue
\mciteSetBstMidEndSepPunct{\mcitedefaultmidpunct}
{\mcitedefaultendpunct}{\mcitedefaultseppunct}\relax
\EndOfBibitem
\bibitem{Sirunyan:2020qir}
CMS collaboration, A.~M. Sirunyan {\em et~al.},
  \ifthenelse{\boolean{articletitles}}{\emph{{Observation of
  the~$\decay{\Bs}{\PX(3872)\Pphi}$~decay}},
  }{}\href{https://doi.org/10.1103/PhysRevLett.125.152001}{Phys.\ Rev.\ Lett.\
  \textbf{125} (2020) 152001},
  \href{http://arxiv.org/abs/2005.04764}{{\normalfont\ttfamily
  arXiv:2005.04764}}\relax
\mciteBstWouldAddEndPuncttrue
\mciteSetBstMidEndSepPunct{\mcitedefaultmidpunct}
{\mcitedefaultendpunct}{\mcitedefaultseppunct}\relax
\EndOfBibitem
\bibitem{PDG2020}
Particle Data Group, P.~A. Zyla {\em et~al.},
  \ifthenelse{\boolean{articletitles}}{\emph{{\href{http://pdg.lbl.gov/}{Review
  of particle physics}}}, }{}\href{https://doi.org/10.1093/ptep/ptaa104}{Prog.\
  Theor.\ Exp.\ Phys.\  \textbf{2020} (2020) 083C01}\relax
\mciteBstWouldAddEndPuncttrue
\mciteSetBstMidEndSepPunct{\mcitedefaultmidpunct}
{\mcitedefaultendpunct}{\mcitedefaultseppunct}\relax
\EndOfBibitem
\bibitem{LHCb-PAPER-2015-033}
LHCb collaboration, R.~Aaij {\em et~al.},
  \ifthenelse{\boolean{articletitles}}{\emph{{Observation of the
  \mbox{\decay{\Bs}{\jpsi\Pphi\Pphi}} decay}},
  }{}\href{https://doi.org/10.1007/JHEP03(2016)040}{JHEP \textbf{03} (2016)
  040}, \href{http://arxiv.org/abs/1601.05284}{{\normalfont\ttfamily
  arXiv:1601.05284}}\relax
\mciteBstWouldAddEndPuncttrue
\mciteSetBstMidEndSepPunct{\mcitedefaultmidpunct}
{\mcitedefaultendpunct}{\mcitedefaultseppunct}\relax
\EndOfBibitem
\bibitem{LHCb-DP-2008-001}
LHCb collaboration, A.~A. Alves~Jr.\ {\em et~al.},
  \ifthenelse{\boolean{articletitles}}{\emph{{The \lhcb detector at the LHC}},
  }{}\href{https://doi.org/10.1088/1748-0221/3/08/S08005}{JINST \textbf{3}
  (2008) S08005}\relax
\mciteBstWouldAddEndPuncttrue
\mciteSetBstMidEndSepPunct{\mcitedefaultmidpunct}
{\mcitedefaultendpunct}{\mcitedefaultseppunct}\relax
\EndOfBibitem
\bibitem{LHCb-DP-2014-002}
LHCb collaboration, R.~Aaij {\em et~al.},
  \ifthenelse{\boolean{articletitles}}{\emph{{LHCb detector performance}},
  }{}\href{https://doi.org/10.1142/S0217751X15300227}{Int.\ J.\ Mod.\ Phys.\
  \textbf{A30} (2015) 1530022},
  \href{http://arxiv.org/abs/1412.6352}{{\normalfont\ttfamily
  arXiv:1412.6352}}\relax
\mciteBstWouldAddEndPuncttrue
\mciteSetBstMidEndSepPunct{\mcitedefaultmidpunct}
{\mcitedefaultendpunct}{\mcitedefaultseppunct}\relax
\EndOfBibitem
\bibitem{LHCb-DP-2014-001}
R.~Aaij {\em et~al.}, \ifthenelse{\boolean{articletitles}}{\emph{{Performance
  of the LHCb Vertex Locator}},
  }{}\href{https://doi.org/10.1088/1748-0221/9/09/P09007}{JINST \textbf{9}
  (2014) P09007}, \href{http://arxiv.org/abs/1405.7808}{{\normalfont\ttfamily
  arXiv:1405.7808}}\relax
\mciteBstWouldAddEndPuncttrue
\mciteSetBstMidEndSepPunct{\mcitedefaultmidpunct}
{\mcitedefaultendpunct}{\mcitedefaultseppunct}\relax
\EndOfBibitem
\bibitem{LHCb-DP-2013-003}
R.~Arink {\em et~al.}, \ifthenelse{\boolean{articletitles}}{\emph{{Performance
  of the LHCb Outer Tracker}},
  }{}\href{https://doi.org/10.1088/1748-0221/9/01/P01002}{JINST \textbf{9}
  (2014) P01002}, \href{http://arxiv.org/abs/1311.3893}{{\normalfont\ttfamily
  arXiv:1311.3893}}\relax
\mciteBstWouldAddEndPuncttrue
\mciteSetBstMidEndSepPunct{\mcitedefaultmidpunct}
{\mcitedefaultendpunct}{\mcitedefaultseppunct}\relax
\EndOfBibitem
\bibitem{LHCb-DP-2017-001}
P.~d'Argent {\em et~al.}, \ifthenelse{\boolean{articletitles}}{\emph{{Improved
  performance of the LHCb Outer Tracker in LHC Run 2}},
  }{}\href{https://doi.org/10.1088/1748-0221/12/11/P11016}{JINST \textbf{12}
  (2017) P11016}, \href{http://arxiv.org/abs/1708.00819}{{\normalfont\ttfamily
  arXiv:1708.00819}}\relax
\mciteBstWouldAddEndPuncttrue
\mciteSetBstMidEndSepPunct{\mcitedefaultmidpunct}
{\mcitedefaultendpunct}{\mcitedefaultseppunct}\relax
\EndOfBibitem
\bibitem{LHCb-PAPER-2012-048}
LHCb collaboration, R.~Aaij {\em et~al.},
  \ifthenelse{\boolean{articletitles}}{\emph{{Measurements of the \Lb, \Xibm,
  and \Omegab baryon masses}},
  }{}\href{https://doi.org/10.1103/PhysRevLett.110.182001}{Phys.\ Rev.\ Lett.\
  \textbf{110} (2013) 182001},
  \href{http://arxiv.org/abs/1302.1072}{{\normalfont\ttfamily
  arXiv:1302.1072}}\relax
\mciteBstWouldAddEndPuncttrue
\mciteSetBstMidEndSepPunct{\mcitedefaultmidpunct}
{\mcitedefaultendpunct}{\mcitedefaultseppunct}\relax
\EndOfBibitem
\bibitem{LHCb-PAPER-2013-011}
LHCb collaboration, R.~Aaij {\em et~al.},
  \ifthenelse{\boolean{articletitles}}{\emph{{Precision measurement of \D meson
  mass differences}}, }{}\href{https://doi.org/10.1007/JHEP06(2013)065}{JHEP
  \textbf{06} (2013) 065},
  \href{http://arxiv.org/abs/1304.6865}{{\normalfont\ttfamily
  arXiv:1304.6865}}\relax
\mciteBstWouldAddEndPuncttrue
\mciteSetBstMidEndSepPunct{\mcitedefaultmidpunct}
{\mcitedefaultendpunct}{\mcitedefaultseppunct}\relax
\EndOfBibitem
\bibitem{LHCb-DP-2012-003}
M.~Adinolfi {\em et~al.},
  \ifthenelse{\boolean{articletitles}}{\emph{{Performance of the \lhcb RICH
  detector at the LHC}},
  }{}\href{https://doi.org/10.1140/epjc/s10052-013-2431-9}{Eur.\ Phys.\ J.\
  \textbf{C73} (2013) 2431},
  \href{http://arxiv.org/abs/1211.6759}{{\normalfont\ttfamily
  arXiv:1211.6759}}\relax
\mciteBstWouldAddEndPuncttrue
\mciteSetBstMidEndSepPunct{\mcitedefaultmidpunct}
{\mcitedefaultendpunct}{\mcitedefaultseppunct}\relax
\EndOfBibitem
\bibitem{LHCb-DP-2020-001}
A.~Betata {\em et~al.}, \ifthenelse{\boolean{articletitles}}{\emph{{Calibration
  and performance of the~LHCb calorimeters in Run~1 and~2 at the~LHC}},
  }{}\href{http://arxiv.org/abs/2008.11556}{{\normalfont\ttfamily
  arXiv:2008.11556}}, {submitted to JINST}\relax
\mciteBstWouldAddEndPuncttrue
\mciteSetBstMidEndSepPunct{\mcitedefaultmidpunct}
{\mcitedefaultendpunct}{\mcitedefaultseppunct}\relax
\EndOfBibitem
\bibitem{LHCb-DP-2012-002}
A.~A. Alves~Jr.\ {\em et~al.},
  \ifthenelse{\boolean{articletitles}}{\emph{{Performance of the LHCb muon
  system}}, }{}\href{https://doi.org/10.1088/1748-0221/8/02/P02022}{JINST
  \textbf{8} (2013) P02022},
  \href{http://arxiv.org/abs/1211.1346}{{\normalfont\ttfamily
  arXiv:1211.1346}}\relax
\mciteBstWouldAddEndPuncttrue
\mciteSetBstMidEndSepPunct{\mcitedefaultmidpunct}
{\mcitedefaultendpunct}{\mcitedefaultseppunct}\relax
\EndOfBibitem
\bibitem{LHCb-DP-2012-004}
R.~Aaij {\em et~al.}, \ifthenelse{\boolean{articletitles}}{\emph{{The \lhcb
  trigger and its performance in 2011}},
  }{}\href{https://doi.org/10.1088/1748-0221/8/04/P04022}{JINST \textbf{8}
  (2013) P04022}, \href{http://arxiv.org/abs/1211.3055}{{\normalfont\ttfamily
  arXiv:1211.3055}}\relax
\mciteBstWouldAddEndPuncttrue
\mciteSetBstMidEndSepPunct{\mcitedefaultmidpunct}
{\mcitedefaultendpunct}{\mcitedefaultseppunct}\relax
\EndOfBibitem
\bibitem{Sjostrand:2007gs}
T.~Sj\"{o}strand, S.~Mrenna, and P.~Skands,
  \ifthenelse{\boolean{articletitles}}{\emph{{A brief introduction to
  {\sc{Pythia\,8.1}}}},
  }{}\href{https://doi.org/10.1016/j.cpc.2008.01.036}{Comput.\ Phys.\ Commun.\
  \textbf{178} (2008) 852},
  \href{http://arxiv.org/abs/0710.3820}{{\normalfont\ttfamily
  arXiv:0710.3820}}\relax
\mciteBstWouldAddEndPuncttrue
\mciteSetBstMidEndSepPunct{\mcitedefaultmidpunct}
{\mcitedefaultendpunct}{\mcitedefaultseppunct}\relax
\EndOfBibitem
\bibitem{LHCb-PROC-2010-056}
I.~Belyaev {\em et~al.}, \ifthenelse{\boolean{articletitles}}{\emph{{Handling
  of the generation of primary events in {\sc{Gauss}}, the~LHCb simulation
  framework}}, }{}\href{https://doi.org/10.1088/1742-6596/331/3/032047}{J.\
  Phys.\ Conf.\ Ser.\  \textbf{331} (2011) 032047}\relax
\mciteBstWouldAddEndPuncttrue
\mciteSetBstMidEndSepPunct{\mcitedefaultmidpunct}
{\mcitedefaultendpunct}{\mcitedefaultseppunct}\relax
\EndOfBibitem
\bibitem{Lange:2001uf}
D.~J. Lange, \ifthenelse{\boolean{articletitles}}{\emph{{The~{\sc{EvtGen}}
  particle decay simulation package}},
  }{}\href{https://doi.org/10.1016/S0168-9002(01)00089-4}{Nucl.\ Instrum.\
  Meth.\  \textbf{A462} (2001) 152}\relax
\mciteBstWouldAddEndPuncttrue
\mciteSetBstMidEndSepPunct{\mcitedefaultmidpunct}
{\mcitedefaultendpunct}{\mcitedefaultseppunct}\relax
\EndOfBibitem
\bibitem{Golonka:2005pn}
P.~Golonka and Z.~Was,
  \ifthenelse{\boolean{articletitles}}{\emph{{{\sc{Photos}} Monte Carlo: A
  precision tool for QED corrections in $\Z$~and $\W$~decays}},
  }{}\href{https://doi.org/10.1140/epjc/s2005-02396-4}{Eur.\ Phys.\ J.\
  \textbf{C45} (2006) 97},
  \href{http://arxiv.org/abs/hep-ph/0506026}{{\normalfont\ttfamily
  arXiv:hep-ph/0506026}}\relax
\mciteBstWouldAddEndPuncttrue
\mciteSetBstMidEndSepPunct{\mcitedefaultmidpunct}
{\mcitedefaultendpunct}{\mcitedefaultseppunct}\relax
\EndOfBibitem
\bibitem{LHCb-PAPER-2015-015}
LHCb collaboration, R.~Aaij {\em et~al.},
  \ifthenelse{\boolean{articletitles}}{\emph{{Quantum numbers of
  the~$\PX(3872)$ state and orbital angular momentum in its $\rhoz\jpsi$
  decays}}, }{}\href{https://doi.org/10.1103/PhysRevD.92.011102}{Phys.\ Rev.\
  \textbf{D92} (2015) 011102(R)},
  \href{http://arxiv.org/abs/1504.06339}{{\normalfont\ttfamily
  arXiv:1504.06339}}\relax
\mciteBstWouldAddEndPuncttrue
\mciteSetBstMidEndSepPunct{\mcitedefaultmidpunct}
{\mcitedefaultendpunct}{\mcitedefaultseppunct}\relax
\EndOfBibitem
\bibitem{Gottfried:1977gp}
K.~Gottfried, \ifthenelse{\boolean{articletitles}}{\emph{{Hadronic transitions
  between quark\nobreakdash-antiquark bound states}},
  }{}\href{https://doi.org/10.1103/PhysRevLett.40.598}{Phys.\ Rev.\ Lett.\
  \textbf{40} (1978) 598}\relax
\mciteBstWouldAddEndPuncttrue
\mciteSetBstMidEndSepPunct{\mcitedefaultmidpunct}
{\mcitedefaultendpunct}{\mcitedefaultseppunct}\relax
\EndOfBibitem
\bibitem{Voloshin:1978hc}
M.~B. Voloshin, \ifthenelse{\boolean{articletitles}}{\emph{{On dynamics of
  heavy quarks in non-perturbative QCD vacuum}},
  }{}\href{https://doi.org/10.1016/0550-3213(79)90037-3}{Nucl.\ Phys.\
  \textbf{B154} (1979) 365}\relax
\mciteBstWouldAddEndPuncttrue
\mciteSetBstMidEndSepPunct{\mcitedefaultmidpunct}
{\mcitedefaultendpunct}{\mcitedefaultseppunct}\relax
\EndOfBibitem
\bibitem{Peskin:1979va}
M.~E. Peskin, \ifthenelse{\boolean{articletitles}}{\emph{{Short-distance
  analysis for heavy-quark systems: (I).\,Diagrammatics}},
  }{}\href{https://doi.org/10.1016/0550-3213(79)90199-8}{Nucl.\ Phys.\
  \textbf{B156} (1979) 365}\relax
\mciteBstWouldAddEndPuncttrue
\mciteSetBstMidEndSepPunct{\mcitedefaultmidpunct}
{\mcitedefaultendpunct}{\mcitedefaultseppunct}\relax
\EndOfBibitem
\bibitem{Bhanot:1979vb}
G.~Bhanot and M.~E. Peskin,
  \ifthenelse{\boolean{articletitles}}{\emph{{Short-distance analysis for
  heavy-quark systems: (II).\,Applications}},
  }{}\href{https://doi.org/10.1016/0550-3213(79)90200-1}{Nucl.\ Phys.\
  \textbf{B156} (1979) 391}\relax
\mciteBstWouldAddEndPuncttrue
\mciteSetBstMidEndSepPunct{\mcitedefaultmidpunct}
{\mcitedefaultendpunct}{\mcitedefaultseppunct}\relax
\EndOfBibitem
\bibitem{Allison:2006ve}
Geant4 collaboration, J.~Allison {\em et~al.},
  \ifthenelse{\boolean{articletitles}}{\emph{{{\sc{Geant4}} developments and
  applications}}, }{}\href{https://doi.org/10.1109/TNS.2006.869826}{IEEE
  Trans.\ Nucl.\ Sci.\  \textbf{53} (2006) 270}\relax
\mciteBstWouldAddEndPuncttrue
\mciteSetBstMidEndSepPunct{\mcitedefaultmidpunct}
{\mcitedefaultendpunct}{\mcitedefaultseppunct}\relax
\EndOfBibitem
\bibitem{Agostinelli:2002hh}
Geant4 collaboration, S.~Agostinelli {\em et~al.},
  \ifthenelse{\boolean{articletitles}}{\emph{{{\sc{Geant4}}: A simulation
  toolkit}}, }{}\href{https://doi.org/10.1016/S0168-9002(03)01368-8}{Nucl.\
  Instrum.\ Meth.\  \textbf{A506} (2003) 250}\relax
\mciteBstWouldAddEndPuncttrue
\mciteSetBstMidEndSepPunct{\mcitedefaultmidpunct}
{\mcitedefaultendpunct}{\mcitedefaultseppunct}\relax
\EndOfBibitem
\bibitem{LHCb-PROC-2011-006}
M.~Clemencic {\em et~al.}, \ifthenelse{\boolean{articletitles}}{\emph{{The
  \lhcb simulation application, {\sc{Gauss}}: Design, evolution and
  experience}}, }{}\href{https://doi.org/10.1088/1742-6596/331/3/032023}{J.\
  Phys.\ Conf.\ Ser.\  \textbf{331} (2011) 032023}\relax
\mciteBstWouldAddEndPuncttrue
\mciteSetBstMidEndSepPunct{\mcitedefaultmidpunct}
{\mcitedefaultendpunct}{\mcitedefaultseppunct}\relax
\EndOfBibitem
\bibitem{LHCb-DP-2013-002}
LHCb collaboration, R.~Aaij {\em et~al.},
  \ifthenelse{\boolean{articletitles}}{\emph{{Measurement of the track
  reconstruction efficiency at LHCb}},
  }{}\href{https://doi.org/10.1088/1748-0221/10/02/P02007}{JINST \textbf{10}
  (2015) P02007}, \href{http://arxiv.org/abs/1408.1251}{{\normalfont\ttfamily
  arXiv:1408.1251}}\relax
\mciteBstWouldAddEndPuncttrue
\mciteSetBstMidEndSepPunct{\mcitedefaultmidpunct}
{\mcitedefaultendpunct}{\mcitedefaultseppunct}\relax
\EndOfBibitem
\bibitem{LHCb-PAPER-2013-047}
LHCb collaboration, R.~Aaij {\em et~al.},
  \ifthenelse{\boolean{articletitles}}{\emph{{Observation of the decay
  \mbox{\decay{\Bc}{\jpsi\Kp\Km\pip}}}},
  }{}\href{https://doi.org/10.1007/JHEP11(2013)094}{JHEP \textbf{11} (2013)
  094}, \href{http://arxiv.org/abs/1309.0587}{{\normalfont\ttfamily
  arXiv:1309.0587}}\relax
\mciteBstWouldAddEndPuncttrue
\mciteSetBstMidEndSepPunct{\mcitedefaultmidpunct}
{\mcitedefaultendpunct}{\mcitedefaultseppunct}\relax
\EndOfBibitem
\bibitem{LHcb-PAPER-2015-060}
LHCb collaboration, R.~Aaij {\em et~al.},
  \ifthenelse{\boolean{articletitles}}{\emph{{Observation of
  \mbox{\decay{\Lb}{\psitwos\proton\Km}} and
  \mbox{\decay{\Lb}{\jpsi\pip\pim\proton\Km}} decays and a measurement of the
  \Lb baryon mass}}, }{}\href{https://doi.org/10.1007/JHEP05(2016)132}{JHEP
  \textbf{05} (2016) 132},
  \href{http://arxiv.org/abs/1603.06961}{{\normalfont\ttfamily
  arXiv:1603.06961}}\relax
\mciteBstWouldAddEndPuncttrue
\mciteSetBstMidEndSepPunct{\mcitedefaultmidpunct}
{\mcitedefaultendpunct}{\mcitedefaultseppunct}\relax
\EndOfBibitem
\bibitem{LHCb-PAPER-2019-023}
LHCb collaboration, R.~Aaij {\em et~al.},
  \ifthenelse{\boolean{articletitles}}{\emph{{Observation of the
  \mbox{\decay{\Lb}{\Pchi_{\cquark1}(3872)\proton\Km}} decay}},
  }{}\href{https://doi.org/10.1007/JHEP09(2019)028}{JHEP \textbf{09} (2019)
  028}, \href{http://arxiv.org/abs/1907.00954}{{\normalfont\ttfamily
  arXiv:1907.00954}}\relax
\mciteBstWouldAddEndPuncttrue
\mciteSetBstMidEndSepPunct{\mcitedefaultmidpunct}
{\mcitedefaultendpunct}{\mcitedefaultseppunct}\relax
\EndOfBibitem
\bibitem{LHCb-PROC-2011-008}
A.~Powell {\em et~al.}, \ifthenelse{\boolean{articletitles}}{\emph{{Particle
  identification at LHCb}}, }{}PoS \textbf{ICHEP2010} (2010) 020,
  \href{https://cdsweb.cern.ch/record/1322666?ln=en}{LHCb-PROC-2011-008}\relax
\mciteBstWouldAddEndPuncttrue
\mciteSetBstMidEndSepPunct{\mcitedefaultmidpunct}
{\mcitedefaultendpunct}{\mcitedefaultseppunct}\relax
\EndOfBibitem
\bibitem{LHCb-PAPER-2011-013}
LHCb collaboration, R.~Aaij {\em et~al.},
  \ifthenelse{\boolean{articletitles}}{\emph{{Observation of \jpsi-pair
  production in \proton\proton collisions at \mbox{$\sqs=$7\tev}}},
  }{}\href{https://doi.org/10.1016/j.physletb.2011.12.015}{Phys.\ Lett.\
  \textbf{B707} (2012) 52},
  \href{http://arxiv.org/abs/1109.0963}{{\normalfont\ttfamily
  arXiv:1109.0963}}\relax
\mciteBstWouldAddEndPuncttrue
\mciteSetBstMidEndSepPunct{\mcitedefaultmidpunct}
{\mcitedefaultendpunct}{\mcitedefaultseppunct}\relax
\EndOfBibitem
\bibitem{Skwarnicki:1986xj}
T.~Skwarnicki, {\em {A study of the radiative cascade transitions between
  the~$\PUpsilon^{\prime}$ and $\PUpsilon$~resonances}}, PhD thesis, Institute
  of Nuclear Physics, Krakow, 1986,
  {\href{http://inspirehep.net/record/230779/}{DESY-F31-86-02}}\relax
\mciteBstWouldAddEndPuncttrue
\mciteSetBstMidEndSepPunct{\mcitedefaultmidpunct}
{\mcitedefaultendpunct}{\mcitedefaultseppunct}\relax
\EndOfBibitem
\bibitem{Hulsbergen:2005pu}
W.~D. Hulsbergen, \ifthenelse{\boolean{articletitles}}{\emph{{Decay chain
  fitting with a Kalman filter}},
  }{}\href{https://doi.org/10.1016/j.nima.2005.06.078}{Nucl.\ Instrum.\ Meth.\
  \textbf{A552} (2005) 566},
  \href{http://arxiv.org/abs/physics/0503191}{{\normalfont\ttfamily
  arXiv:physics/0503191}}\relax
\mciteBstWouldAddEndPuncttrue
\mciteSetBstMidEndSepPunct{\mcitedefaultmidpunct}
{\mcitedefaultendpunct}{\mcitedefaultseppunct}\relax
\EndOfBibitem
\bibitem{Hanhart:2007yq}
C.~Hanhart, Y.~S. Kalashnikova, A.~E. Kudryavtsev, and A.~V. Nefediev,
  \ifthenelse{\boolean{articletitles}}{\emph{{Reconciling
  the~\mbox{$\PX(3872)$} with the~near\nobreakdash-threshold enhancement in
  the~\mbox{$\Dz\Dstarzb$}~final state}},
  }{}\href{https://doi.org/10.1103/PhysRevD.76.034007}{Phys.\ Rev.\
  \textbf{D76} (2007) 034007},
  \href{http://arxiv.org/abs/0704.0605}{{\normalfont\ttfamily
  arXiv:0704.0605}}\relax
\mciteBstWouldAddEndPuncttrue
\mciteSetBstMidEndSepPunct{\mcitedefaultmidpunct}
{\mcitedefaultendpunct}{\mcitedefaultseppunct}\relax
\EndOfBibitem
\bibitem{Stapleton:2009ey}
E.~Braaten and J.~Stapleton,
  \ifthenelse{\boolean{articletitles}}{\emph{{Analysis of
  \mbox{$\jpsi\pip\pim$} and \mbox{$\Dz\Dzb\piz$}~decays of
  the~\mbox{$\PX(3872)$}}},
  }{}\href{https://doi.org/10.1103/PhysRevD.81.014019}{Phys.\ Rev.\
  \textbf{D81} (2010) 014019},
  \href{http://arxiv.org/abs/0907.3167}{{\normalfont\ttfamily
  arXiv:0907.3167}}\relax
\mciteBstWouldAddEndPuncttrue
\mciteSetBstMidEndSepPunct{\mcitedefaultmidpunct}
{\mcitedefaultendpunct}{\mcitedefaultseppunct}\relax
\EndOfBibitem
\bibitem{Kalashnikova:2009gt}
Y.~S. Kalashnikova and A.~V. Nefediev,
  \ifthenelse{\boolean{articletitles}}{\emph{{Nature of \mbox{$\PX(3872)$} from
  data}}, }{}\href{https://doi.org/10.1103/PhysRevD.80.074004}{Phys.\ Rev.\
  \textbf{D80} (2009) 074004},
  \href{http://arxiv.org/abs/0907.4901}{{\normalfont\ttfamily
  arXiv:0907.4901}}\relax
\mciteBstWouldAddEndPuncttrue
\mciteSetBstMidEndSepPunct{\mcitedefaultmidpunct}
{\mcitedefaultendpunct}{\mcitedefaultseppunct}\relax
\EndOfBibitem
\bibitem{Hanhart:2011jz}
C.~Hanhart, Y.~S. Kalashnikova, and A.~V. Nefediev,
  \ifthenelse{\boolean{articletitles}}{\emph{{Interplay of quark and meson
  degrees of freedom in a~near\nobreakdash-threshold resonance:
  multi\nobreakdash-channel case}},
  }{}\href{https://doi.org/10.1140/epja/i2011-11101-9}{Eur.\ Phys.\ J.\
  \textbf{A47} (2011) 101},
  \href{http://arxiv.org/abs/1106.1185}{{\normalfont\ttfamily
  arXiv:1106.1185}}\relax
\mciteBstWouldAddEndPuncttrue
\mciteSetBstMidEndSepPunct{\mcitedefaultmidpunct}
{\mcitedefaultendpunct}{\mcitedefaultseppunct}\relax
\EndOfBibitem
\bibitem{LHCb-PAPER-2020-008}
LHCb collaboration, R.~Aaij {\em et~al.},
  \ifthenelse{\boolean{articletitles}}{\emph{{Study of the line shape of the
  $\chicone(3872)$ meson}},
  }{}\href{https://doi.org/10.1103/PhysRevD.102.092005}{Phys.\ Rev.\
  \textbf{D102} (2020) 092005},
  \href{http://arxiv.org/abs/2005.13419}{{\normalfont\ttfamily
  arXiv:2005.13419}}\relax
\mciteBstWouldAddEndPuncttrue
\mciteSetBstMidEndSepPunct{\mcitedefaultmidpunct}
{\mcitedefaultendpunct}{\mcitedefaultseppunct}\relax
\EndOfBibitem
\bibitem{Byckling}
E.~Byckling and K.~Kajantie, {\em Particle kinematics}, John Wiley \& Sons
  Inc., New York, 1973\relax
\mciteBstWouldAddEndPuncttrue
\mciteSetBstMidEndSepPunct{\mcitedefaultmidpunct}
{\mcitedefaultendpunct}{\mcitedefaultseppunct}\relax
\EndOfBibitem
\bibitem{Wilks:1938dza}
S.~S. Wilks, \ifthenelse{\boolean{articletitles}}{\emph{{The large-sample
  distribution of the likelihood ratio for testing composite hypotheses}},
  }{}\href{https://doi.org/10.1214/aoms/1177732360}{Ann.\ Math.\ Stat.\
  \textbf{9} (1938) 60}\relax
\mciteBstWouldAddEndPuncttrue
\mciteSetBstMidEndSepPunct{\mcitedefaultmidpunct}
{\mcitedefaultendpunct}{\mcitedefaultseppunct}\relax
\EndOfBibitem
\bibitem{Pivk:2004ty}
M.~Pivk and F.~R. Le~Diberder,
  \ifthenelse{\boolean{articletitles}}{\emph{{sPlot: A statistical tool to
  unfold data distributions}},
  }{}\href{https://doi.org/10.1016/j.nima.2005.08.106}{Nucl.\ Instrum.\ Meth.\
  \textbf{A555} (2005) 356},
  \href{http://arxiv.org/abs/physics/0402083}{{\normalfont\ttfamily
  arXiv:physics/0402083}}\relax
\mciteBstWouldAddEndPuncttrue
\mciteSetBstMidEndSepPunct{\mcitedefaultmidpunct}
{\mcitedefaultendpunct}{\mcitedefaultseppunct}\relax
\EndOfBibitem
\bibitem{LHCb-DP-2018-001}
R.~Aaij {\em et~al.}, \ifthenelse{\boolean{articletitles}}{\emph{{Selection and
  processing of calibration samples to measure the particle identification
  performance of the LHCb experiment in Run 2}},
  }{}\href{https://doi.org/10.1140/epjti/s40485-019-0050-z}{Eur.\ Phys.\ J.\
  Tech.\ Instr.\  \textbf{6} (2018) 1},
  \href{http://arxiv.org/abs/1803.00824}{{\normalfont\ttfamily
  arXiv:1803.00824}}\relax
\mciteBstWouldAddEndPuncttrue
\mciteSetBstMidEndSepPunct{\mcitedefaultmidpunct}
{\mcitedefaultendpunct}{\mcitedefaultseppunct}\relax
\EndOfBibitem
\bibitem{LHCb-PAPER-2012-040}
LHCb collaboration, R.~Aaij {\em et~al.},
  \ifthenelse{\boolean{articletitles}}{\emph{{Amplitude analysis and branching
  fraction measurement of \mbox{\decay{\Bsb}{\jpsi\Kp\Km}}}},
  }{}\href{https://doi.org/10.1103/PhysRevD.87.072004}{Phys.\ Rev.\
  \textbf{D87} (2013) 072004},
  \href{http://arxiv.org/abs/1302.1213}{{\normalfont\ttfamily
  arXiv:1302.1213}}\relax
\mciteBstWouldAddEndPuncttrue
\mciteSetBstMidEndSepPunct{\mcitedefaultmidpunct}
{\mcitedefaultendpunct}{\mcitedefaultseppunct}\relax
\EndOfBibitem
\bibitem{LHCb-PAPER-2017-008}
LHCb collaboration, R.~Aaij {\em et~al.},
  \ifthenelse{\boolean{articletitles}}{\emph{{Resonances and \CP-violation in
  \Bs and \mbox{\decay{\Bsb}{\jpsi\Kp\Km}} decays in the mass region above the
  $\phiz(1020)$}}, }{}\href{https://doi.org/10.1007/JHEP08(2017)037}{JHEP
  \textbf{08} (2017) 037},
  \href{http://arxiv.org/abs/1704.08217}{{\normalfont\ttfamily
  arXiv:1704.08217}}\relax
\mciteBstWouldAddEndPuncttrue
\mciteSetBstMidEndSepPunct{\mcitedefaultmidpunct}
{\mcitedefaultendpunct}{\mcitedefaultseppunct}\relax
\EndOfBibitem
\bibitem{Anashin:2015rca}
KEDR collaboration, V.~V. Anashin {\em et~al.},
  \ifthenelse{\boolean{articletitles}}{\emph{{Final analysis of KEDR data on
  $\jpsi$ and $\psitwos$~masses}},
  }{}\href{https://doi.org/10.1016/j.physletb.2015.07.057}{Phys.\ Lett.\
  \textbf{B749} (2015) 50}\relax
\mciteBstWouldAddEndPuncttrue
\mciteSetBstMidEndSepPunct{\mcitedefaultmidpunct}
{\mcitedefaultendpunct}{\mcitedefaultseppunct}\relax
\EndOfBibitem
\bibitem{LHCb-PAPER-2014-002}
LHCb collaboration, R.~Aaij {\em et~al.},
  \ifthenelse{\boolean{articletitles}}{\emph{{Study of beauty hadron decays
  into pairs of charm hadrons}},
  }{}\href{https://doi.org/10.1103/PhysRevLett.112.202001}{Phys.\ Rev.\ Lett.\
  \textbf{112} (2014) 202001},
  \href{http://arxiv.org/abs/1403.3606}{{\normalfont\ttfamily
  arXiv:1403.3606}}\relax
\mciteBstWouldAddEndPuncttrue
\mciteSetBstMidEndSepPunct{\mcitedefaultmidpunct}
{\mcitedefaultendpunct}{\mcitedefaultseppunct}\relax
\EndOfBibitem
\bibitem{LHCb-PAPER-2019-037}
LHCb collaboration, R.~Aaij {\em et~al.},
  \ifthenelse{\boolean{articletitles}}{\emph{{Precision measurement of the
  $\Xiccpp$ mass}}, }{}\href{https://doi.org/10.1007/JHEP02(2020)049}{JHEP
  \textbf{02} (2020) 049},
  \href{http://arxiv.org/abs/1911.08594}{{\normalfont\ttfamily
  arXiv:1911.08594}}\relax
\mciteBstWouldAddEndPuncttrue
\mciteSetBstMidEndSepPunct{\mcitedefaultmidpunct}
{\mcitedefaultendpunct}{\mcitedefaultseppunct}\relax
\EndOfBibitem
\bibitem{LHcb-PAPER-2020-003}
LHCb collaboration, R.~Aaij {\em et~al.},
  \ifthenelse{\boolean{articletitles}}{\emph{{Precision measurement of the \Bc
  meson mass}}, }{}\href{https://doi.org/10.1007/JHEP07(2020)123}{JHEP
  \textbf{07} (2020) 123},
  \href{http://arxiv.org/abs/2004.08163}{{\normalfont\ttfamily
  arXiv:2004.08163}}\relax
\mciteBstWouldAddEndPuncttrue
\mciteSetBstMidEndSepPunct{\mcitedefaultmidpunct}
{\mcitedefaultendpunct}{\mcitedefaultseppunct}\relax
\EndOfBibitem
\bibitem{Ebert:2008kb}
D.~Ebert, R.~N. Faustov, and V.~O. Galkin,
  \ifthenelse{\boolean{articletitles}}{\emph{{Excited heavy tetraquarks with
  hidden charm}},
  }{}\href{https://doi.org/10.1140/epjc/s10052-008-0754-8}{Eur.\ Phys.\ J.\
  \textbf{C58} (2008) 399},
  \href{http://arxiv.org/abs/0808.3912}{{\normalfont\ttfamily
  arXiv:0808.3912}}\relax
\mciteBstWouldAddEndPuncttrue
\mciteSetBstMidEndSepPunct{\mcitedefaultmidpunct}
{\mcitedefaultendpunct}{\mcitedefaultseppunct}\relax
\EndOfBibitem
\bibitem{Blatt:1952ije}
J.~M. Blatt and V.~F. Weisskopf, {\em {Theoretical nuclear physics}},
  \href{https://doi.org/10.1007/978-1-4612-9959-2}{ Springer, New York,
  1952}\relax
\mciteBstWouldAddEndPuncttrue
\mciteSetBstMidEndSepPunct{\mcitedefaultmidpunct}
{\mcitedefaultendpunct}{\mcitedefaultseppunct}\relax
\EndOfBibitem
\bibitem{Flatte:1976xu}
S.~M. Flatt\'e, \ifthenelse{\boolean{articletitles}}{\emph{{Coupled-channel
  analysis of the~\mbox{$\Ppi\Peta$} and \mbox{$\kaon\bar{\kaon}$}~systems near
  \mbox{$\kaon\bar{\kaon}$}~threshold}},
  }{}\href{https://doi.org/10.1016/0370-2693(76)90654-7}{Phys.\ Lett.\
  \textbf{B63} (1976) 224}\relax
\mciteBstWouldAddEndPuncttrue
\mciteSetBstMidEndSepPunct{\mcitedefaultmidpunct}
{\mcitedefaultendpunct}{\mcitedefaultseppunct}\relax
\EndOfBibitem
\bibitem{Ablikim:2004wn}
BES collaboration, M.~Ablikim {\em et~al.},
  \ifthenelse{\boolean{articletitles}}{\emph{{Resonances in
  \mbox{$\decay{\jpsi}{\Pphi\pip\pim}$} and \mbox{$\Pphi\Kp\Km$} }},
  }{}\href{https://doi.org/10.1016/j.physletb.2004.12.041}{Phys.\ Lett.\
  \textbf{B607} (2005) 243},
  \href{http://arxiv.org/abs/hep-ex/0411001}{{\normalfont\ttfamily
  arXiv:hep-ex/0411001}}\relax
\mciteBstWouldAddEndPuncttrue
\mciteSetBstMidEndSepPunct{\mcitedefaultmidpunct}
{\mcitedefaultendpunct}{\mcitedefaultseppunct}\relax
\EndOfBibitem
\bibitem{LHCb-PAPER-2011-002}
LHCb collaboration, R.~Aaij {\em et~al.},
  \ifthenelse{\boolean{articletitles}}{\emph{{First observation of
  \mbox{\decay{\Bs}{\jpsi {\mathrm{f}}_0(980)}} decays}},
  }{}\href{https://doi.org/10.1016/j.physletb.2011.03.006}{Phys.\ Lett.\
  \textbf{B698} (2011) 115},
  \href{http://arxiv.org/abs/1102.0206}{{\normalfont\ttfamily
  arXiv:1102.0206}}\relax
\mciteBstWouldAddEndPuncttrue
\mciteSetBstMidEndSepPunct{\mcitedefaultmidpunct}
{\mcitedefaultendpunct}{\mcitedefaultseppunct}\relax
\EndOfBibitem
\bibitem{LHCb-PAPER-2012-010}
LHCb collaboration, R.~Aaij {\em et~al.},
  \ifthenelse{\boolean{articletitles}}{\emph{{Measurement of relative branching
  fractions of \B~decays to \psitwos and \jpsi~mesons}},
  }{}\href{https://doi.org/10.1140/epjc/s10052-012-2118-7}{Eur.\ Phys.\ J.\
  \textbf{C72} (2012) 2118},
  \href{http://arxiv.org/abs/1205.0918}{{\normalfont\ttfamily
  arXiv:1205.0918}}\relax
\mciteBstWouldAddEndPuncttrue
\mciteSetBstMidEndSepPunct{\mcitedefaultmidpunct}
{\mcitedefaultendpunct}{\mcitedefaultseppunct}\relax
\EndOfBibitem
\bibitem{DeBruyn:2012wj}
K.~De~Bruyn {\em et~al.}, \ifthenelse{\boolean{articletitles}}{\emph{{Branching
  ratio measurements of \Bs~decays}},
  }{}\href{https://doi.org/10.1103/PhysRevD.86.014027}{Phys.\ Rev.\
  \textbf{D86} (2012) 014027},
  \href{http://arxiv.org/abs/1204.1735}{{\normalfont\ttfamily
  arXiv:1204.1735}}\relax
\mciteBstWouldAddEndPuncttrue
\mciteSetBstMidEndSepPunct{\mcitedefaultmidpunct}
{\mcitedefaultendpunct}{\mcitedefaultseppunct}\relax
\EndOfBibitem
\bibitem{LHCb-PAPER-2011-035}
LHCb collaboration, R.~Aaij {\em et~al.},
  \ifthenelse{\boolean{articletitles}}{\emph{{Measurement of \bquark-hadron
  masses}}, }{}\href{https://doi.org/10.1016/j.physletb.2012.01.058}{Phys.\
  Lett.\  \textbf{B708} (2012) 241},
  \href{http://arxiv.org/abs/1112.4896}{{\normalfont\ttfamily
  arXiv:1112.4896}}\relax
\mciteBstWouldAddEndPuncttrue
\mciteSetBstMidEndSepPunct{\mcitedefaultmidpunct}
{\mcitedefaultendpunct}{\mcitedefaultseppunct}\relax
\EndOfBibitem
\bibitem{LHCb-PAPER-2018-018}
LHCb collaboration, R.~Aaij {\em et~al.},
  \ifthenelse{\boolean{articletitles}}{\emph{{Observation of the decay
  \mbox{\decay{\Bsb}{\Pchi_{\cquark2}\Kp\Km}}}},
  }{}\href{https://doi.org/10.1007/JHEP08(2018)191}{JHEP \textbf{08} (2018)
  191}, \href{http://arxiv.org/abs/1806.10576}{{\normalfont\ttfamily
  arXiv:1806.10576}}\relax
\mciteBstWouldAddEndPuncttrue
\mciteSetBstMidEndSepPunct{\mcitedefaultmidpunct}
{\mcitedefaultendpunct}{\mcitedefaultseppunct}\relax
\EndOfBibitem
\bibitem{LHCb-PAPER-2018-046}
LHCb collaboration, R.~Aaij {\em et~al.},
  \ifthenelse{\boolean{articletitles}}{\emph{{Observation of
  \mbox{\decay{\BdorBs}{\jpsi \proton\antiproton}} decays and precision
  measurements of the \BdorBs masses}},
  }{}\href{https://doi.org/10.1103/PhysRevLett.122.191804}{Phys.\ Rev.\ Lett.\
  \textbf{122} (2019) 191804},
  \href{http://arxiv.org/abs/1902.05588}{{\normalfont\ttfamily
  arXiv:1902.05588}}\relax
\mciteBstWouldAddEndPuncttrue
\mciteSetBstMidEndSepPunct{\mcitedefaultmidpunct}
{\mcitedefaultendpunct}{\mcitedefaultseppunct}\relax
\EndOfBibitem
\bibitem{Lyons:1988rp}
L.~Lyons, D.~Gibaut, and P.~Clifford,
  \ifthenelse{\boolean{articletitles}}{\emph{{How to combine correlated
  estimates of a~single physical quantity}},
  }{}\href{https://doi.org/10.1016/0168-9002(88)90018-6}{Nucl.\ Instrum.\
  Meth.\  \textbf{A270} (1988) 110}\relax
\mciteBstWouldAddEndPuncttrue
\mciteSetBstMidEndSepPunct{\mcitedefaultmidpunct}
{\mcitedefaultendpunct}{\mcitedefaultseppunct}\relax
\EndOfBibitem
\end{mcitethebibliography}
